\documentclass[12pt]{article}

\usepackage{verbatim,amssymb}
\usepackage{xcolor}
\usepackage{amsmath}					
\usepackage{amsthm}					
\usepackage{natbib}
\usepackage{bibunits}  
\usepackage{multirow}
\usepackage{setspace}
\usepackage[mathscr]{euscript}
\usepackage{fancyhdr}
\usepackage{enumitem}
\usepackage{graphicx}
\usepackage{lineno}
\usepackage{hyperref}
\usepackage{array,booktabs}
\usepackage{subcaption}
\usepackage{arydshln}  
\usepackage{rotating}   

\usepackage[section]{placeins}  

\usepackage{lscape}

\usepackage{setspace}
\usepackage{tikz}
\usetikzlibrary{arrows}
\usepackage{multirow}
\usepackage{wrapfig}

\usepackage{Sarkar_macros}

\allowdisplaybreaks

\begin{document}
\thispagestyle{empty}
\baselineskip=28pt

\begin{center}
    {\LARGE{\bf Bayesian Semiparametric\\ 
    \vspace{-1ex} Multivariate Density Regression\\ with Coordinate-Wise Predictor Selection}}
\end{center}
\baselineskip=12pt

\vskip 2mm
\begin{center}
    Giovanni Toto$^{1}$, Peter M\"uller$^{1,2}$, and Abhra Sarkar$^{1}$\\[.2cm]
    giovanni.toto@austin.utexas.edu, pmueller@math.utexas.edu, and abhra.sarkar@utexas.edu \\[.2cm]
    $^{1}$Department of Statistics and Data Sciences,
    The University of Texas at Austin\\
    $^{2}$Department of Mathematics, The University of Texas at Austin\\
\end{center}

\vskip 8mm
\begin{center}
{\Large{\bf Abstract}} 
\end{center}
We propose a flexible Bayesian approach for estimating the joint density of a multivariate random variable of interest in the presence of categorical covariates. 
Leveraging a Gaussian copula framework, our method captures the dependence structure across different coordinates of the multivariate response. 
The conditional (on covariates) marginal (across outcomes) distributions are modeled as flexible mixtures with shared atoms across coordinates.
The mixture weights are allowed to vary with covariates through a novel Tucker tensor factorization-based structure which enables the identification of coordinate-specific subsets of influential covariates. 
In particular, we replace the traditional mode matrices with coordinate-specific random partition models on the covariate levels, 
offering a flexible mechanism to aggregate covariate levels that exhibit similar effects on the response. 
We introduce a Markov chain Monte Carlo algorithm that scales with the number of aggregated levels rather than the original levels, significantly reducing memory requirements and improving computational efficiency. 
We demonstrate the method's numerical performance through simulation experiments and its practical applicability through the analysis of NHANES dietary data.

\baselineskip=12pt

\vskip 8mm
\baselineskip=12pt
\noindent\underline{\bf Keywords}: 
Common atoms models, 
Copula, 
Dietary intakes,
Density regression, 
Markov chain Monte Carlo, 
Variable selection, 
Tensor factorization.

\clearpage\pagebreak\newpage
\pagenumbering{arabic}
\newlength{\gnat}
\setlength{\gnat}{26pt}  
\baselineskip=\gnat

\section{Introduction}
\paragraph{Problem Statement.} 
We address the problem of flexibly estimating the joint density of a continuous multivariate random variable in the presence of categorical covariates, allowing each coordinate to be determined by a different subset of covariates, while also facilitating information sharing across coordinates and covariate combinations. 
Specifically, for each coordinate of the multivariate response, we infer a partition over the covariate space such that a different conditional marginal density (marginal for each coordinate, conditional on the covariates) is obtained for each cluster rather than for every possible combination of covariate levels.
To achieve this, we leverage the representational power of mixture models and conditional probability tensors, integrated with a copula dependence structure, while addressing significant computational challenges. 
This problem is motivated by real-world applications, as described below. 
While some literature exists on multivariate density estimation and regression, to our knowledge, the specific problem addressed here has not been explored previously.

\paragraph{Motivation.} 
For any (univariate or multivariate) outcome variable in a regression, estimating its entire (marginal, conditional, and joint) distributions, rather than relying on summaries, offers a more comprehensive understanding. 
For instance, two groups of observations may have similar means, suggesting little difference between them. However, their underlying distributions could differ substantially in shape, spread, or modality, revealing important distinctions that summary statistics alone may obscure.

{This issue is particularly salient in our motivating application involving dietary intake data from the National Health and Nutrition Examination Survey (NHANES). NHANES collects detailed information on the amounts of various dietary components consumed by participants over two 24-hour dietary recall interviews. Dietary intake is recorded for multiple components (most continuous), and extensive demographic information (all categorical) accompanies each participant’s record. In such settings, examining the full distribution of dietary intake, rather than focusing solely on averages, can reveal differences in dietary patterns across demographic groups, identify subpopulations with unique consumption profiles, and inform targeted nutritional interventions.} 

\begin{figure}[!ht]
    \centering
    \begin{subfigure}[b]{0.49\textwidth}
        \centering
        \includegraphics[width=\textwidth]{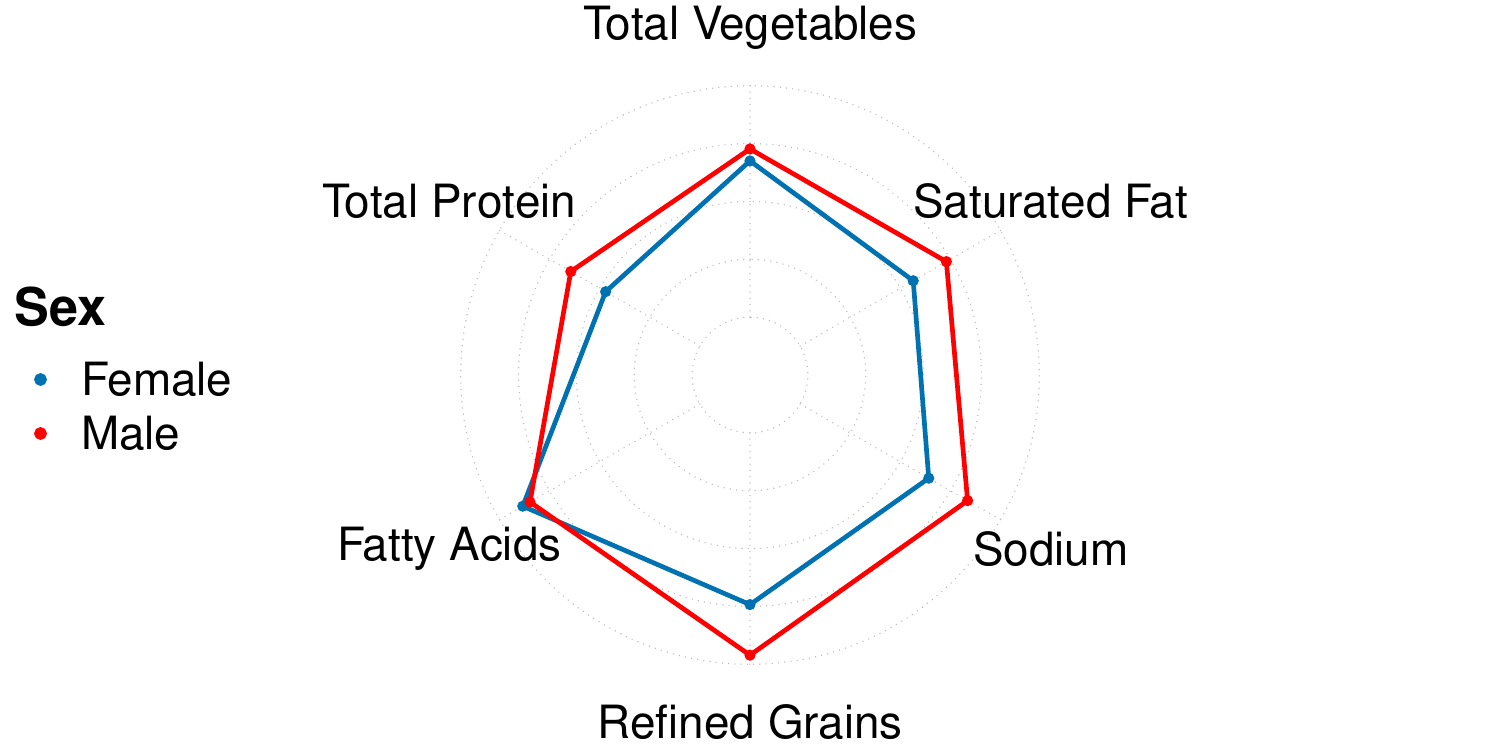}
    \end{subfigure}
    \begin{subfigure}[b]{0.49\textwidth}
        \centering
        \includegraphics[width=\textwidth]{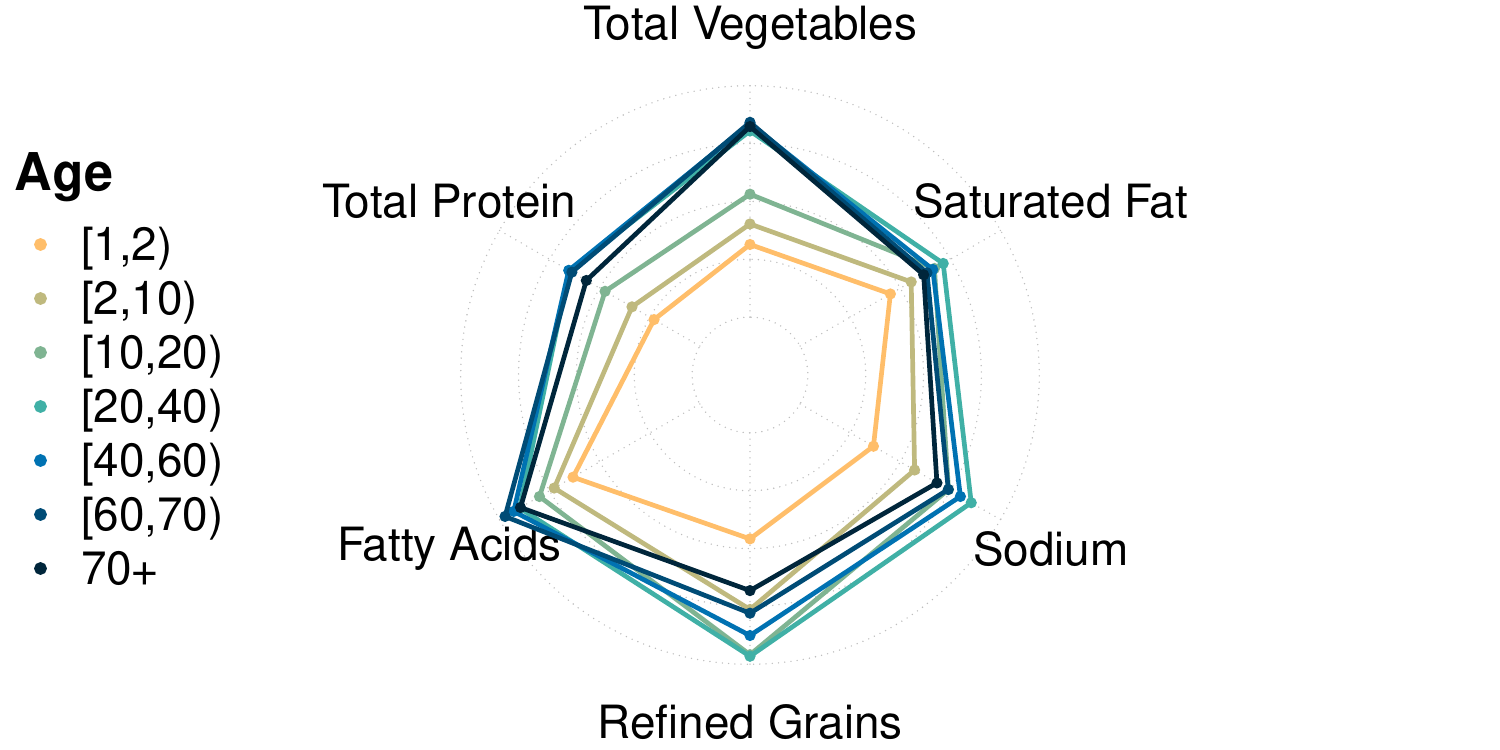}
    \end{subfigure}
    
    \vspace{0.1cm}
    \begin{subfigure}[b]{0.49\textwidth}
        \centering
        \includegraphics[width=\textwidth]{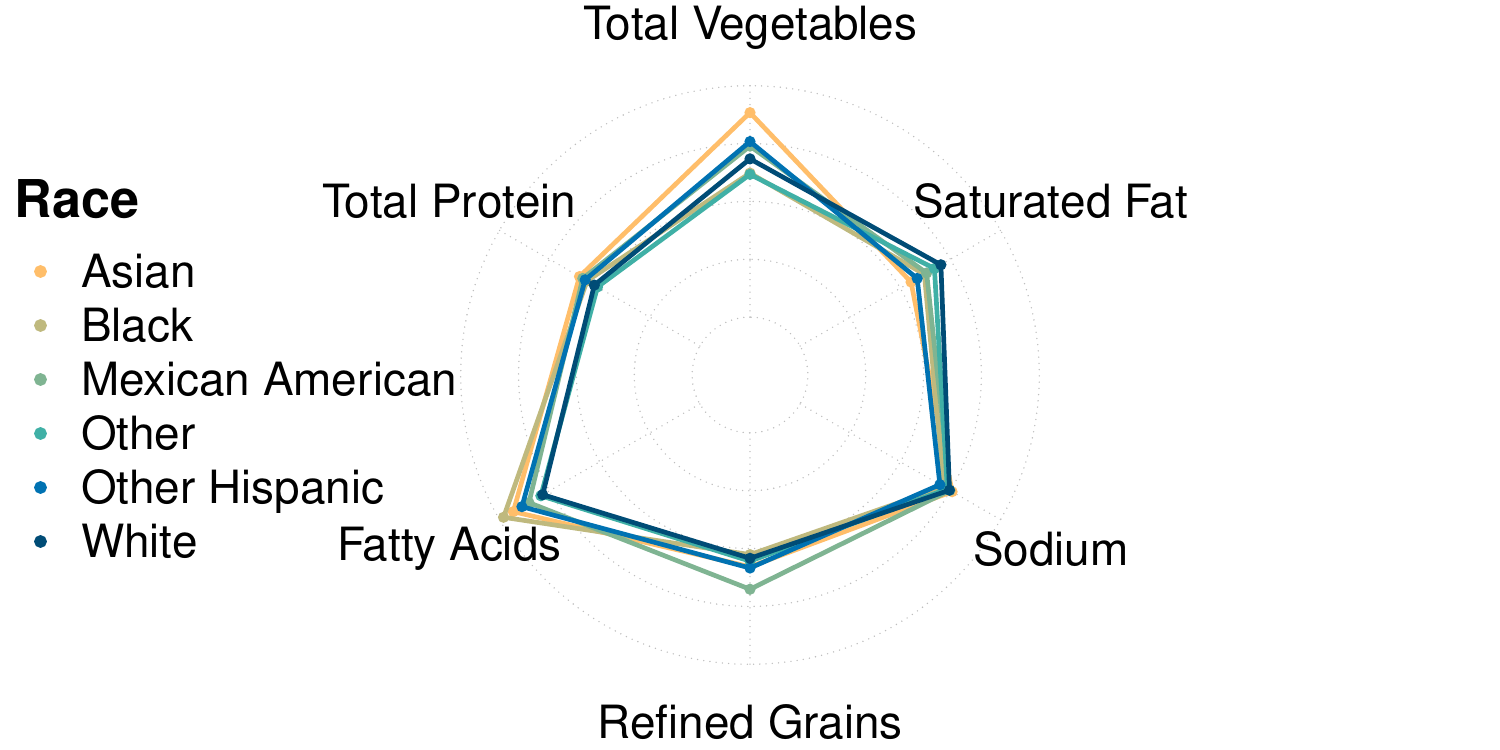}
    \end{subfigure}
    \begin{subfigure}[b]{0.49\textwidth}
        \centering
        \includegraphics[width=\textwidth]{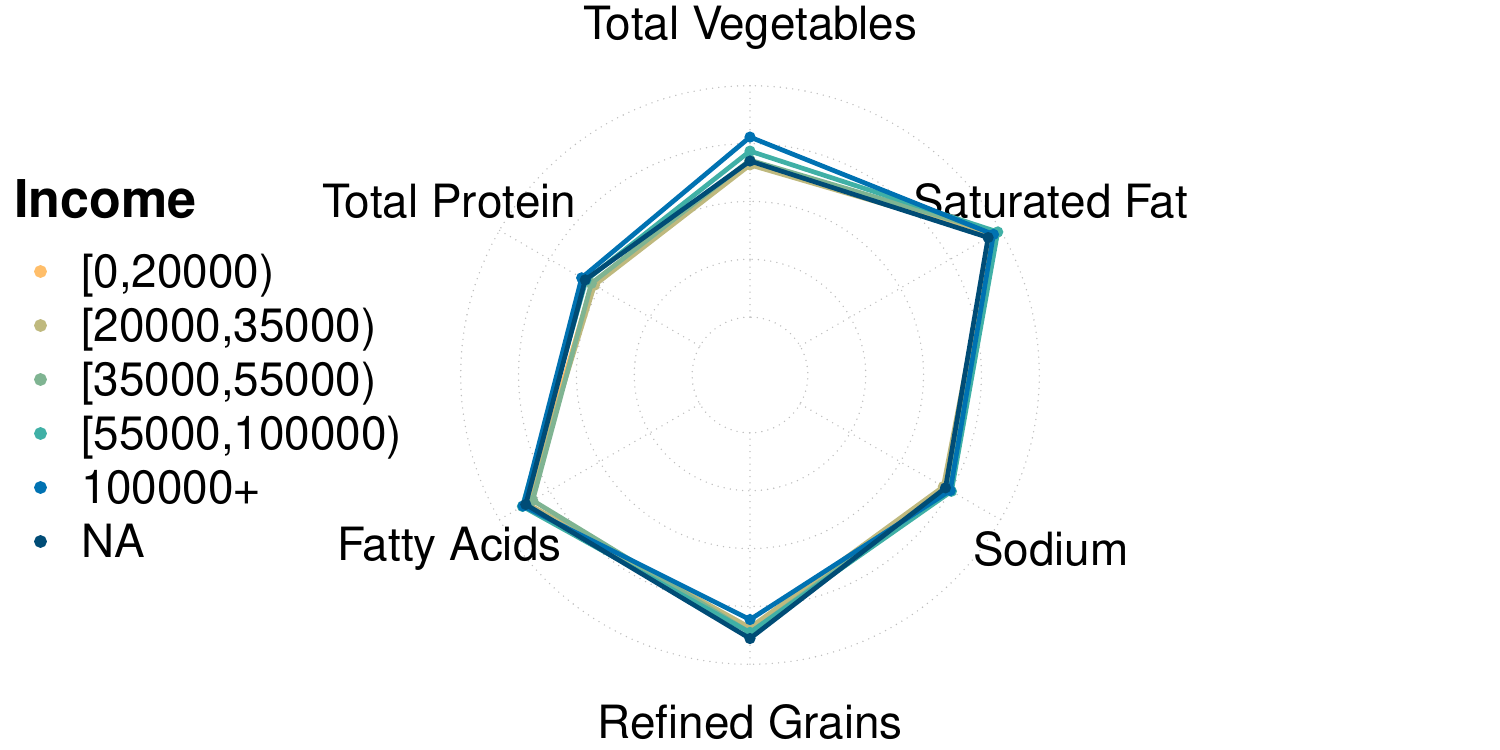}
    \end{subfigure}
    \caption{Radar plots comparing six dietary components across subpopulations, with each panel corresponding to a different stratifying demographic characteristic.}
    \label{fig:nhanes_radar}
\end{figure}

As illustrated in \autoref{fig:nhanes_radar}
, participants from different demographic groups may exhibit similar average intakes for certain dietary components (e.g., sex for Fatty Acids), while showing marked differences for others (e.g., sex for Sodium), highlighting how demographic factors can have varying impacts depending on the specific component being considered. 
While \autoref{fig:nhanes_radar} only concerns average consumptions, as we will see later in Section \ref{sec:application}, the conditional (on covariates) marginal (across outcomes) densities also show similar heterogeneous patterns across components and demographic groups. 
Conversely, modeling the densities separately for every possible covariate combination is neither desired nor feasible. 
The NHANES dataset includes four covariates: sex, race, age, and income, all categorical, with 2, 6, 7, and 6 levels, respectively. 
A fully stratified approach would thus require estimating $2 \times 6 \times 7 \times 6 = 504$ distinct distributions, ignoring their shared structural similarities across different demographic groups.
This motivates a parsimonious framework that clusters similar densities across different demographic groups, striking a balance between model flexibility and computational tractability.

Beyond our focus on the NHANES application, estimating dietary intake as a function of demographic and socio-economic covariates remains a central problem in nutritional epidemiology with broad implications for public health. 
Various frameworks have been developed to address this objective \citep{hutchinson2025advances}. 
However, in the presence of covariates, multiple linear regression remains the most prevalent tool for estimating mean intakes \citep[e.g.,][]{kirkpatrick2012income, hiza2013diet}, whereas quantile regression has also been used to characterize distributional tails \citep[e.g.,][]{variyam2002characterizing}. 
Another popular approach employs Latent Profile Analysis (LPA) first to identify population subgroups with similar dietary patterns and then regresses these class labels on associated covariates \citep{affret2017socio, farmer2020cooking}. 
However, such methods are limited by their reliance on discrete categorization, which introduces additional uncertainty while also obscuring the continuous variability and complex distributional shifts inherent in data on dietary intakes. 
Such rigidity again underscores the broader need for robust multivariate approaches capable of simultaneously modeling the full, non-Gaussian distributions of multiple dietary components as they vary flexibly with associated covariates, to capture both their shared structures and their distinct variations without the loss of information inherent in categorical clustering, especially when complex interactions between the demographic factors defy standard parametric assumptions.

\paragraph{Existing Methods for Density Regression.} 
A substantial statistical literature exists for univariate density estimation in the presence of covariates, a problem often referred to as density regression. 
Bayesian mixture models have been a popular tool for density estimation and density regression. 
With a proper choice of kernels, large classes of densities can be approximated \citep{dunson_BayesianDensityRegression_2007}. 
Several authors extend finite mixtures of Gaussians \citep{mclachlan_FiniteMixtureModels_2000,fruhwirth-schnatter_FiniteMixtureMarkov_2006} by allowing means, variances, and mixture probabilities to depend on covariates.
\citet{wood_BayesianMixtureSplines_2002} and \citet{geweke_SmoothlyMixingRegressions_2007} model covariate-dependent means via basis expansion methods based on splines and polynomials, and use a multinomial probit model for the covariate-dependent mixture weights.
\citet{villani_RegressionDensityEstimation_2009} show that assuming homoscedastic variances limits model performance, regardless of the number of components. 
To address this, \citet{villani_RegressionDensityEstimation_2009,nott_RegressionDensityEstimation_2012} and \citet{tran_SimultaneousVariableSelection_2012} model both means and variances as functions of covariates, adopt a multinomial logistic model for efficient inference, and incorporate model and variable selection strategies. 

Mixture models can handle the presence of covariates by explicitly setting up an augmented model, including a distribution for the covariates \citep{muller_BayesianCurveFitting_1996,taddy_BayesianNonparametricApproach_2010,norets_BayesianModelingJoint_2012} 
or by introducing covariate-dependent prior distributions for the mixing proportions and/or the atoms in the mixture. 
The latter approaches often rely on extensions of the Dirichlet Process \citep[DP,][]{ferguson_BayesianAnalysisNonparametric_1973}. 
\citet{maceachern_DependentNonparametricProcesses_1999} introduced the Dependent DP (DDP), where atoms and weights vary with covariates through a stochastic process.
Some popular approaches, often referred to as common-weight processes, consider the special case with covariate-independent weights, e.g.,
spatial DP \citep{gelfand_BayesianNonparametricSpatial_2005},
ANOVA DDP \citep{deiorio_ANOVAModelDependent_2004} for categorical covariates, 
its extension to mixed covariates \citep{deiorio_BayesianNonparametricNonproportional_2009}. 
Several extensions using covariate-dependent stick-breaking probabilities have been proposed \citep{griffin_OrderBasedDependentDirichlet_2006,dunson_BayesianDensityRegression_2007,dunson_KernelStickBreakingProcesses_2008,dunson_NonparametricBayesianModels_2011,chung_NonparametricBayesConditional_2009}. 
See \cite{quintana_DependentDirichletProcess_2022} for a review.

Beyond the popular DDP constructions, other methods have been developed relying on
logistic Gaussian processes \citep{tokdar_BayesianDensityRegression_2010,payne_ConditionalDensityEstimation_2020}, 
extensions of product partition models depending on covariates \citep{park_BayesianGeneralizedProduct_2010,muller_ProductPartitionModel_2011}, 
dependent tail-free processes \citep{jara_ClassMixturesDependent_2011}, 
dependent beta processes \citep{trippa_MultivariateBetaProcess_2011}, 
latent factor models \citep{kundu_LatentFactorModels_2014}, 
transformation models \citep{hothorn_ConditionalTransformationModels_2014,hothorn_PredictiveDistributionModeling_2021},
tensor product of B-splines \citep{shen_AdaptiveBayesianDensity_2016},
dependent Bernstein polynomials \citep{barrientos_FullyNonparametricRegression_2017},
generalized P\'olya trees \citep{ma_RecursivePartitioningMultiscale_2017},
tree-based stick-breaking construction \citep{stefanucci_MultiscaleStickbreakingMixture_2021,horiguchi_TreePerspectiveStickBreaking_2025},
and Bayesian Additive Regression Trees \citep[BART,][]{orlandi_DensityRegressionBayesian_2021,li_AdaptiveConditionalDistribution_2023}.

In contrast to univariate density regression, the multivariate problem remains relatively underexplored. 
A natural multivariate extension of univariate mixture models replaces the univariate mixture kernels with multivariate ones, e.g., \citet{dao_FlexibleMultivariateRegression_2021}. 
Alternatively, copula models offer greater flexibility by modeling marginals and dependence structures separately.
\citet{pitt_EfficientBayesianInference_2006} proposed a Bayesian copula approach with non-Gaussian regression marginals, later formalized by \citet{song_JointRegressionAnalysis_2009} as the Gaussian Copula Regression (GCR) or Vector Generalized Linear Model (VGLM), where marginals are Generalized Linear Models \citep[GLM,][]{nelder_GeneralizedLinearModels_1972} linked via a Gaussian copula \citep{song_MultivariateDispersionModels_2000}.
Some approaches embed distributional parameters within parametric families as smooth functions of covariates, ensuring similar conditional densities for similar covariate profiles \citep{yee_VectorGeneralizedLinear_2015,kock_TrulyMultivariateStructured_2024}. 
An exception is Multivariate Conditional Transformation Model \citep[MCTM,][]{klein_MultivariateConditionalTransformation_2022}, which sets up transformations to the response in such a way that the transformed response follows a convenient distribution.
If the latter is a zero-mean multivariate Gaussian distribution, MCTM can be seen as a Gaussian copula with univariate conditional transformation models \citep{hothorn_ConditionalTransformationModels_2014} as marginals.
MCTM allows for flexible estimation of marginal densities; however, it does not provide a way to merge similar densities, making visual inspection and interpretation of the group-specific densities daunting.

\paragraph{Our Proposed Approach.} 
To address the lack of flexible methods for multivariate density regression, particularly in the challenging setting of categorical covariates, we propose a Bayesian framework that integrates the representational richness of mixture models and conditional probability tensors with the information sharing and dependence modeling capabilities of common atoms models and copulas.
Specifically, we model the marginal densities using flexible mixtures of truncated normal kernels, sharing atoms across coordinates to facilitate borrowing of strength. 
Dependence across coordinates is captured through a Gaussian copula, allowing for flexible joint modeling while maintaining interpretable marginal structures. 
The mixture weights are modeled as functions of the covariates via a unique tensor factorization approach, which is set up to include selection of the most important covariates.

Our approach builds on the framework for deconvolving the multivariate density of latent vectors observed with measurement error introduced by \citet{sarkar_BayesianSemiparametricCovariate_2022}. 
However, we address two key limitations of their method. 
First, their method does not allow for coordinate-specific level aggregation or covariate selection. 
Second, there is a limitation on the nature of the covariate-driven clusters.
Overcoming these limitations presents significant challenges. 
To establish the foundational framework, here we focus on the more tractable problem of multivariate density regression, ignoring measurement error.

Specifically, to efficiently construct component weights for the truncated normal mixtures across multiple categorical predictors, we adapt the Higher-Order Singular Value Decomposition (HOSVD) framework for conditional probability tensors \citep{yang_BayesianConditionalTensor_2016}. 
Rather than specifying unstructured mixture weights for every covariate combination, we arrange these conditional probability vectors into a tensor, which we factor into a latent core probability tensor and covariate-specific weight matrices. 
Under this framework, each conditional weight vector is represented as a convex combination of lower-dimensional core probability vectors, weighted by products of the corresponding mode-matrix elements.
This low-rank tensor construction offers three primary advantages. 
First, by building complex category combinations from a compact set of shared baseline distributions, represented by the core tensor, the total number of parameters to be estimated is significantly reduced. 
Second, if a predictor has no meaningful impact on the outcome variable, its influence weights naturally remain uniform across all of its levels, effectively removing that variable from the model.
Third, despite the simplified parameterization, the underlying structure is mathematically general enough to represent any valid conditional probability relationship without sacrificing predictive accuracy.

However, implementing this HOSVD conditional probability model is computationally prohibitive, particularly if a separate HOSVD needs to be set up for each outcome coordinate. 
In particular, inferring the multirank of the latent core probability tensor, the key parameter that governs both approximation quality and covariate selection, is challenging even in simpler settings, as the model size varies with it. 
To circumvent this, \citet{sarkar_BayesianSemiparametricCovariate_2022} treated the component label as an additional artificial categorical covariate, constructing a single larger combined mixture probability tensor and then applying HOSVD only once, using an MCMC algorithm that jointly infers the multirank and other model parameters.
While this greatly simplified computation, it imposed the restriction that all components share the same set of influential covariates -- a limitation that is clearly at odds with our motivating application (Figure \ref{fig:nhanes_radar}). 
Furthermore, by constructing covariate partitions as the product of marginal partitions, their approach precluded the possibility of representing all possible covariate-driven clusters.

We address the first issue by considering a separate HOSVD-induced partition for each component of the multivariate response. 
For each component, we partition the levels of each covariate using constrained probability vectors, as discussed earlier and illustrated in Figure \ref{fig:tensor_comparison}. 
We then address the second issue by introducing a second layer to further partition the joint cluster labels of the covariate level combinations induced by the first layer.
At the cost of introducing partitions whose size depends on the number of clusters in the first layer, this construction replaces the multiway core tensor of the single-layer formulation \citep{yang_BayesianConditionalTensor_2016,sarkar_BayesianCopulaDensity_2021} with a core vector, thereby avoiding the challenging problem of multirank selection.
The latter is replaced with the problem of selecting the maximum number of clusters in each second-layer partition, which also corresponds to the maximum number of different conditional marginal densities allowed for each coordinate of the response.
Following the commonly adopted strategy of over-specifying mixture components, we set those numbers to values that are sufficiently large to preserve model flexibility, yet moderate enough to maintain computational feasibility.
Similar ideas were first introduced in \citet{paulon_BayesianSemiparametricHidden_2024} to factorize covariate-dependent real-valued regression effect tensors to achieve covariate-dependent local clustering of latent trajectories in a functional longitudinal data setting; we adapt the ideas here to probability tensors in a mixture model framework instead.

We address the associated computational challenges by developing a memory-efficient MCMC strategy that enables fast mixing posterior exploration through trans-dimensional Metropolis-Hastings (M-H) steps employing the Hamming ball sampler \citep{titsias_HammingBallSampler_2017}. 
Exploiting the model's structural properties, our algorithm begins with all covariate levels in a single cluster, then dynamically allocates memory to second-layer partitions only as new clusters emerge, creating the image of a budding flower of cluster allocations as in \autoref{fig:tensor_evolution}. 
Consequently, memory requirements scale with the total number of covariate levels and the number of active clusters rather than the total number of covariate combinations, facilitating flexible inference in high-dimensional spaces.

Applied to our motivating NHANES dataset, our analysis provides novel insights into how dietary intake distributions vary across demographic subgroups.
By adaptively partitioning covariate-level combinations, we identify component-specific subpopulations, allowing the intake patterns of different nutrients to depend on distinct demographic structures.
This also pinpoints the demographic characteristics most influential for each dietary component and enables us to estimate their conditional joint densities through the copula framework.

\section{Background on Tensor Factorization}
Consider a categorical outcome with $K$ categories given $p$ categorical predictors. 
Let $d_{h}$ denote the number of levels $c_{h}\in\C_{h} = \{1,\ldots,d_{h}\}$ for the $h\th$ categorical covariate for $h = 1, \ldots, p$. 
For each covariate combination $\bc = (c_{1}, \ldots, c_{p})\trans \in \C = \C_{1} \times \dots \times \C_{p}$, we define a probability vector $\bp_{\bc} = \{p_{\bc}(1),\ldots,p_{\bc}(K)\}\trans \in \Delta^{K-1}$, where $\Delta^{K-1}$ denotes the $(K-1)$-dimensional simplex. 
Collecting these vectors across all covariate configurations then yields a $(d_{1} \times \cdots \times d_{p})$-dimensional conditional probability tensor $\bP = \{p_{c_{1},\ldots,c_{p}}(k): k=1,\ldots,K, \, c_{h}=1,\ldots,d_{h}, \, h=1,\ldots,p\} = \{\bp_{c_{1},\ldots,c_{p}} : c_{h}=1,\ldots,d_{h}, \, h=1,\ldots,p\} = \{\bp_{\bc}: \bc \in \C\}$. 
\citet{yang_BayesianConditionalTensor_2016} leveraged Higher-Order Singular Value Decomposition (HOSVD) \citep{tucker_MathematicalNotesThreeMode_1966, delathauwer_MultilinearSingularValue_2000} to obtain a representation of $\bP$ in terms of a $(K_{1} \times \cdots \times K_{p})$-dimensional core probability tensor $\blambda$, and a collection $\bpi$ of $d_{h}\times K_{h}$ mode matrices $\bpi_{h}$ as 
\be
\bp_{c_{1},\ldots,c_{p}} = \sum_{k_{1}=1}^{K_{1}} \cdots \sum_{k_{p}=1}^{K_{p}} \left\{ \blambda_{k_{1},\ldots,k_{p}} \prod_{h=1}^p \pi_{h}^{(c_{h})}(k_{h}) \right\}, \label{eq:tensor_dec}
\ee
where $\blambda_{k_{1},\ldots,k_{p}} = \{\lambda_{k_{1},\ldots,k_{p}}(1),\ldots,\lambda_{k_{1},\ldots,k_{p}}(K)\}\trans\in\Delta^{K-1}$ is a probability vector for each $\bk=(k_{1},\ldots,k_{p})\trans\in\{1,\ldots,K_{1}\}\times\cdots\times\{1,\ldots,K_{p}\}$; the $c_{h}\th$ row of the $h\th$ mode matrix,  $\bpi_{h}^{(c_{h})}=\{\pi_{h}^{(c_{h})}(1),\ldots,\pi_{h}^{(c_{h})}(K_{h})\}\trans\in\Delta^{K_{h}-1}$, is a probability vector which specifies the effect of the $d_h$ levels of the $h\th$ covariate. 
In other words, rather than specifying an unstructured probability vector $\bp_{c_{1},\ldots,c_{p}}$ for every covariate combination $(c_{1},\ldots,c_{p})$, we define base probability vectors $\blambda_{k_{1},\ldots,k_{p}}$ associated with latent core tensor indices $(k_{1},\ldots,k_{p})$. 
The products of the mode matrix elements, $\prod_{h=1}^p \pi_{h}^{(c_{h})}(k_{h})$, then serve as weights to express each $\bp_{c_{1},\ldots,c_{p}}$ as a convex combination of these core vectors $\blambda_{k_{1},\ldots,k_{p}}$. 
By specifying $1 \le K_{h} \le d_{h}$ for each $h = 1, \ldots, p$, the number of core parameters required to characterize $\mathbf{P}$ is substantially reduced from $(K-1)\prod_{h=1}^p d_{h}$ to the significantly smaller $(K-1)\prod_{h=1}^{p} K_{h}$. 
Beyond providing effective dimensionality reduction, this formulation naturally facilitates covariate selection: when $K_{h} = 1$, we have $\pi_{h}^{(c_{h})}(1) = 1$ for all levels $c_{h}$, implying that the $h\th$ covariate exerts no influence on $\mathbf{P}$.
Crucially, any conditional probability tensor can be represented in this form, yielding a highly flexible modeling strategy, as long as each element in $\bP$ and $\blambda$ belongs to the same simplex.

\begin{figure}[!ht]
    \begin{center}
    \includegraphics[width=0.9\textwidth]{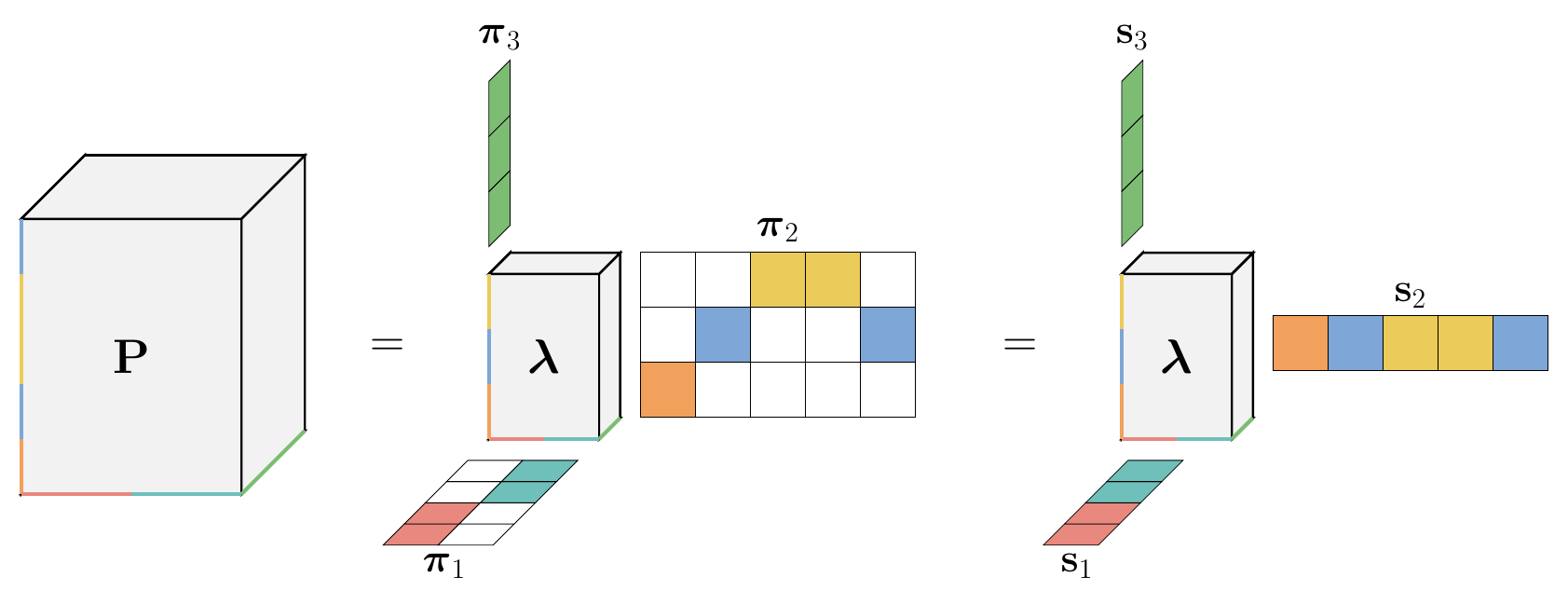}
    \end{center}
    \vspace*{-4ex}
    \caption{Cluster-inducing tensor factorization structure in \cite{sarkar_BayesianSemiparametricCovariate_2022}. 
    Colors represent entries equal to 1 in the probability vectors. 
    Example with 3 covariates with $(d_{1},d_{2},d_{3})=(4,5,3)$ levels clustered respectively into $(K_{1},K_{2},K_{3}) = (2,3,1)$ groups.}
    \label{fig:tensor_comparison}
\end{figure}

Adapting the above idea to model mixture weights in a kernel mixture framework $f(x \mid \bc) = \sum_{k} p_{\bc}(k)\K(x \mid \bvartheta_{k})$ for density deconvolution, where $\K(x \mid \bvartheta_{k})$ is a density kernel characterized by parameters $\bvartheta_{k}$, \citet{sarkar_BayesianSemiparametricCovariate_2022} imposed additional binary constraints on each $d_{h} \times K_{h}$ mode matrix $\bpi_{h}$. 
Specifically, he let exactly one element in each row $c_{h}$ equal 1, and all other elements equal 0, i.e., $\pi_{h}^{(c_{h})}(k_{h}) = 1$ for a single column $k_{h}$, and $\pi_{h}^{(c_{h})}(k_{h}) = 0$ otherwise.
Let $\bs_{h} = \{s_{h}^{(c_{h})}:  c_{h} \in \C_{h}\}$ be a latent random allocation vector such that $s_{h}^{(c_{h})} \in \{1,\ldots,K_{h}\}$ denotes the allocation index of the column in $\bpi_{h}$ in row $c_{h}$ whose element equals to one, i.e., $\pi_{h}^{(c_{h})}(k_{h})=1$ if and only if $s_{h}^{(c_{h})}=k_{h}$, and $\pi_{h}^{(c_{h})}(k_{h})=0$ otherwise. 
Given $s_{1}^{(c_{1})},\ldots,s_{p}^{(c_{p})}$, model \eqref{eq:tensor_dec} can then be reformulated in terms of these latent allocation variables as 
\be
\bp_{c_{1},\ldots,c_{p}} = \sum_{k_{1}=1}^{K_{1}} \cdots \sum_{k_{p}=1}^{K_{p}} \left\{ \blambda_{k_{1},\ldots,k_{p}} \prod_{h=1}^p \Ind\left(s_{h}^{(c_{h})}=k_{h}\right) \right\} = \blambda_{s_{1}^{(c_{1})},\ldots,s_{p}^{(c_{p})}}. \label{eq:tensor_dec_cons}
\ee
This many-to-one mapping from $\bP$ to the core tensor $\blambda$ is illustrated in \autoref{fig:tensor_comparison}, where (possibly many) fibers in $\bP$ corresponding to the same cluster share a color and map to the matching colored (unique) fiber in $\blambda$. 

\vspace*{-2ex}

In essence, these binary restrictions establish a hard clustering mechanism: the $d_{h} \times K_{h}$ mode matrix $\bpi_{h}$ partitions the $d_{h}$ covariate levels in $\C_{h}$ into a partition of $K_{h}$ latent clusters via the allocation variables $s_{h}^{(c_h)}$. 
Specifically, assigning $s_{h}^{(c_h)} = k_{h}$ places covariate level $c_{h}$ into the $k_{h}\th$ cluster. 
Consequently, the fibers in the probability tensor $\bP$ indexed by the covariate levels $(h,c_{h})$ also get uniquely mapped to a single core tensor fiber in $\blambda$ indexed by the labels $(h, s_{h}^{(c_{h})})$. 
Also, as before, when $K_{h}=1$, all $d_{h}$ levels of the $h\th$ predictor get clustered together, $\bpi_{h}$ reduces to a row vector with $\pi_{h}^{(c_{h})}(1) = 1$ for all $c_{h}$, and hence has no influence on $\bP$. 
{The Cartesian product of these allocation variables, $s_{h}^{(c_{h})}$, for $c_{h}=1,\ldots,d_{h}$, $h=1,\ldots,p$, thus induces a partitioning $\calP$ of the joint covariate space $\C$, which by extension, also induces a clustering of the associated probability vectors $\bp_{c_{1},\ldots,c_{p}}$ into core probability vectors $\blambda_{s_{1}^{(c_{1})},\ldots,s_{p}^{(c_{p})}}$.} 

A key advantage of partitioning the joint covariate space $\mathcal{C} = \mathcal{C}_{1} \times \dots \times \mathcal{C}_{p}$ via HOSVD-based Cartesian products of the marginal allocations $s_{h}^{(c_{h})}$ is that redundant covariates are automatically removed when $K_{h}=1$. 
However, this structure also restricts the induced joint partition $\mathcal{S}$ to a Cartesian product of the marginal partitions, $\mathcal{S} = \mathcal{S}_{1} \times \dots \times \mathcal{S}_{p}$. 
Consequently, whenever $K_{h} > 1$ for two or more modes $h$, the joint partition lacks full flexibility in that it cannot represent non-product partitioning structures.
Additionally, the model is computationally challenging, especially in multivariate outcome settings where implementing it for each coordinate of a $d$-dimensional outcome variable $\bx=(x_{1},\ldots,x_{d})\trans$ becomes computationally prohibitive.

To address this computational bottleneck, \citet{sarkar_BayesianSemiparametricCovariate_2022} introduced an artificial auxiliary $0\th$ categorical predictor, $c_{0} \in \{1,\dots,d\}$, representing the coordinate index of the outcome vector $\bx=(x_{1},\dots,x_{d})\trans$. 
By operating on the resulting expanded probability tensor $\bp_{c_{0},c_{1},\ldots,c_{p}}$, their approach required performing HOSVD only once across all $d$ coordinates of $\bx$, rendering the model computationally tractable. 
However, this efficiency gain came at the cost of an implicit restriction that the same set of covariates influences every coordinate of $\bx$. 
Because their main motivating application selected only a single covariate, further refinement of the Cartesian product partition structure was neither needed nor pursued. 

To factorize covariate-dependent real-valued regression effect tensors for local clustering in longitudinal functional data, \citet{paulon_BayesianSemiparametricHidden_2024} extended the ideas to construct an unrestricted partition of the joint covariate space $\mathcal{C} = \mathcal{C}_{1} \times \dots \times \mathcal{C}_{p}$ using a two-layer partition model that relaxes the first-layer product-of-marginals structure using a second layer while preserving interpretable covariate combinations:
\begin{enumerate} 
\item 
First layer $\bs_{1},\ldots,\bs_{p}$: Marginal allocation vectors $\bs_{h} = \{s_{h}^{(c_{h})}: c_{h} \in \mathcal{C}_{h}\}$ map the $d_{h}$ original levels of each covariate $h$ into $K_{h}$ aggregated clusters, as before. This projects $\mathcal{C}$ onto a lower-dimensional space $\mathcal{S} = \{1,\ldots,K_{1}\}\times\cdots\times\{1,\ldots,K_{p}\}$.
\item 
Second layer $\bs^{\star}$: A $(K_{1}\times\cdots\times K_{p})$-dimensional latent allocation tensor $\bs^{\star} = \{s^{\star}_{k_{1},\ldots,k_{p}}: k_{h} = 1,\ldots, K_{h}; h=1,\ldots,p\} \in \{1,\ldots,K^{\star}\}$ further groups combinations of these aggregated levels, defining an unrestricted partition $\S^{\star}$ over $\mathcal{S}$.
\end{enumerate} 
Although first-layer covariate clusters could, in principle, be re-aggregated in the second layer, introducing a minor form of redundancy, this was never observed in practice. 
In any case, since inference relies solely on the final joint partition, it had no impact on practical outcomes.

In this article, we extend the two-layer partitioning idea described above to probability tensors. 
Let $\widetilde{s}_{c_{1},\ldots,c_{p}} = s^{\star}_{s_{1}^{(c_{1})},\ldots,s_{p}^{(c_{p})}}$ denote the element in the partition tensor $\bs^{\star}$ corresponding to the covariate combination $\bc = (c_{1},\ldots,c_{p})\trans$, where the index vector $(s_{1}^{(c_{1})},\ldots,s_{p}^{(c_{p})})\trans \in \mathcal{S}$ specifies its spatial position within $\bs^{\star}$. 
Given $\wt{s}_{c_1,\dots,c_{p}}$, the fibers of the probability tensor $\bP$ under the two-layer hierarchy are then given by 
\be
\bp_{c_{1},\ldots,c_{p}} = \blambda_{\wt{s}_{c_1,\dots,c_{p}}}. \label{eq:tensor_dec_2layers}
\ee 
Specifically, to perform multivariate density regression with coordinate-wise variable selection, we apply \eqref{eq:tensor_dec_2layers} to the mixture weights $\bp_{\ell,c_{1},\ldots,c_{p}}$ within a kernel mixture model framework for each coordinate $x_{\ell}$ of an outcome vector $\bx = (x_{1},\dots,x_{d})\trans$. 
This allows us to identify a different unrestricted partition $\S^{\star}$ over the joint covariate space $\C$ for each $x_{\ell}$ within a highly flexible conditional mixture model framework, identifying the set of important predictors for each $x_{\ell}$ as well as the nature of their influence on $x_{\ell}$.

\begin{figure}[!ht]
    \begin{center}
    \includegraphics[width=0.85\textwidth]{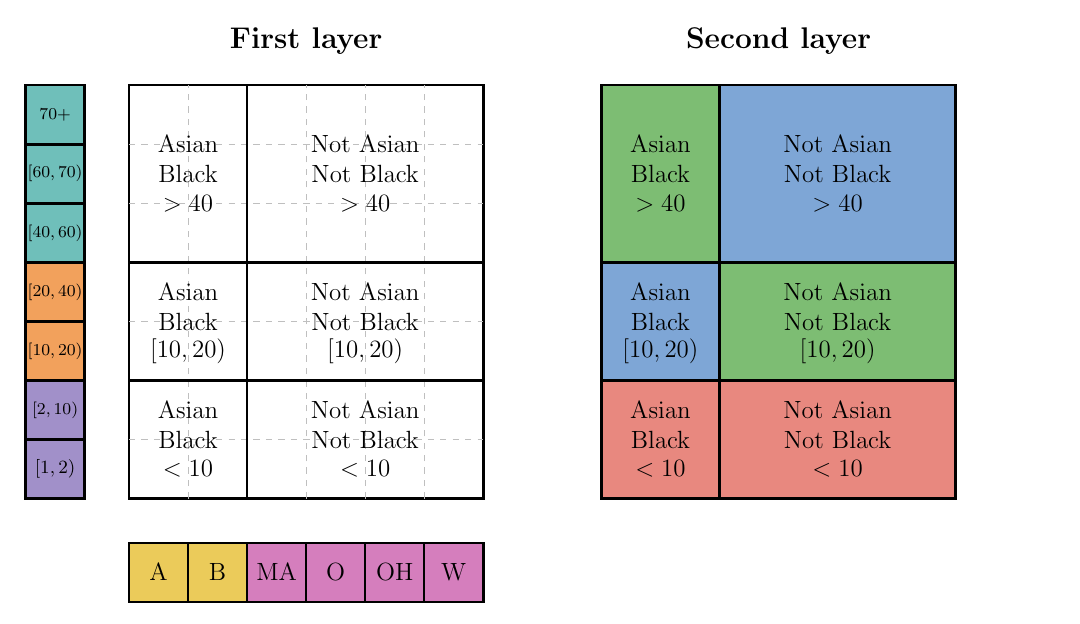}
    \end{center}
    \vspace*{-3ex}
    \caption{Two-layer construction of the partition of the joint covariate space, shown for two covariates, Age and Race.} 
    \label{fig:tensor_example}
\end{figure}

We conclude the section with a synthetic visual illustration of how the two layers induce a partition of the joint covariate space in \autoref{fig:tensor_example}. 
For simplicity, we consider only two covariates, Age and Race, with race levels denoted by their initials (A = Asian, B = Black, MA = Mexican American, O = Other, OH = Other Hispanic, W = White).
The colored vectors (left and bottom) represent the first-layer covariate-specific partitions, and the colored matrix (right) represents the second-layer partition; throughout, entries sharing a color belong to the same cluster.
The first-layer partitions aggregate the 7 age levels into 3 groups and the 6 race levels into 2 groups.
Their Cartesian product defines a reduced joint covariate space of only $3\cdot2=6$ combinations, far fewer than the original $7\cdot6=42$.
Each combination represents a group of the original covariate combinations; e.g., the bottom-left group `Asian Black $<10$' contains `Asian $[1,2)$', `Asian $[2,10)$', `Black $[1,2)$', and `Black $[2,10)$'.
Under \eqref{eq:tensor_dec_cons}, this 6-cell partition of the joint covariate space is final; 
while under \eqref{eq:tensor_dec_2layers}, it serves as an intermediate space that the second-layer partition tensor regroups.
This additional clustering step lifts the Cartesian-product constraint, which requires a chosen race grouping to hold across all age groups; e.g., the two race groups merge for age $<10$ (red) but remain split for $>10$ (blue and green), making race relevant only for ages $10+$.
This non-product partition is attainable under \eqref{eq:tensor_dec_2layers}, thanks to the second-layer partition, but not under \eqref{eq:tensor_dec_cons}.

\newpage
\section{The (Bouquet of Flowers) Model} \label{sec:model}
Let $\bx_{i}=(x_{1,i},\ldots,x_{d,i})\trans$ be a multivariate response and $\bc_{i}=(c_{1,i},\ldots,c_{p,i})\trans$ covariates for $n$ observational units $i=1,\ldots,n$.
Consistent with our motivating application, we assume all covariates are categorical with $d_{1},\ldots,d_{p}$ levels, respectively. 
Additionally, all response coordinates are assumed to be continuous and supported on a common interval $[A,B]$. 
If the components of $\bx$ have different supports and units of measurement to begin with, we can rescale to unit-free coordinates with a common support, as detailed in Section S.1 of the Supplementary Materials.

We use a copula model to characterize the correlation across different coordinates while allowing for complex covariate-dependent marginal distributions.
The joint density of the multivariate response $\bx_{i}$ given a covariate vector $\bc_{i}$, $f_{\bx\mid\bc}(\bx_{i};\bc_{i})$, is specified as 
\be
f_{\bx\mid\bc}(\bx_{i};\bc_{i}) = C(\bx_{i}) \prod_{\ell=1}^{d} f_{x_{\ell}\mid\bc}(x_{\ell,i}; \bc_{i}), \label{eq:joint}
\ee
where $C(\cdot)$ is a copula function, and $f_{x_{\ell}\mid\bc}(\cdot\,;\bc)$ is the $\ell\th$ marginal distribution.

In the remainder of the section, we introduce the hierarchical model for $f_{x_{\ell}\mid\bc}(\cdot\,;\bc)$, which uses ideas from the partition-based tensor factorizations \citep{yang_BayesianConditionalTensor_2016,sarkar_BayesianSemiparametricCovariate_2022,paulon_BayesianSemiparametricHidden_2024} to include covariate information in the model, and then a Gaussian copula for $C(\cdot)$ that connects these marginals to build the joint distribution.

\subsection{Conditional Marginal Distributions}
The conditional (on covariates) marginal (across outcomes) densities $f_{x_{\ell}\mid\bc}(\cdot\,;\bc_{i})$ are modeled as mixtures of truncated normal distributions with atoms shared across the different coordinates and associated mixture probabilities varying with the covariates:
\be
f_{x_{\ell}\mid\bc}(x_{\ell,i}\mid\bmu,\bsigma^{2};\bc_{i}) = \sum_{k=1}^{K} p_{\ell,\bc_{i}}(k) \TN\left(x_{\ell,i}; \mu_{k}, \sigma^{2}_{k}, [A,B]\right), \label{eq:marginal}
\ee
where $\{(\mu_{k},\sigma^{2}_{k})\}_{k=1}^{K}$ are the atoms shared across the different coordinates, collected in $\bmu=(\mu_{1},\ldots,\mu_{K})\trans\in\rR^{K}$ and $\bsigma^{2}=(\sigma^{2}_{1},\ldots,\sigma^{2}_{K})\trans\in(0,\infty)^{K}$; $\TN\left(x_{\ell,i}; \mu_{k}, \sigma^2_{k}, [A,B]\right)$ denotes a truncated normal distribution with mean $\mu_{k}$, variance $\sigma^{2}_{k}$ and support $[A,B]\subseteq\rR$ evaluated at $x_{\ell,i}$. 
Truncated normal kernel-based mixture models allow bounded support across different components, while being highly flexible and computationally tractable.
Sharing atoms across different coordinates and covariate combinations allows information to be borrowed. 
For prior distributions on the shared atoms, we consider 
\be
\mu_{k} \sim \TN(m_{0},s^{2}_{0},[A,B]), \quad \sigma_{k}^{2} \sim \IG(a_{\sigma},b_{\sigma}), \label{eq:atoms_prior}
\ee
where $\IG(a_{\sigma},b_{\sigma})$ denotes an inverse Gamma distribution with shape $a_{\sigma}>0$ and scale $b_{\sigma}>0$.

In \eqref{eq:marginal}, a vector of mixture weights 
$\bp_{\ell,c_{1},\ldots,c_{p}} = \{p_{\ell,c_{1},\ldots,c_{p}}(1), \ldots, p_{\ell,c_{1},\ldots,c_{p}}(K)\}\trans\in\Delta^{K-1}$ is defined for each covariate combination $\bc=(c_{1},\ldots,c_{p})\trans\in\C$ and for each coordinate $\ell$.
Without any additional structure, we would need to estimate $d\prod_{h=1}^{p}d_{h}$ probability vectors, which corresponds to $(K-1)d\prod_{h=1}^{p}d_{h}$ parameters.
This may be problematic in several scenarios, e.g., the response variable is high-dimensional (high $d$) or a large number of covariates with many levels (high $\prod_{h=1}^{p}d_{h}$).
To reduce this number, $(d_{1}\times\ldots\times d_{p})$-dimensional probability tensors, $\bP_{\ell} = \{\bp_{\ell,c_{1},\ldots,c_{p}}, c_{h}=1,\ldots,d_h\}$, are introduced and modeled using $d$ partition-based tensor factorizations.
Our proposed probability tensor factorization follows the form in \eqref{eq:tensor_dec_2layers}, which allows us to obtain $d$ different unrestricted partitions over the joint covariate space $\C$.

For each response coordinate $\ell$, we introduce a core vector, $\blambda_{\ell}=\{\blambda_{\ell,k^{\star}}\}_{k^{\star}=1,\ldots,K_{\ell}^{\star}}$, and two layers of partitions, $(\bs_{\ell},\bs_{\ell}^{\star})$, which allows us to define the elements in the probability tensor $\bP_{\ell}$ in terms of the ones in $\blambda_{\ell}$.
The first layer $\bs_{\ell}=\{\bs_{\ell,1},\ldots,\bs_{\ell,p}\}$ collects $p$ partitions over covariate levels, $\bs_{\ell,h}=(s_{\ell,h}^{(1)},\ldots,s_{\ell,h}^{(d_{h})})\trans\in\{1,\ldots,d_{h}\}^{d_h}$, one for each categorical covariate.
Each partition $\bs_{\ell,h}$ maps the original $d_h$ levels to a simplified representation in which the number of aggregated levels coincides with the number of observed clusters $K_{\ell,h}$. 
The second layer is composed of a single partition defined as a $(K_{\ell,1}\times\cdots\times K_{\ell,p})$-dimensional tensor, $\bs^{\star}_{\ell} = \{s^{\star}_{\ell,k_{1},\ldots,k_{p}}, k_{h}=1,\ldots,K_{\ell,h}\}$, where $s^{\star}_{\ell,k_{1},\ldots,k_{p}}\in\{1,\ldots,K^{\star}_{\ell}\}$ with $1\leq K^{\star}_{\ell}\leq \prod_{h=1}^{p}d_{h}$ fixed.

Using the partitions in $\bs_{\ell}$, each covariate combination $\bc$ is associated with an element of the second layer partition $\bs_{\ell}^{\star}$ through $\bs_{\ell}^{(\bc)} = (s_{\ell,1}^{(c_{1})},\ldots,s_{\ell,p}^{(c_{p})})\trans\in\S_{\ell}=\{1,\ldots,K_{\ell,1}\}\times\cdots\times\{1,\ldots,K_{\ell,p}\}$; the element $s_{\ell,\bs_{\ell}^{(\bc)}}^{\star}$ then specifies which one of the $K_{\ell}^{\star}$ mixture weights is considered for all units $i$ with that observed covariate combination $(c_{1},\dots,c_{p})\trans$.
Thus, for the $\ell\th$ response coordinate $x_{\ell}$, the joint use of the two partition layers $(\bs_{\ell},\bs_{\ell}^{\star})$ induces a partition over the covariate space $\C$.
Let $\wt{\bs}_{\ell}=\{\wt{s}_{\ell,c_{1},\ldots,c_{p}}, c_{h}=1,\ldots,d_{h}\}$, with $\wt{s}_{\ell,\bc} = s_{\ell,\bs_{\ell}^{(\bc)}}^{\star} \in \{1,\ldots,K_{\ell}^{\star}\}$, be the partition for coordinate $\ell$ of $\C$ into $K_{\ell}^{\star}$ clusters, we define the mixture weight assigned to $(\ell,\bc)$ as $\bp_{\ell,\bc} = \blambda_{\ell,\wt{s}_{\ell,\bc}}$, which allows replacing $p_{\ell,\bc_{i}}(k)$ in \eqref{eq:marginal} with $\lambda_{\ell,\wt{s}_{\ell,\bc_{i}}}(k)$.
This leads to
\be
f_{x_{\ell}\mid\bc}(x_{\ell,i} \mid \bs_{\ell},\bs_{\ell}^{\star},\blambda_{\ell},\bmu,\bsigma^{2};\bc_{i}) = \sum_{k=1}^{K} \lambda_{\ell,\wt{s}_{\ell,\bc_{i}}}(k) \TN\left(x_{\ell,i}; \mu_{k}, \sigma^{2}_{k}, [A,B]\right). \label{eq:marginal_part1}
\ee

We consider a Dirichlet-multinomial prior for the allocation variables in both first and second layers.
Each cluster allocation variable $s_{\ell,h}^{(c_{h})}$ follows a categorical distribution with a coordinate- and covariate-specific probability vector, which in turn follows a Dirichlet distribution as 
\be
s_{\ell,h}^{(c_{h})} \mid \etam_{\ell,h} \sim \Cat_{d_{h}}(\etam_{\ell,h}), \quad \etam_{\ell,h} \mid \phi \sim \Dir_{d_{h}}(\phi/d_{h}), \quad \phi \sim \Ga(a_{\phi},b_{\phi}),  \label{eq:s_prior}
\ee
where $\Dir_{d_{h}}(\phi/d_{h})$ is a symmetric Dirichlet distribution with parameter $\phi/d_{h}$, and $\Ga(a_{\phi},b_{\phi})$ is a Gamma distribution with shape $a_{\phi}>0$ and scale $b_{\phi}>0$.
Analogously, we assume that each cluster allocation $s^{\star}_{\ell,\bk}$ follows a categorical distribution with a coordinate-specific probability vector that, in turn, follows a Dirichlet distribution as 
\be
s^{\star}_{\ell,\bk} \mid \bs_{\ell}, \etam^{\star}_{\ell} \sim \Cat_{K_{\ell}^{\star}}(\etam_{\ell}^{\star}), \quad \etam^{\star}_{\ell} \mid \bs_{\ell} \sim \Dir_{K_{\ell}^{\star}}(\phi^{\star}/K_{\ell}^{\star}), \label{eq:s2_prior}
\ee
where $\phi^{\star}>0$ is a fixed hyperparameter, and $1\leq K_{\ell}^{\star}\leq\prod_{h=1}^{p}d_h$ is the maximum number of different conditional marginal densities allowed for coordinate $\ell$. 
Because the cardinality of $\bs^{\star}_{\ell}$ (specifically, $\prod_{h=1}^{p}K_{\ell,h}$) is determined by the number of observed clusters in the first-layer partitions, transitions between latent spaces require a trans-dimensional M-H step. 
To ensure computational tractability, we assume a fixed maximum number of densities, $K_{\ell}^{\star}$.
While an unrestricted version, where $s^{\star}_{\ell,\bk} \in \{1,\ldots,\prod_{h=1}^{p}K_{\ell,h}\}$, is theoretically possible, a sufficiently large $K^{\star}_{\ell}$ preserves the model's ability to identify the underlying number of densities in practice.

We complete the specification of the marginal distributions by assigning priors to the mixture weights in $\blambda_{\ell}$, for $\ell=1,\ldots,d$.
We assign a hierarchical prior as
\be
\blambda_{\ell,k^{\star}} \mid \blambda_{\ell,0}, \alpha \sim \Dir_{K}(\alpha\blambda_{\ell,0}), \quad \blambda_{\ell,0} \sim \Dir_{K}(\alpha_{0}/K), \quad \alpha \sim \Ga(a_{\alpha},b_{\alpha}), \label{eq:lambda_prior}
\ee
where in this case $\Dir_{K}(\alpha\blambda_{\ell,0})$ denotes a Dirichlet distribution with parameter $\alpha\blambda_{\ell,0}$.

\subsection{Copula}
We consider a Gaussian copula model \citep{song_MultivariateDispersionModels_2000}.
Let $\bR$ denote a $d\times d$ correlation matrix, let $y_{\ell,i} = \Phi^{-1}\left\{F_{x_{\ell}\mid\bc}(x_{\ell,i} ; \bc)\right\}$, with $\Phi(\cdot)$ and $F_{x_{\ell}\mid\bc}(\cdot\,;\bc)$ being the cumulative distribution function of a standard normal distribution and the $\ell\th$ marginal distribution $f_{x_{\ell}\mid\bc}$, respectively.
The copula is defined as
\be
C(\bx_{i}) = |\bR|^{-\frac{1}{2}} \exp\left\{-\frac{1}{2}\by_{i}\trans(\bR^{-1}- \bI_{d})\by_{i}\right\},  \label{eq:copula}
\ee
implying that $\by_{i} = (y_{1,i},\dots,y_{d,i})\trans$ follows a multivariate normal distribution with mean $\bzero_{d}$ and covariance matrix $\bR$, where $\bzero_{d}$ is a $d$-dimensional vector of zeros and $\bI_{d}$ is a $d\times d$ identity matrix.
The correlation matrix, $\bR$, is assumed to be fixed across covariate combinations and modeled using a spherical coordinate representation of Cholesky factorization
\citep{zhang_NewMultivariateMeasurement_2011,sarkar_BayesianCopulaDensity_2021,sarkar_BayesianSemiparametricCovariate_2022}. 
This representation allows us to characterize any correlation matrix through two sets of parameters, $\bb\in(-1,1)^{d-1}$ and $\btheta\in(-\pi,\pi)^{(d^2-3d+2)/2}$, that can be easily and independently updated in an MCMC algorithm via Metropolis-Hastings steps.
Let $\bR=\bV\bV\trans$, where $\bV$ is a $d\times d$ lower triangular matrix whose elements $v_{\ell,\ell'}$, for $\ell,\ell'=1,\ldots,d$, are such that $v_{\ell,\ell'}=0$ if $\ell<\ell'$, and
\bse
&& \hspace{-0.8cm}\textstyle v_{1,1}=1, \\
&& \hspace{-0.8cm}\textstyle v_{2,1}=b_{1}, ~ v_{2,2}=\sqrt{1-b_{1}^{2}},\\
&& \hspace{-0.8cm}\textstyle v_{3,1} =b_{2}\sin\theta_{1}, ~v_{3,2}=b_{2}\cos\theta_{1},~v_{3,3}=\sqrt{1-b_{2}^{2}},\\
&& \hspace{-0.8cm}\textstyle v_{\ell,1}=b_{\ell-1}\sin\theta_{i_{1}(\ell)},\\
&& \hspace{-0.8cm}\textstyle v_{\ell,\ell'}=b_{\ell-1}\cos\theta_{i_{1}(\ell)}\cos\theta_{i_{1}(\ell)+1}\dots\cos\theta_{i_{1}(\ell)+\ell'-2}\sin\theta_{i_{1}(\ell)+\ell'-1}, \quad \hbox{for}~\ell'=2,\dots,(\ell-2),\\
&& \hspace{-0.8cm}\textstyle v_{\ell,\ell-1}=b_{\ell-1}\cos\theta_{i_{1}(\ell)}\cos\theta_{i_{1}(\ell)+1}\dots \cos\theta_{i_{2}(\ell)-1}\cos\theta_{i_{2}(\ell)},~~~~v_{\ell,\ell}=\sqrt{1-b_{\ell-1}^{2}},
\ese
where $i_1(\ell) = (\ell^2-5\ell+8)/2$ and $i_2(\ell) = (\ell^2-3\ell+2)/2$, for $\ell=1,\ldots,d$.
The specification of the model for $\bR$ is completed by assigning uniform prior distribution to both set of parameters, $b_{s'} \sim \Unif(-1,1)$ and $\theta_{s''} \sim \Unif(-\pi,\pi)$, where $\Unif(a,b)$ denotes a uniform distribution with support $(a,b)$.

This completes the model construction, including the copula \eqref{eq:joint} and \eqref{eq:copula}, the mixture of truncated normal marginal models \eqref{eq:marginal_part1}, with the prior on the atoms as in \eqref{eq:atoms_prior}, and a multivariate probability tensor on the weights defined by way of the hierarchical partition in \eqref{eq:s_prior} through \eqref{eq:lambda_prior}.
For easy reference, the completed model is stated in Section S.2 of the Supplementary Materials.

Drawing on the structural evolution of the tensor factorization, we conceptualize our model through the analogy of blooming flowers. 
By representing first-layer partitions as constrained mode matrices, the model functions like a flower with a central stigma (representing the core tensor $\blambda$ or $\bs^{\star}$) encircled by $p$ petals (representing the covariate-specific partition allocation variables $\bpi_{h}$ or $\bs_{h}$) (Figure \ref{fig:tensor_comparison}). 
Consider then a sequence of increasingly complex models starting with a single distribution shared across all covariate levels, and adding complexity by splitting by different covariate levels.
The model originates as a bud, then blooms as new clusters emerge across covariate levels in the first layer (Figure \ref{fig:tensor_evolution}).
This way, for our multivariate density regression problem, we have $d$ flowers, one for each coordinate of the $d$-dimensional response. 
We thus refer to our proposed framework as the `Bouquet of Flowers' model, or, in short, just the `Flower' model throughout the remainder of this work.

\newpage
\section{Posterior Computation} \label{sec:inference}
Our inference is based on samples drawn from the posterior using an MCMC algorithm. 
Copula models often suffer from poor mixing and computational inefficiency due to the disruption of the conjugacy of the marginal distributions by the joint likelihood. 
To address this, we follow the iterative approach of \citet{silva_CopulaMarginalDistributions_2008}: marginal parameters are first updated via a pseudo-likelihood which ignores the contribution of the copula, followed by an exact likelihood update for the copula parameters.

We describe our MCMC algorithm below. 
Achieving computational efficiency for our multivariate density regression model, particularly with many covariate levels, requires careful attention to implementation details. 
We leverage specific model properties and rigorous memory management strategies, which we highlight later in Section \ref{sec:comp_strategies}.

\subsection{MCMC Details}
To simplify posterior inference for quantities in \eqref{eq:s_prior} and \eqref{eq:s2_prior}, we introduce latent mixture allocations $z_{\ell,i}\in\{1,\ldots,K\}$ to link each $x_{\ell,i}$ to one of the $K$ components in \eqref{eq:marginal_part1}. Conditional on these, we can analytically marginalize out the core vector elements $\blambda_{\ell,k^{\star}}$; we further marginalize out the parameters with conditionally conjugate priors, namely $\etam_{\ell,h}$ and $\etam^{\star}_{\ell}$, to improve mixing.
The resulting algorithm provides samples from the posterior distribution $\Pr(\bz,\bs,\bs^{\star},\blambda_0,\alpha,\phi,\bmu,\bsigma^{2},\bb,\btheta \mid \bx;\bc)$.
More details in Section S.2 of the Supplementary Materials.

\subsubsection{Updating the Parameters of the Marginal Distribution}
Throughout the section, we will denote the full conditional distribution of a random variable $y$ as $\Pr(y\mid-)$; we further use the more compact notation $p_{y}(\cdot)$ to denote its unnormalized version in the Metropolis-Hastings (M-H) proposals.

The full conditional distribution for $z_{\ell,i}$ is
\bse
\Pr(z_{\ell,i}=k\mid-) \propto \left\{ \alpha\lambda_{\ell,0}(k) + n^{(-i)}_{\ell,\wt{s}_{\ell,\bc_{i}}}(k) \right\} \times \TN(x_{\ell,i}; \mu_{k},\sigma^{2}_{k}, [A,B]),
\ese
where $n^{(-i)}_{\ell,k^{\star}}(k)$ denotes the count $n_{\ell,k^{\star}}(k) = \sum_{i=1}^{n}\Ind\{\wt{s}_{\ell,\bc_{i}}=k^{\star}\}\Ind\{z_{\ell,i}=k\}$ without considering the $i\th$ unit.

To update $\blambda_{\ell,0}$, following ideas from \citet{sarkar_BayesianHigherOrder_2019}, we first sample an auxiliary variable $\omega_{\ell,j,k}^{(k^{\star})}$ as
\bse
\omega_{\ell,j,k}^{(k^{\star})}\mid- \sim \Bern\left( \frac{\alpha\lambda_{\ell,0}(k)}{j-1+\alpha\lambda_{\ell,0}(k)} \right),
\ese
for $j=1,\ldots,n_{\ell,k^{\star}}(k)$, $k^{\star}\in\{1,\ldots,K_{\ell}^{\star}: n_{\ell,k^{\star}}>0\}$, $k=1,\ldots,K$, $\ell=1,\ldots,d$, and then sample a new value for $\blambda_{\ell,0}$ as
\bse
\blambda_{\ell,0} \mid- \sim \Dir_{K}\left(\frac{\alpha_0}{K} + \sum_{k^{\star}=1}^{K_{\ell}^{\star}}\sum_{j=1}^{n_{\ell,k^{\star}}(1)}\omega_{\ell,j,1}^{(k^{\star})}, \ldots, \frac{\alpha_0}{K} + \sum_{k^{\star}=1}^{K_{\ell}^{\star}}\sum_{j=1}^{n_{\ell,k^{\star}}(K)}\omega_{\ell,j,K}^{(k^{\star})}\right).
\ese

The full conditional distributions of $\alpha$ and $\phi$ do not have closed forms; however, they can be easily updated via M-H steps.
A new value $\alpha^{(new)}$ is proposed as $\alpha^{(new)} = \exp\{\wt{A}\}$, with $\wt{A}\sim\Normal(\log(\alpha),\sigma^{2}_{\alpha})$; analogously, a new value $\phi^{(new)}$ is proposed as $\phi^{(new)} = \exp\{\wt{Q}\}$, with $\wt{Q}\sim\Normal(\log(\phi),\sigma^{2}_{\phi})$.
These values are respectively accepted with probability
\bse
\min\left\{1, \frac{p_\alpha(\alpha^{(new)})}{p_\alpha(\alpha)} \cdot \frac{\alpha^{(new)}}{\alpha} \right\}, \quad \min\left\{1, \frac{p_\phi(\phi^{(new)})}{p_\phi(\phi)} \cdot \frac{\phi^{(new)}}{\phi} \right\},
\ese
where
\bse
& p(\alpha\mid-) \propto p_\alpha(\alpha) = \Ga(\alpha;a_\alpha,b_\alpha) \prod_{\ell=1}^d \prod_{k^{\star}=1}^{K_{\ell}^{\star}} \left\{ \frac{\Gamma(\alpha)}{\prod_{k=1}^{K}\Gamma(\alpha\lambda_{\ell,0}(k))} \cdot \frac{\prod_{k=1}^{K}\Gamma(\alpha\lambda_{\ell,0}(k)+n_{\ell,k^{\star}}(k))}{\Gamma(\alpha+n_{\ell,k^{\star}})} \right\}, \\
& p(\phi\mid-) \propto p_\phi(\phi) = \Ga(\phi;a_\phi,b_\phi) \prod_{\ell=1}^d\prod_{h=1}^p \left\{ \frac{\Gamma(\phi)}{\prod_{q_h=1}^{d_h}\Gamma(\phi/d_h)} \cdot \frac{\prod_{q_h=1}^{d_h}\Gamma(\phi/d_h+m_{\ell,h}(q_{h}))}{\Gamma(\phi+d_h)} \right\},
\ese
with $m_{\ell,h}(q_{h}) = \sum_{c_{h}=1}^{d_{h}} \Ind\{s_{\ell,h}^{(c_{h})}=q_h\}$.
The variances $\sigma^{2}_{\alpha}$ and $\sigma^{2}_{\phi}$ are updated adaptively throughout the burn-in phase of the algorithm to maintain the acceptance probability near the optimal value of 0.44 \citep{roberts_OptimalScalingVarious_2001}. Specifically, every 50 iterations, each variance is adjusted by $\min\{0.01, b^{-0.5}\}$: it is increased if the acceptance probability exceeds 0.44, and decreased if it falls below 0.44, where $b$ denotes the current MCMC iteration.

Similarly, no closed form is available for the atoms of the truncated normal mixtures; hence we update them via M-H steps.
A new value $\mu^{(new)}$ is proposed from $\TN(\mu_{k},\sigma_{\mu}^{2},[A,B])$, with fixed $\sigma_{\mu}^{2}$; analogously, a new value $\sigma^{2(new)}$ is proposed from $\TN(\sigma_{k}^{2}, \sigma^{2}_{\sigma}, [\max\{10^{-6},\sigma_{k}^{2}-1\},\sigma_{k}^{2}+1])$, with fixed $\sigma^{2}_{\sigma}$.
These values are respectively accepted with probability
\bse
& \min\left\{1, \frac{p_{\mu_{k}}(\mu^{(new)})}{p_{\mu_{k}}(\mu_{k})} \cdot \frac{\text{TN}(\mu_{k};\mu^{(new)},\sigma_{\mu}^{2},[A,B])}{\text{TN}(\mu^{(new)};\mu_{k},\sigma_{\mu}^{2},[A,B])} \right\}, \\
& \min\left\{1, \frac{p_{\sigma_{k}}(\sigma^{2(new)})}{p_{\sigma_{k}}(\sigma^{2}_{k})} \cdot \frac{\text{TN}(\sigma^{2}_{k};\sigma^{2(new)}, \sigma^{2}_{\sigma}, [\max\{10^{-6},\sigma^{2(new)}-1\},\sigma^{2(new)}+1])}{\text{TN}(\sigma^{2(new)}; \sigma^{2}_{k}, \sigma^{2}_{\sigma}, [\max\{10^{-6},\sigma_{k}^{2}-1\},\sigma_{k}^{2}+1])} \right\},
\ese
where
\bse
& p(\mu_{k}=\mu\mid-) \propto p_{\mu_{k}}(\mu) = \TN(\mu; m_{0},s_{0}^{2}, [A,B]) \prod_{(\ell,i):z_{\ell,i}=k} \TN(x_{\ell,i}; \mu,\sigma_{k}^{2},[A,B]), \\
& p(\sigma^{2}_{k}=\sigma^{2}\mid-) \propto p_{\sigma_{k}}(\sigma^{2}) = \IG(\sigma^{2};a_{\sigma},b_{\sigma}) \prod_{(\ell,i):z_{\ell,i}=k} \TN(x_{\ell,i}; \mu_{k},\sigma^{2},[A,B]).
\ese

The dimensionality of the latent space depends on the number of clusters $K_{\ell,h}$, identified in the first layer, since the number of clusters allocations in $\bs^{\star}_{\ell}$ is given by $\prod_{h=1}^{p}K_{\ell,h}$, for $\ell=1,\ldots,d$.
Therefore, a joint update for the cluster allocations in $\bs_{\ell}$ and $\bs_{\ell}^{\star}$ is required.
We employ a trans-dimensional M-H transition step which jointly updates $(\bs_{\ell,h},\bs^{\star}_{\ell})$, for $\ell=1,\ldots,d$, $h=1,\ldots,p$, resulting in the partitions in the first layer being potentially updated one time at each MCMC iteration, while the partitions in the second layer potentially $p$ times.
We consider a two-step proposal where, first, $\bs_{\ell,h}^{(new)}$ is randomly sampled from a Hamming ball of radius 1 around the current partition $\bs_{\ell,h}$, $\text{HB}_1(\bs_{\ell,h})$, and then the cluster allocations in $\bs^{\star (new)}_{\ell}$, whose dimension now depends on the number of clusters $K_{\ell,h}^{(new)}$ in $\bs_{\ell,h}^{(new)}$, are sampled randomly.
Let $q_1$ and $q_2$ be the proposal distributions for the two steps, then the two-step proposal $q$ is given by
\bse
q(\bs^{(new)}_{\ell,h},\bs^{\star (new)}_{\ell}) = q_1(\bs^{(new)}_{\ell,h}\mid\bs_{\ell,h}) \times q_2(\bs^{\star (new)}_{\ell} \mid \bs^{(new)}_{\ell,h}, \bs_{\ell,-h}) \qquad\qquad\qquad\quad\;\; \\
= \frac{1}{|\text{HB}_1(\bs_{\ell,h})|} \Ind\left\{\bs^{(new)}_{\ell,h}\in\text{HB}_1(\bs_{\ell,h})\right\} \times \prod_{\bk\in\bK^{(h)}_{\ell}} \Cat_{K^{\star}_{\ell}}\left(\frac{1}{K^{\star}_{\ell}}\right),
\ese
where $\bK^{(h)}_{\ell} = \{(k_{1},\ldots,k_{p}): k_{h}\in\{1,\ldots,K^{(new)}_{\ell,h}\}, k_{h'}\in\{1,\ldots,K_{\ell,h'}\}, h'\neq h \}$, and $\bs_{\ell,-h} = \{\bs^{(new)}_{\ell,h'}\}_{h':h'\neq h}$. 
A new value is accepted with probability
\bse
\min\left\{1, \frac{p_{\bs_{\ell,h},\bs_{\ell}^{\star}}(\bs^{(new)}_{\ell,h},\bs^{\star (new)}_{\ell})}{p_{\bs_{\ell,h},\bs_{\ell}^{\star}}(\bs_{\ell,h},\bs^{\star}_{\ell})} \cdot \frac{q_1(\bs_{\ell,h}\mid\bs^{(new)}_{\ell,h}) q_2(\bs^{\star}_{\ell} \mid \bs_{\ell,h}, \bs_{\ell,-h})}{q_1(\bs^{(new)}_{\ell,h}\mid\bs_{\ell,h}) q_2(\bs^{\star (new)}_{\ell} \mid \bs^{(new)}_{\ell,h}, \bs_{\ell,-h})} \right\},
\ese
where $p_{\bs_{\ell,h},\bs_{\ell}^{\star}}(\cdot)$ is the unnormalized full conditional for $(\bs_{\ell,h},\bs^{\star}_{\ell})$,
\bse
& \Pr(\bs_{\ell,h},\bs^{\star}_{\ell}\mid-) \propto p_{\bs_{\ell,h},\bs_{\ell}^{\star}}(\bs_{\ell,h},\bs^{\star}_{\ell}) = \prod_{k^{\star}=1}^{K_{\ell}^{\star}} \frac{\prod_{k=1}^{K}\Gamma(\alpha\lambda_{\ell,0}(k)+n_{\ell,k^{\star}}(k))}{\Gamma(\alpha+n_{\ell,k^{\star}})} \\
& \times \prod_{q_h=1}^{d_h}\Gamma(\phi/d_h+m_{\ell,h}(q_{h}))
\times \frac{\prod_{k^{\star}=1}^{K_{\ell}^{\star}}\Gamma(\phi^{\star}/K_{\ell}^{\star}+m^{\star}_{\ell}(k^{\star}))}{\Gamma(\phi^{\star}+\prod_{h=1}^{p}K_{\ell,h})},
\ese
and the remaining part is the proposal ratio
\bse
\frac{q_1(\bs_{\ell,h}\mid\bs^{(new)}_{\ell,h}) q_2(\bs^{\star}_{\ell} \mid \bs_{\ell,h}, \bs_{\ell,-h})}{q_1(\bs^{(new)}_{\ell,h}\mid\bs_{\ell,h}) q_2(\bs^{\star (new)}_{\ell} \mid \bs^{(new)}_{\ell,h}, \bs_{\ell,-h})}
= (K^{\star}_{\ell})^{K_{\ell,h}^{(new)}\prod_{h'\neq h}K_{\ell,h'} - \prod_{h'=1}^{p}K_{\ell,h'}}.
\ese

The full conditional distribution for $s^{\star}_{\ell,\bk}$ is
\bse
& \Pr(s^{\star}_{\ell,\bk} = s\mid-) \propto \left\{ \frac{\phi^{\star}}{K^{\star}_{\ell}} + m^{\star (-\bk)}_{\ell}(s) \right\} \times \prod_{k^{\star}=1}^{K^{\star}_{\ell}} \frac{\prod_{k=1}^{K}\prod_{q=0}^{n_{\ell,k^{\star}}^{(\bk)}(k)-1} \left\{\alpha\lambda_{\ell,0}(k) + n_{\ell,k^{\star}}^{(-\bk)}(k) + q\right\}}{\prod_{q=0}^{\sum_{k=1}^{K}n_{\ell,k^{\star}}^{(\bk)}(k)-1} \left\{\alpha + \sum_{k=1}^{K}n_{\ell,k^{\star}}^{(-\bk)}(k) + q\right\}},
\ese
where $m^{\star (-\bk)}_{\ell}(k^{\star})$ denotes the count $m^\star_\ell(k^\star)=\sum_{\bk'\in\S_\ell}\Ind\{s^\star_{\ell,\bk'}=k^\star\}$ without considering the combination $\bk$, and the counts are defined as $n_{\ell,k^{\star}}^{(\bk)}(k) = \sum_{i=1}^{n}\Ind\{\bs_{\ell}^{(\bc_i)}=\bk\}\Ind\{\wt{s}_{\ell,\bc_{i}}=k^{\star}\}\Ind\{z_{\ell,i}=k\}$, and $n_{\ell,k^{\star}}^{(-\bk)}(k) = \sum_{i=1}^{n}\Ind\{\bs_{\ell}^{(\bc_i)}\neq\bk\}\Ind\{\wt{s}_{\ell,\bc_{i}}=k^{\star}\}\Ind\{z_{\ell,i}=k\}$.

\subsubsection{Updating the Copula Parameters}
The full conditional distributions of the copula parameters do not have closed forms; however, they can be easily updated via M-H steps. We discretize the values of $b_{s'}$ and $\theta_{s''}$ to grids of $M_{b}$ and $M_{\theta}$ values, respectively. The $m\th$ value of the grids is defined as $\{-0.99+2\times0.99(m-1)/(M_{b}-1)\}$ and $\{-3.14+2\times3.14(m-1)/(M_{\theta}-1)\}$, where $M_{b}$ and $M_{\theta}$ are fixed integers.
A new value $b^{(new)}$, or $\theta^{(new)}$, is proposed at random from the set comprising the current value of the parameter and its neighbors on the grid.
These are respectively accepted with probability
\bse
\min\left\{1, \frac{p_{b_{s'}}(b^{(new)})}{p_{b_{s'}}(b_{s'})} \cdot \frac{q(b_{s'} \mid b^{(new)})}{q(b^{(new)} \mid b_{s'})} \right\}, \quad \min\left\{1, \frac{p_{\theta_{s''}}(\theta^{(new)})}{p_{\theta_{s''}}(\theta_{s''})} \cdot \frac{q(\theta_{s''} \mid \theta^{(new)})}{q(\theta^{(new)} \mid \theta_{s''})} \right\},
\ese
with
\bse
q(b_{s'} \mid b^{(new)}) = 
\begin{cases}
    1/2 & \text{if $|b^{(new)}|=0.99 $} \\
    1/3 & \text{otherwise}
\end{cases}, \quad
q(\theta_{s''} \mid \theta^{(new)}) = 
\begin{cases}
    1/2 & \text{if $|\theta^{(new)}|=3.14 $} \\
    1/3 & \text{otherwise}
\end{cases}
\ese
and
\bse
& p(b\mid-) \propto p_{b_{s'}}(b) = (1-b^{2})^{-\frac{n}{2}} \prod_{i=1}^{n}\exp\left\{-\frac{1}{2}\by_{i}\trans\bR_{s',\cdot}(b)^{-1}\by_{i}\right\}, \\
& p(\theta\mid-) \propto p_{\theta_{s''}}(\theta) = \prod_{i=1}^{n}\exp\left\{-\frac{1}{2}\by_{i}\trans\bR_{\cdot,s''}(\theta)^{-1}\by_{i}\right\},
\ese
where $\bR_{s',\cdot}(b)$ denotes the correlation matrix obtained considering $\bb$ and $\btheta$ with $b_{s'}=b$, and $\bR_{\cdot,s''}(\theta)$ denotes the correlation matrix obtained considering $\bb$ and $\btheta$ with $\theta_{s''}=\theta$. The $\ell\th$ element of $\by_{i}$, $y_{\ell,i}$, depends on the atoms $\bmu$, $\bsigma^2$ and the mixture weights $\blambda_{\ell}$, and is given by the cumulative distribution function of a mixture of truncated normal distributions with means $\mu_{k}$, variances $\sigma^{2}_{k}$, $k=1,\ldots,K$, and support $[A,B]\subseteq\rR$ evaluated at $x_{\ell,i}$, $y_{\ell,i} = \Phi^{-1}\left\{ \sum_{k=1}^{K} \wh{\lambda}_{\ell,\wt{s}_{\ell,\bc_{i}}}(k) F_{\text{TN}}\left(x_{\ell,i}; \mu_{k}, \sigma^{2}_{k}, [A,B]\right) \right\}$. The mixture weight $\wh{\lambda}_{\ell,\wt{s}_{\ell,\bc_{i}}}(k)$ is the conditional expected value of $\lambda_{\ell,\wt{s}_{\ell,\bc_{i}}}(k)$ at the current MCMC iteration; see Section S.2 of the Supplementary Materials for its definition.

\subsection{Strategies for Improving Speed and Scalability}\label{sec:comp_strategies}

To improve computational efficiency, we integrated some elements of streamlined memory management into our model and the MCMC implementation, as detailed below.

(1) Rather than using a variable maximum for the number of distinct densities ($\prod_{h=1}^{p}K_{\ell,h}$), which fluctuates as clusters emerge or vanish during MCMC, we adopt a fixed, sufficiently large maximum $K_{\ell}^{\star}$. This scalar serves a role similar to the multirank in the tensor factorization model \eqref{eq:tensor_dec} but is computationally much simpler to manage.

(2) Drawing on the common practice of over-specifying mixture components, we set $K_{1}^{\star},\ldots,K_{d}^{\star}$ to sufficiently large values that preserve model flexibility yet maintain computational and memory efficiency. 
This rationale similarly motivates our use of finite mixtures, rather than infinite processes, for the marginal distributions.

(3) We explicitly integrate the mixture weights $\blambda_{\ell,k^{\star}}$ out of the model and work with the marginalized model at all stages of the analysis. 
By further marginalizing the remaining parameters with conjugate priors, i.e., the probability vectors $\etam_{\ell,h}$ and $\etam_{\ell}^{\star}$, we improve the mixing of the Markov chain algorithm. 

\begin{figure}[!ht]
   \begin{center}
   \includegraphics[width=\textwidth, trim=0cm 0.35cm 0cm 0.35cm, clip=true]{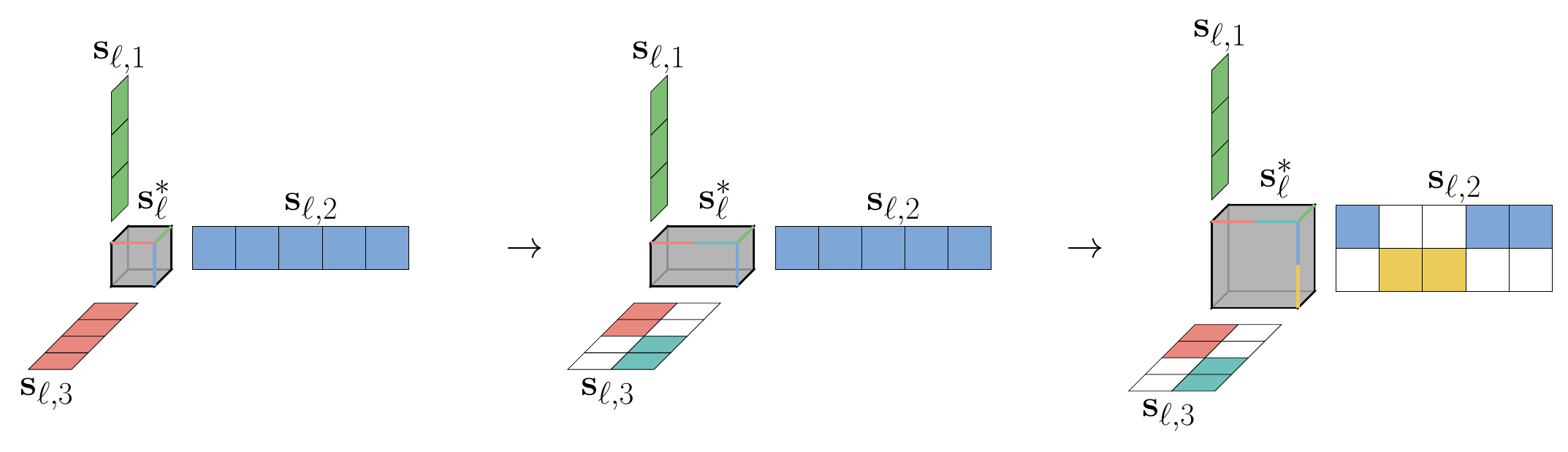}
   \end{center}
   \vspace*{-3ex}
   \caption{Evolution of memory-allocated second-layer partition, $\bs_{\ell}^{\star}$, as the number of clusters increases in the partitions of covariate levels. 
   Here, each cluster is represented with a different color.
   {The cardinality of a partition (number of distinct colors) determines the size of the corresponding dimension of the second-layer partition.}
   Example with 3 covariates with $(d_{1},d_{2},d_{3})=(3,5,4)$.}
   \label{fig:tensor_evolution}
\end{figure}

(4) Finally, to minimize memory usage and encourage fewer partitions of the covariate levels, we initialize all covariate partitions 
$\bs_{\ell,h}$ with a single cluster, so each $\bs^{\star}_{\ell}$ is a $(1\times\cdots\times1)$-dimensional tensor. 
New partition slices are added only when new clusters emerge in the first layer. 
As illustrated in \autoref{fig:tensor_evolution}, where partitions in the first layer are shown in their equivalent representation as constrained mode matrices, every increase of one in $K_{\ell,h}$ adds a new slice containing $\prod_{h'\neq h}K_{\ell,h'}$ count vectors to $\bs^{\star}_{\ell}$.

Overall, these strategies enable efficient inference of varying parameter tensor sizes and facilitate the exploration of complex posterior spaces.
The algorithm’s memory requirements now scale with the total number of covariate levels and the number of clusters identified by the model, rather than with the total number of covariate combinations.

\vspace*{-2ex}
\subsection{Runtime and Other Details}\label{sec:runtime}
The MCMC sampler is implemented in C++ and integrated into R \citep{RCoreTeam2024} via the Rcpp \citep{RcppJournal2011,RcppBook2013,RcppJournal2018,RcppPackage2025} and RcppArmadillo frameworks \citep{RcppArmadilloJournal2014,RcppArmadilloManual2025}. 
For the NHANES dataset analyzed in Section \ref{sec:application}, the MCMC sampler takes 130 minutes to run on a machine equipped with two AMD EPYC 7763 processors (2.45~GHz base, 128 physical cores, 256 threads total) and approximately 881~GiB of RAM.

\section{Analysis of NHANES Dietary Data}\label{sec:application}
We discuss here the results obtained by fitting the flower model to dietary data collected in the What We Eat in America (WWEIA) nutritional assessment component of the National Health and Nutrition Examination Survey (NHANES), conducted by the Centers for Disease Control and Prevention (CDC). 
The survey serves as a tool to evaluate eating habits in the United States and aims to gain a better understanding of the relationship between diet, nutrition, and health.
Furthermore, NHANES is fundamental to evaluating the impact of program changes, including welfare reform,
legislation, food fortification policy, and child nutrition programs\footnote{Source: ``Dietary Recall'' PDF, available at the bottom of this \href{https://wwwn.cdc.gov/nchs/nhanes/continuousnhanes/questionnaires.aspx?BeginYear=2017}{webpage}.}.
The nutritional assessment consists of two 24-hour dietary recall interviews, taking place three to ten days apart, in which participants reported the types and amounts of food consumed. 
Although interviews are conducted on a continuous basis, they are grouped into two-year cycles from 1999. 
See \citet{ho2024hei} for more details.
We focus on the 2017–2018 cycle, which includes 24-hour dietary recall data for $8704$ participants, along with their categorized demographic information on sex, age, race, and family income.
Following NHANES guidelines, here we only consider individuals who were at least one year old and participated in both recalls, resulting in a reduced dataset comprising $n=6087$ individuals. 
For the analysis presented here, we use the mean of the two dietary recalls for $d=6$ regularly consumed dietary components 
as our multivariate response and the above-mentioned $p = 4$ associated demographic variables as our covariates. Covariate levels are reported in the first column of \autoref{tab:nhanes_partitions1}, and the names of the dietary components are in the first row.

We fit our model using the MCMC algorithm described in \autoref{sec:inference}.
Posterior inference is based on 2,000 samples, retained after 30,000 total iterations, discarding the first 20,000, and then thinning by 5.
For each coordinate $\ell$, we obtain the MAP estimate for the partitions in $(\bs_{\ell},\bs_{\ell}^{\star})$ and the conditional marginal density $f_{x_{\ell}\mid\bc}(\cdot\,;\bc)$ for each covariate combination $\bc\in\C$, which are jointly used to build the figures reported below.
Additional details are reported in Sections S.2 and S.3 of the Supplementary Materials.

Posterior checks to evaluate the goodness of fit of the model are reported in Section S.4 of the Supplementary Materials.
Comparing empirical means and standard deviations against their estimated counterparts (Figure S.1) shows that the conditional marginal densities are consistent with the original covariate combinations. For a representative covariate combination, the observations fall well within the high-density regions of the estimated conditional marginal and bivariate densities (Figure S.2). Additionally, a unit-by-unit comparison with a replicated dataset drawn from the posterior predictive (Figure S.3) shows differences centered and symmetric around zero, confirming that the copula and shared atoms capture the cross-component dependence.

\begin{figure}[!ht]
    \centering
    \includegraphics[width=0.98\linewidth]{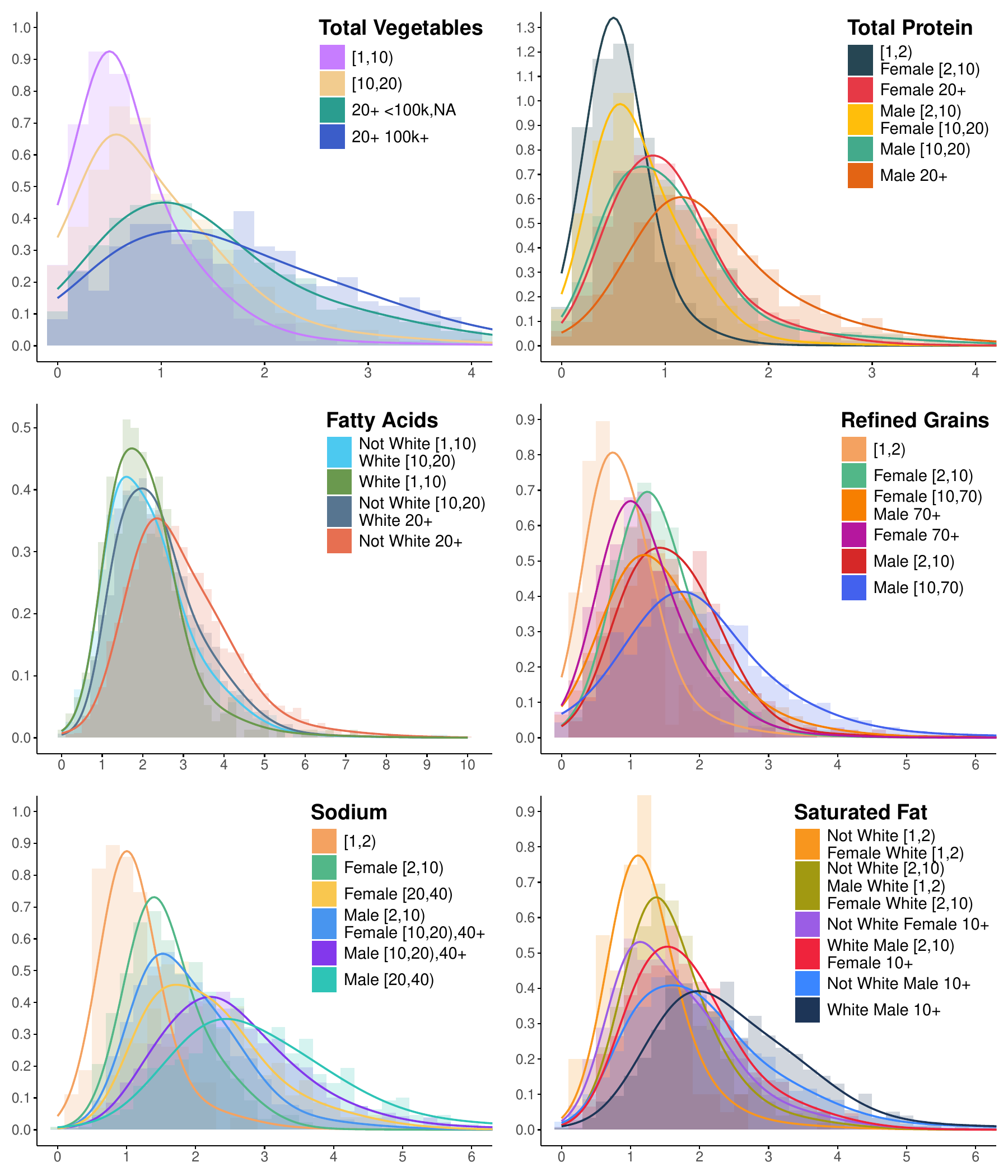}
    \caption[Estimated conditional marginal densities for any group of units identified by the posterior point estimates of the partitions.]{Results for NHANES data: The panels, one for each coordinate of the response, show estimated conditional densities, identified by the posterior partitions of covariate level combinations, overlaid on histograms for the corresponding empirical distributions.}
    \label{fig:nhanes_marginals}
\end{figure}

First, we examine the effects of each covariate on each response coordinate individually.
Recall that our model achieves a parsimonious representation by defining conditional marginal densities over clustered covariate levels. 
This effectively reduces the number of unique distributions per coordinate from $\prod_{h=1}^{p}d_{h}=504$ possible covariate combinations to a maximum of $K^{\star}_{1}=\cdots=K^{\star}_{d}=K^{\star}=20$ distinct densities. 
Aggregated levels are first induced marginally for each covariate via the first partition layer $\bs_{\ell,1}, \dots, \bs_{\ell,p}$. These levels are then further refined by a second partition layer $\bs_{\ell}^{\star}$, which clusters covariate combinations to define the final group structure.
Aggregated covariate levels determined by the first layer are reported in \autoref{tab:nhanes_partitions1}, while groups of covariate combinations can be found in \autoref{fig:nhanes_marginals}, where the corresponding posterior estimates of the conditional marginal densities $f_{x_{\ell}\mid\bc}(\cdot\,;\bc)$ are shown.
A color is assigned to each group of covariate combinations, which can be defined in multiple lines in the legend of \autoref{fig:nhanes_marginals}; for instance, the first group of the Total Protein component represents any participant aged 1-2 and female participants aged 2-10.

\begin{table}[!ht]
    \footnotesize
    \centering
    \begin{tabular}{ccccccc}
        \hline
                         & Total            & Total          & Fatty             & Refined          & Sodium           & Saturated     \\
                         & Vegetables       & Protein        & Acids             & Grains           &                  & Fat           \\
        \hline
        Female           & All              & Female         & All               & Female           & Female           & Female        \\
        Male             &                  & Male           &                   & Male             & Male             & Male          \\
        \hdashline\noalign{\vskip 1pt}
        $[1,2)$          & $[1,10)$         & $[1,2)$        & $[1,10)$          & $[1,2)$          & $[1,2)$          & $[1,2)$       \\
        $[2,10)$         & $[10,20)$        & $[2,10)$       & $[10,20)$         & $[2,10)$         & $[2,10)$         & $[2,10)$      \\
        $[10,20)$        & $20+$            & $[10,20)$      & $20+$             & $[10,70)$        & $[10,20),40+$    & $10+$         \\
        $[20,40)$        &                  & $20+$          &                   & $70+$            & $[20,40)$        &               \\
        $[40,60)$        &                  &                &                   &                  &                  &               \\
        $[60,70)$        &                  &                &                   &                  &                  &               \\
        $70+$            &                  &                &                   &                  &                  &               \\
        \hdashline\noalign{\vskip 1pt}
        Asian            & All              & All            & White             & All              & All              & White         \\
        Black            &                  &                & Not White         &                  &                  & Not White     \\
        Mexican American &                  &                &                   &                  &                  &               \\
        Other            &                  &                &                   &                  &                  &               \\
        Other Hispanic   &                  &                &                   &                  &                  &               \\
        White            &                  &                &                   &                  &                  &               \\
        \hdashline\noalign{\vskip 1pt}
        $[0,20000)$      & $100\text{k}+$            & All            & All               & All              & All              & All           \\
        $[20000,35000)$  & $<100\text{k}$ and NA      &                &                   &                  &                  &               \\
        $[35000,55000)$  &                  &                &                   &                  &                  &               \\
        $[55000,100000)$ &                  &                &                   &                  &                  &               \\
        $100000+$        &                  &                &                   &                  &                  &               \\
        NA               &                  &                &                   &                  &                  &               \\
        \hline
    \end{tabular}
    \caption{Results for NHANES data: Original (first column) and aggregated levels of the covariates sex, age, race, and income, induced by the first partition layer; the unique group `All' implies that the covariate is not influential for that dietary component.}
    \label{tab:nhanes_partitions1}
\end{table}

\begin{figure}[!ht]
    \centering
    \includegraphics[width=\linewidth]{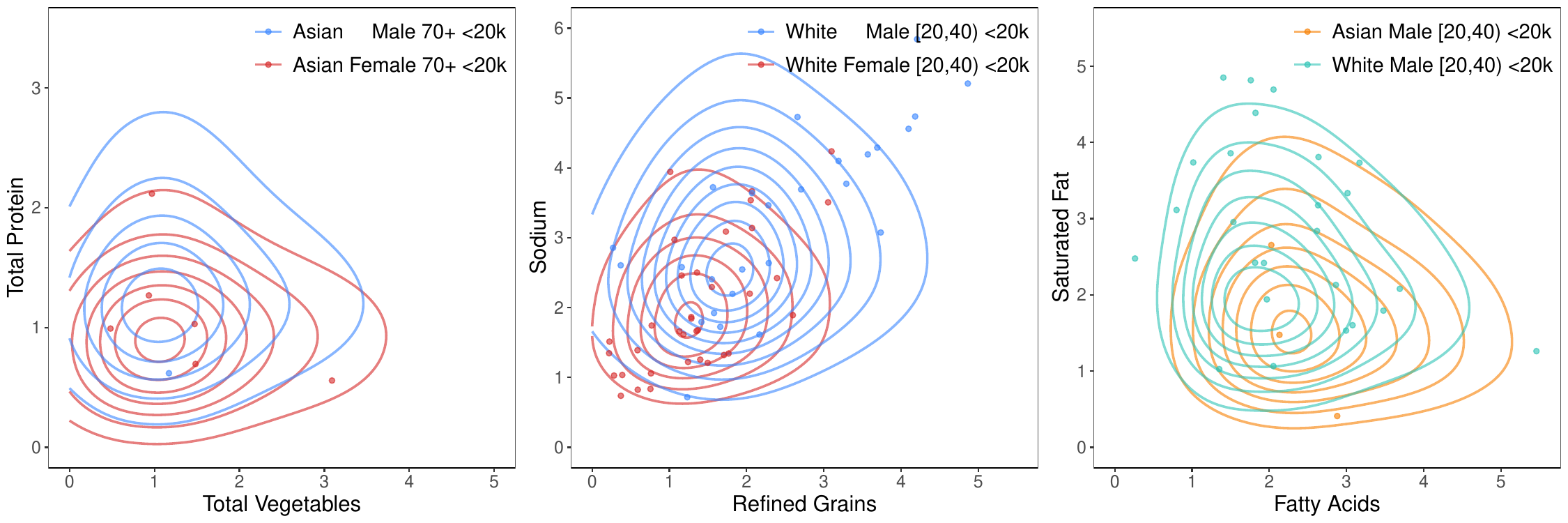}
    \caption{Results for NHANES data: Estimated contours of conditional bivariate densities for a few partitions of covariate level combinations identified a posteriori.}
    \label{fig:nhanes_contours}
\end{figure}

Age is the only covariate that is influential across all six dietary components, although the age aggregations differ by dietary component.
The remaining covariates exhibit selective relevance, with sex being influential for four components, race for two, and income for one.
Across all components, the estimated conditional marginal densities are unimodal and right-skewed.
The location of the modes is primarily driven by age, with additional shifts attributable to sex, for Total Protein, Refined Grains, Sodium, and Saturated Fat components.
For the remaining components, age and race/income jointly influence the distributions' centers and shapes, showing systematic intake differences within comparable age groups.
For instance, Non-White children aged 1-9 display Fatty Acids intake levels comparable to those of White children and teenagers aged 10-19.
Notably, the Total Vegetables component shows the importance of the second layer.
Income is relevant only when individuals are 20 years old or older.

We report contour plots of posterior point estimates of conditional bivariate densities in \autoref{fig:nhanes_contours}.
Each plot compares the joint densities of two covariate combinations that differ in only one level (sex, race).
The left panel shows virtually no difference between females and males in Total Vegetables intake, but a higher Total Protein intake for males. This aligns with the estimated partitions, where sex is influential for the Total Protein component only.
The center panel displays a clear separation between females and males, with males exhibiting higher intake of both components.
The right panel highlights distinct dietary patterns between Asian and White individuals, with the former showing higher Saturated Fat but lower Fatty Acids intake, and the latter exhibiting the opposite pattern.

Finally, we evaluate the uncertainty surrounding the number of distinct conditional marginal densities in \autoref{fig:nhanes_marginals} by examining their posterior distributions. 
Figure S.4a in the Supplementary Materials indicates that all posterior distributions concentrate their mass between 3 and 9 distinct densities—far fewer than the total number of covariate combinations. 
Total Protein and Saturated Fat exhibit slightly higher uncertainty, with posterior mass split almost evenly between two values, whereas the remaining response components concentrate most of their mass on a single value.
Additionally, Figure S.4b in the Supplementary Materials displays the posterior distribution for the number of distinct conditional joint densities $f_{\bx\mid\bc}(\cdot\,;\bc)$. 
This distribution ranges between 32 and 48, with most of its mass concentrated between 33 and 36. 
This underscores the model's ability to parsimoniously summarize the joint behavior of the six dietary components across all demographic groups using fewer than 50 distinct joint densities.

\section{Simulation Study} \label{sec:simulations}
We evaluate the performance of the proposed flower model against other methods using two simulation scenarios. 
The main objectives are to assess the model's ability to recover the true conditional marginal densities and to identify the most influential covariates for each response coordinate $\ell$. 
Additionally, we demonstrate the model's ability to accurately reconstruct the latent partition structure across response coordinates.

We first considered a simpler scenario (Scenario A) with $n\in\{1000,2000,3000\}$ and 3-dimensional responses in the presence of 5 covariates, and then a more complex scenario (Scenario B) that emulates the characteristics of NHANES dietary data, consisting of $n=6087$ units with 6-dimensional responses and 4 covariates.
Covariates are randomly sampled in Scenario A, while they coincide with those in the NHANES data in Scenario B. 
The multivariate response is assumed to be distributed as the model introduced in \eqref{eq:joint}, \eqref{eq:copula}, \eqref{eq:marginal_part1} without the common atoms, leading to $f_{\bx\mid\bc}(\bx_{i};\bc_{i}) = C(\bx_{i})\prod_{\ell=1}^{d}f_{x_{\ell}\mid\bc}(x_{\ell,i};\bc_{i})$ with 
\bse
& C(\bx_{i}) = |\bR|^{-\frac{1}{2}} \exp\left\{-\frac{1}{2}\by_{i}\trans(\bR^{-1}- \bI_{d})\by_{i}\right\}, \\ 
& f_{x_{\ell}\mid\bc}(x_{\ell,i};\bc_{i}) = \sum_{k=1}^{K} \lambda_{\ell,\wt{s}_{\ell,\bc_{i}}}(k) \TN\left(x_{\ell,i}; \mu_{\ell,k}, \sigma^{2}_{\ell,k}, [A,B]\right).
\ese
In Scenario A, we sample the core vector elements $\blambda_{\ell,k^{\star}}$ from a symmetric Dirichlet distribution, and fix the correlation matrix $\bR$, the atoms $(\bmu,\bsigma^2)$ and the partitions $(\bs,\bs^{\star})$; in Scenario B, we use the posterior point estimates of $\bR$ and $f_{x_{\ell}\mid\bc}(\cdot\,;\bc)$. 
The atoms in Scenario A are chosen such that the resulting marginal densities exhibit more complex behaviors than the skewed unimodal distributions observed in NHANES data. 
Details are reported in Section S.5 of the Supplementary Materials.

We fit our flower model and compare it with two univariate and one multivariate density regression methods.
The univariate approaches, independently estimated for each coordinate of the response, are a regression-scale Pitman-Yor Mixture Model \citep[PYMM,][]{corradin_BNPmixPackageBayesian_2021}, and a Box-Cox-inspired non-linear regression model \citep{hothorn_ConditionalTransformationModels_2014} with parameters partitioned over the covariate space through a tree \citep[BoxCoxTree,][]{hothorn_PredictiveDistributionModeling_2021}.
The multivariate approach is the Multivariate Conditional Transformation Model \citep[MCTM,][]{klein_MultivariateConditionalTransformation_2022}. 
For the latter, we consider three different formulations of the marginal distributions: linear regression (MCTM-Lm), Box-Cox-inspired non-linear regression \citep[MCTM-BoxCox,][]{hothorn_ConditionalTransformationModels_2014}, and continuous outcome logistic regression \citep[MCTM-Colr,][]{lohse_ContinuousOutcomeLogistic_2017}.

To evaluate the similarity between the posterior estimates of the marginal densities, $\widehat{f}_{x_{\ell}\mid\bc}(\cdot\,;\bc)$, and their true counterparts, $f_{x_{\ell}\mid\bc}(\cdot\,;\bc)$,  we compute the Integrated Squared Error (ISE) between the two quantities as
\bse
\text{ISE}(\ell,{\bc}) = \sum_{g=1}^{G} \left\{ f_{x_{\ell}\mid\bc}(\widetilde{x}_{g}) - \widehat{f}_{x_{\ell}\mid\bc}(\widetilde{x}_{g};\bc) \right\}^{2} \Delta_{\widetilde{x}},
\ese
where $\Delta_{\widetilde{x}} = \widetilde{x}_{2}-\widetilde{x}_{1}$ is the distance between two consecutive points in the equispaced grid $(\widetilde{x}_{1},\ldots,\widetilde{x}_{G})$ of $G=300$ points over the interval $[A,B]$. 
Specifically, we report an overall summary, computed as $\text{ISE}=\sum_{\ell=1}^{d}\sum_{\bc\in\C}\text{ISE}(\ell,\bc)$, for both scenarios in \autoref{tab:sim_ISE}.
In the simpler Scenario A, our approach consistently significantly outperforms all other alternatives, among which BoxCoxTree performs best.
Notably, BoxCoxTree is the only competitor that partitions the covariate space to define cluster-specific conditional marginal densities. The remaining methods, lacking this mechanism, fail to recover the true underlying densities shared across covariate combinations. 
As expected, the ISE tends to decrease monotonically across all models as the sample size grows. 
In the more complex Scenario B, featuring a higher-dimensional response and larger sample sizes mimicking the NHANES dataset, our model maintains its superior performance, whereas BoxCoxTree yields results comparable to MCTM-BoxCox and MCTM-Colr.

\begin{table}
    \footnotesize
    \centering
    \begin{tabular}{c|ccc|c}
        & \multicolumn{3}{c|}{Scenario A} & Scenario B \\
        model & $n=1000$ & $n=2000$ & $n=3000$ & $n=6087$ \\
        \hline
        PYMM & 0.0348 (0.0044) & 0.0367 (0.0157) & 0.0365 (0.0135) & 0.0098 (0.0006) \\ 
        BoxCoxTree & 0.0066 (0.0010) & 0.0072 (0.0013) & 0.0071 (0.0015) & 0.0066 (0.0023) \\ 
        \hdashline\noalign{\vskip 1pt}
        MCTM-Lm & 0.0375 (0.0089) & 0.0363 (0.0104) & 0.0357 (0.0107) & 0.0233 (0.0011) \\ 
        MCTM-BoxCox & 0.0237 (0.0039) & 0.0227 (0.0035) & 0.0217 (0.0033) & 0.0068 (0.0004) \\ 
        MCTM-Colr & 0.0216 (0.0042) & 0.0202 (0.0033) & 0.0192 (0.0033) & 0.0053 (0.0005) \\ 
        FlowerModel & \textbf{0.0010 (0.0007)} & \textbf{0.0008 (0.0002)} & \textbf{0.0004 (0.0002)} & \textbf{0.0035 (0.0005)} \\ 
    \end{tabular}
    \caption{Results for simulated data. Median ISEs (interquartile ranges in parentheses) between the posterior estimates of the conditional marginal densities and their true counterparts for the two simulated scenarios, computed on 5 replications. The horizontal dashed line separates univariate and multivariate approaches.}
    \label{tab:sim_ISE}
\end{table}

To assess the model's ability to retrieve the true partitions of the covariate combinations, we compute the Adjusted Rand Index \citep[ARI,][]{hubert_ComparingPartitions_1985} between the posterior point estimates of the partitions over the covariate combinations and their true counterparts. 
Again, we report an overall summary for both scenarios in \autoref{tab:sim_ari}.
Neither model seems to be influenced by the sample size, and our approach always performs best.
Although the models do not perfectly retrieve the true partitions of the joint covariate space, they are always able to identify the influential covariates for each coordinate.

\begin{table}
    \small
    \centering
    \begin{tabular}{c|ccc|c}
        & \multicolumn{3}{c|}{Scenario A} & Scenario B \\
        model & $n=1000$ & $n=2000$ & $n=3000$ & $n=6087$ \\
        \hline
        BoxCoxTree & 0.8142 (0.0354) & 0.8877 (0.0741) & 0.8877 (0.0278) & 0.8434 (0.0114) \\ 
        \hdashline\noalign{\vskip 1pt}
        FlowerModel & \textbf{0.9437 (0.0271)} & \textbf{0.9165 (0.0464)} & \textbf{0.9437 (0.0563)} & \textbf{0.8691 (0.0457)} \\ 
    \end{tabular}
    \caption{Results for simulated data. Median ARIs (interquartile ranges in parentheses) between the posterior point estimates of the partitions over the covariate combinations and their true counterparts, computed on 5 replications. The horizontal dashed line separates univariate and multivariate approaches.}
    \label{tab:sim_ari}
\end{table}

\begin{figure}[!ht]
    \begin{center}
    \includegraphics[width=\textwidth, trim=0cm 0cm 0cm 0cm, clip=true]{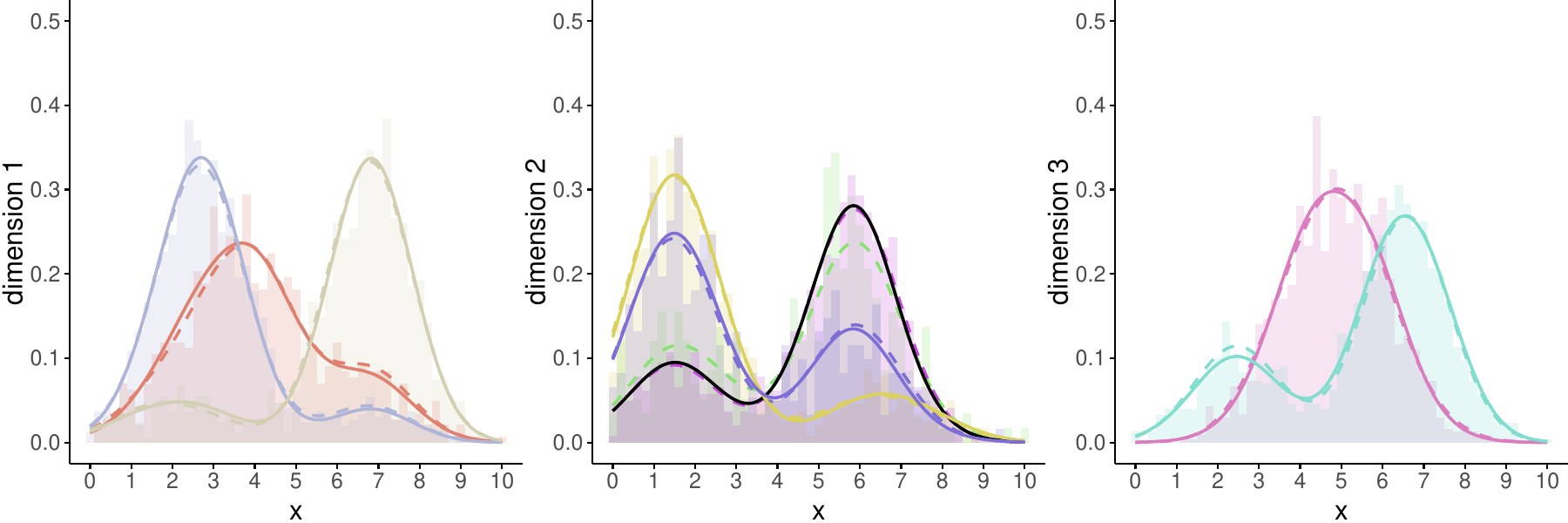}
    \end{center}
    \caption{Results for simulated data: The panels, one for each coordinate of the response, compare the estimated conditional densities for the estimated partitions of the covariate level combinations (solid lines) and the true conditional densities for the true partitions of the covariate level combinations (dashed lines), overlaid on histograms representing the corresponding empirical distributions. 
    Each color refers to a different true underlying distribution.
    Black is used to denote estimated densities of incorrectly identified groups, which typically happens when the model collapses highly similar true distributions or misassigns lower-prevalence groups in the estimated partitions.}
    \label{fig:sim_marginals_toy}
\end{figure}

Finally, we provide a visual comparison of the true and estimated densities.
\autoref{fig:sim_marginals_toy} shows the conditional marginal distributions for Scenario A with $n=3000$.
Posterior estimates of $f_{x_{\ell}\mid\bc}(\cdot\,;\bc)$ given $(\wh{\bs}_{\ell}, \wh{\bs}_{\ell}^{\star})$ for any coordinate $\ell$ are shown as solid lines, while the true marginal distributions used to simulate the responses given the true partitions are shown as dashed lines.
Each color identifies a different true conditional marginal distribution, and the same color is assigned to all estimated densities referring to the same group of covariate combinations.
Black is used to show the estimated density of incorrectly identified groups. 
This usually happens when the model collapses two or more true distributions with highly similar profiles or incorrectly partitions members of lower-prevalence groups. 
For instance, in the middle panel of Figure \ref{fig:sim_marginals_toy}, the model identified a single group for the green and purple true densities, which is reasonable given the very close resemblance of these distributions. 

Similar plots for Scenario B are reported in Figures S.5 and S.6 in Section S.5.2 of the Supplementary Materials.
We recall that in this second configuration, the simulation parameters were specifically calibrated to closely replicate the characteristics of the NHANES dataset. 
Overall, we observe a robust recovery of the underlying true densities, though the accuracy is slightly reduced compared to the simpler Scenario A. 
As expected in such a high-dimensional setting, some discrepancies arise from the complexity of reconstructing a large number of distinct densities.
They generally arise when the model collapses true distributions with very similar profiles or misassigns lower-prevalence groups, resulting in localized deviations from the true covariate-level partitions.
Nevertheless, the high ARI between true and estimated partitions confirms their close agreement, indicating that the reconstruction errors are limited.

In Scenario A, the number of covariates and the underlying true partition are held fixed across replicates.
To examine how the model behaves as these observed and latent characteristics change, Section S.5 of the Supplementary Materials reports additional configurations in which we vary the number of covariates, the number of their levels, and the complexity of the true partition structure.
Across these configurations, the flower model attains the lowest ISE in estimating the conditional marginal densities in all cases but one, 
while remaining competitive in reconstructing the latent partition, achieving the highest ARI whenever the true partition is not overly intricate.

We conclude by examining the computational cost of the sampler, summarized in Table S.7 of the Supplementary Materials. 
The elapsed time grows approximately linearly with the sample size $n$, which is the dominant factor; it also increases with the number of covariates and their levels; however, it is essentially unaffected by the complexity of the latent partition structure.
The memory needed to store the posterior samples of $(\bs, \bs^{\star}, \bmu, \bsigma^2, \bb, \btheta, \wh{\blambda})$ used to estimate the posterior densities is, by contrast, independent of the sample size and is governed mainly by the dimension of the first-layer partitions of the covariates. 
In a minor way, a more complex latent structure further increases the needed memory through the second-layer partitions.
These empirical findings agree with the time and memory complexity derived in Section S.6 of the Supplementary Materials, which shows that the sample size enters the cost only linearly and that neither time nor memory scales with the number of covariate combinations $\prod_{h=1}^{p}d_{h}$.

\section{Discussion} \label{sec:discussion}
In this article, we proposed a flexible Bayesian framework for simultaneous multivariate density regression and coordinate-wise variable selection with categorical covariates. 
The model captures multivariate dependencies via a Gaussian copula, 
while mixtures of truncated normals with shared atoms and covariate-dependent weights represent the marginal distributions. 
Covariate information is integrated through a clustering approach that combines tensor factorization with random partition models. 
By utilizing coordinate-specific partition-based tensor factorizations, the model naturally aggregates covariate combinations and facilitates a comparison of marginal impacts across the multivariate response.
A significant by-product of this aggregation is the identification of influential covariates. 
Through simulation studies and a motivating real-world application, we demonstrated that our model effectively recovers meaningful covariate groupings for each response coordinate, identifying distinct population subgroups characterized by shared marginal distributions.

While our current model captures cross-coordinate dependence via the copula and hierarchical marginal structures, several enhancements are possible.
First, the covariate-independent Gaussian copula could be replaced with more flexible alternatives, such as covariate-dependent correlation matrices \citep{klein_RegressionCopulasMultivariate_2024,kock_TrulyMultivariateStructured_2024} or non-Gaussian copulas that capture tail dependence \citep{czado_AnalyzingDependentData_2019}.
Notably, our two-step estimation approach remains compatible with these complex copula specifications.
Regarding the marginals, one could relax the independence of the tensor factorizations by introducing dependence among mixture weights $\blambda_{\ell,k^{\star}}$, e.g., by enforcing smoothness across covariate levels regardless of response coordinate.
Alternatively, we could encourage similarity across partitions through distance-based penalties to discourage highly dissimilar configurations \citep{paganin_CenteredPartitionProcesses_2021,smith_DemandModelsRandom_2020}.
We also plan to extend the proposed ideas to density deconvolution settings \citep{sarkar_BayesianSemiparametricCovariate_2022,sarkar_BayesianCopulaDensity_2021} to accommodate measurement error in the response variables.


\baselineskip=14pt
\vspace*{-10pt}
\section*{Supplementary Materials}
The supplementary materials provide additional details on posterior computation, hyperparameter selection, and the simulation study, alongside a comparative summary of various univariate and multivariate density regression frameworks.

\vspace*{-20pt}
\section*{Acknowledgments}
The research reported here is supported in part by NSF DMS Grant DMS-2515902. 
We also thank the NHANES investigators, staff, and participants for their dedicated work in conducting the survey and for providing the data used in this research.

\baselineskip=14pt
\bibliographystyle{natbib}
\vspace*{-20pt}
\bibliography{references,references_implementation,references_nutrition}

\clearpage\pagebreak\newpage

\pagestyle{fancy}
\fancyhf{}
\rhead{\bfseries\thepage}
\lhead{\bfseries SUPPLEMENTARY MATERIALS}

\vskip 5mm
\begin{center}
\baselineskip=27pt
{\LARGE Supplementary Materials for\\ 
{\bf Bayesian Semiparametric\\ 
\vspace*{-10pt} Multivariate Density Regression\\ 
with Coordinate-Wise Predictor Selection
}}
\end{center}

\vskip0.25cm

\begin{center}
    Giovanni Toto$^{1}$, Peter M\"uller$^{1,2}$, and Abhra Sarkar$^{1}$\\[.2cm]
    giovanni.toto@austin.utexas.edu, pmueller@math.utexas.edu, and abhra.sarkar@utexas.edu \\[.2cm]
    $^{1}$Department of Statistics and Data Sciences,
    The University of Texas at Austin\\
    $^{2}$Department of Mathematics, The University of Texas at Austin\\
\end{center}

\setcounter{equation}{0}
\setcounter{page}{1}
\setcounter{table}{0}
\setcounter{figure}{0}
\setcounter{section}{0}

\renewcommand{\theequation}{S.\arabic{equation}}
\renewcommand{\thesubsection}{S.\arabic{section}.\arabic{subsection}}
\renewcommand{\thesection}{S.\arabic{section}}
\renewcommand{\thepage}{S.\arabic{page}}
\renewcommand{\thetable}{S.\arabic{table}}
\renewcommand{\thefigure}{S.\arabic{figure}}
\baselineskip=14pt

\makeatletter
\def\hyper@natlinkstart#1{\Hy@backout{#1}\hyper@linkstart{cite}{cite.SM.#1}\def\hyper@nat@current{#1}}
\def\hyper@natlinkbreak#1#2{\hyper@linkend#1\hyper@linkstart{cite}{cite.SM.#2}}
\def\hyper@natanchorstart#1{\Hy@raisedlink{\hyper@anchorstart{cite.SM.#1}}}
\makeatother

\begin{bibunit}[natbib]

\begin{abstract}
\baselineskip=14pt
The supplementary materials provide details on the rescaling of the responses to a common support, posterior computation, hyperparameter choices, posterior checks, the simulation study, and the time and memory complexity of the proposed algorithm, alongside a comparative summary of various univariate and multivariate density regression frameworks. 
\end{abstract}

\section{Transformation to Shared Support}
Let $\bx_{i}=(x_{1,i},\ldots,x_{d,i})\trans\in\rR^{d}$ be a multivariate response and $\bc_{i}=(c_{1,i},\ldots,c_{p,i})\trans\in\mathcal{C}$ covariates for $n$ observational units $i=1,\ldots,n$, with $\mathcal{C}=\{1,\ldots,d_{1}\}\times\cdots\times\{1,\ldots,d_{p}\}$.
Additionally, all response coordinates are assumed continuous and supported on a common interval $[A,B]\subseteq\rR$. 
The common interval is a requirement of the common atoms model, i.e., the mixtures with shared atoms, employed for the marginal distributions.
If the components of $\bx$ have different supports and units of measurement to begin with, one can rescale them to arrive at unit-free coordinates with a common support $[A,B]$ as follows
\bse
A + (B-A) \times \frac{x_{\ell,i} - \min\{x_{\ell,1},\ldots,x_{\ell,n}\}}{\max\{x_{\ell,1},\ldots,x_{\ell,n}\} - \min\{x_{\ell,1},\ldots,x_{\ell,n}\}}, \quad \ell=1,\ldots,d, \quad i=1,\ldots,n.
\ese
Strictly technically speaking, the transformation is data-dependent, since the minima and maxima are functions of the $x_{\ell,i}$'s; but the strategy performs very well in practice.
We consider $[A,B]=[0,10]$ for NHANES dietary data as well as all scenarios of the simulation study.

\vskip 0pt
\section{Posterior Inference - Additional Details}
Our inference is based on samples drawn from the posterior using an MCMC algorithm. 
To avoid the disruption of the conjugacy of the marginal distributions caused by the copula, we follow the iterative approach of \citet{silva_CopulaMarginalDistributions_2008}: first, the latent quantities specifying the marginal distributions are updated using a pseudo-likelihood that ignores the contribution of the copula, and then the copula parameters are updated using the exact likelihood conditionally on the parameters obtained in the first step.
In practice, this means that posterior inference is performed using an MCMC algorithm in which latent quantities specifying the marginal distributions are updated as if the copula did not exist, while the copula parameters are updated exactly.

The model formulated in Section 3 of the manuscript, presented in a compact form, is
\bse
& \textstyle  f_{\bx\mid\bc}(\bx_{i};\bc_{i}) = |\bR|^{-\frac{1}{2}} \exp\left\{-\frac{1}{2}\by_{i}\trans(\bR^{-1}- \bI_{d})\by_{i}\right\} \prod_{\ell=1}^{d} f_{x_{\ell}\mid\bc}(x_{\ell,i}; \bc_{i}), \\
& \textstyle  f_{x_{\ell}\mid\bc}(x_{\ell,i}\mid\bs_{\ell},\bs_{\ell}^{\star},\blambda_{\ell},\bmu,\bsigma^{2};\bc_{i}) = \sum_{k=1}^{K} \lambda_{\ell,\wt{s}_{\ell,\bc_{i}}}(k) \TN\left(x_{\ell,i}; \mu_{k}, \sigma^{2}_{k}, [A,B]\right), \\
& \blambda_{\ell,k^{\star}} \mid \blambda_{\ell,0}, \alpha \sim \Dir_{K}(\alpha\blambda_{\ell,0}), \quad \blambda_{\ell,0} \sim \Dir_{K}(\alpha_{0}/K), \quad \alpha \sim \Ga(a_{\alpha},b_{\alpha}), \\
& s_{\ell,h}^{(c_{h})} \mid \etam_{\ell,h} \sim \Cat_{d_{h}}(\etam_{\ell,h}), \quad \etam_{\ell,h} \mid \phi \sim \Dir_{d_{h}}(\phi/d_{h}), \quad \phi \sim \Ga(a_{\phi},b_{\phi}), \\
& s^{\star}_{\ell,\bk} \mid \bs_{\ell}, \etam^{\star}_{\ell} \sim \Cat_{K_{\ell}^{\star}}(\etam_{\ell}^{\star}), \quad \etam^{\star}_{\ell} \mid \bs_{\ell} \sim \Dir_{K_{\ell}^{\star}}(\phi^{\star}/K_{\ell}^{\star}), \\
& \mu_{k} \sim \TN(m_{0},s^{2}_{0},[A,B]), \quad \sigma_{k}^{2} \sim \IG(a_{\sigma},b_{\sigma}), \\
& b_{s'} \sim \Unif(-1,1), \quad \theta_{s''} \sim \Unif(-\pi,\pi),
\ese
where $\bR$ is defined in terms of $\bb$ and $\btheta$, and $\by_{i} = (y_{1,i},\dots,y_{d,i})\trans$ with $y_{\ell,i} = \Phi^{-1}\left\{F_{x_{\ell}\mid\bc}(x_{\ell,i} \mid \bc)\right\}$.
To simplify posterior inference, we introduce latent mixture allocations $z_{\ell,i}\in\{1,\ldots,K\}$ for each $x_{\ell,i}$, $\ell=1,\ldots,d$, $i=1,\ldots,n$.
The model thus becomes
\bse
& \textstyle  f_{\bx\mid\bc}(\bx_{i};\bc_{i}) = |\bR|^{-\frac{1}{2}} \exp\left\{-\frac{1}{2}\by_{i}\trans(\bR^{-1}- \bI_{d})\by_{i}\right\} \prod_{\ell=1}^{d} \TN\left(x_{\ell,i}; \mu_{z_{\ell,i}}, \sigma^{2}_{z_{\ell,i}}, [A,B]\right), \\
& x_{\ell,i} \mid z_{\ell,i}, \bmu, \bsigma^{2} \sim \TN\left(\mu_{z_{\ell,i}},\sigma^{2}_{z_{\ell,i}}, [A,B]\right), \quad
z_{\ell,i} \mid \bs_{\ell}, \bs^{\star}_{\ell}, \blambda_{\ell}; \bc_{i} \sim \Cat\left( \blambda_{\ell,\wt{s}_{\ell,\bc_{i}}} \right), \\
&\blambda_{\ell,k^{\star}} \mid \blambda_{\ell,0}, \alpha \sim \Dir_{K}(\alpha\blambda_{\ell,0}), \quad \blambda_{\ell,0} \sim \Dir_{K}(\alpha_{0}/K), \quad \alpha \sim \Ga(a_{\alpha},b_{\alpha}), \\
&s_{\ell,h}^{(c_{h})} \mid \etam_{\ell,h} \sim \Cat_{d_{h}}(\etam_{\ell,h}), \quad \etam_{\ell,h} \mid \phi \sim \Dir_{d_{h}}(\phi/d_{h}), \quad \phi \sim \Ga(a_{\phi},b_{\phi}), \\
&s^{\star}_{\ell,\bk} \mid \bs_{\ell}, \etam^{\star}_{\ell} \sim \Cat_{K_{\ell}^{\star}}(\etam_{\ell}^{\star}), \quad \etam^{\star}_{\ell} \mid \bs_{\ell} \sim \Dir_{K_{\ell}^{\star}}(\phi^{\star}/K_{\ell}^{\star}), \\
&\mu_{k} \sim \TN(m_{0},s_{0}^{2},[A,B]), \quad \sigma^{2}_{k} \sim \IG(a_{\sigma},b_{\sigma}), \\
& b_{s'} \sim \Unif(-1,1), \quad \theta_{s''} \sim \Unif(-\pi,\pi),
\ese
with $y_{\ell,i} = \Phi^{-1}\left\{\sum_{k=1}^{K}\wh{\lambda}_{\ell,\wt{s}_{\ell,\bc_{i}}} (k)F_{\text{TN}}(x_{\ell,i};\mu_{k},\sigma^{2}_{k},[A,B])\right\}$, where $F_{\text{TN}}(\cdot)$ denotes the cumulative distribution function of a truncated normal.
This data augmentation allows us to marginalize out the core vector elements $\blambda_{\ell,k^{\star}}$, and the probability vectors $\etam_{\ell,h}$ and $\etam^{\star}_{\ell}$, thus obtaining an MCMC algorithm which provides samples approximately from the posterior distribution $\Pr(\bz,\bs,\bs^{\star},\blambda_0,\alpha,\phi,\bmu,\bsigma^{2},\bb,\btheta\mid\bx;\bc)$.

Inference for the marginalized parameters can be recorded by evaluating  the following conditional means at each MCMC iteration
\bse
& \widehat{\lambda}_{\ell,k^{\star}}(k) &= \frac{ \alpha\lambda_{\ell,0}(k) + n_{\ell,k^{\star}}(k) }{ \sum_{k'=1}^{K}\{\alpha\lambda_{\ell,0}(k')+n_{\ell,k^{\star}}(k')\} }, \quad k=1,\ldots,K, \\
& \widehat{\eta}_{\ell,h}(q_{h}) &= \frac{ \phi/d_{h} + m_{\ell,h}(q_{h}) }{ \sum_{q'_{h}=1}^{d_{h}}\{\phi/d_{h} + m_{\ell,h}(q'_{h})\} }, \quad q_{h}=1,\ldots,d_{h}, \\
& \widehat{\eta}^{\star}_{\ell}(k^{\star}) &= \frac{ \phi^{\star}/K_{\ell}^{\star} + m^{\star}_{\ell}(k^{\star}) }{ \sum_{k^{\star '}=1}^{K_{\ell}^{\star}}\{\phi^{\star}/K_{\ell}^{\star} + m^{\star}_{\ell}(k^{\star '})\} },
\ese
where $n_{\ell,k^{\star}}(k) = \sum_{i=1}^{n}\Ind\{\wt{s}_{\ell,\bc_{i}}=k^{\star}\}\Ind\{z_{\ell,i}=k\}$, $m_{\ell,h}(q_{h}) = \sum_{c_{h}=1}^{d_{h}} \Ind\{s_{\ell,h}^{(c_{h})}=q_h\}$, and $m^{\star}_{\ell}(k^{\star})=\sum_{\bk\in\mathcal{S}_{\ell}}\Ind\{s^{\star}_{\ell,\bk}=k^{\star}\}$ are counts.
Averaging those, we obtain Rao-Blackwellized estimates of their posterior means.

\vskip 2ex
{All posterior estimates reported are computed on a set of 2,000 samples from a single MCMC run, obtained by collecting 30,000 samples, discarding the first 20,000 and then thinning the remaining samples by a factor of 5.}

\subsubsection*{Joint distribution}
The joint distribution of observed and latent variables without the contribution of the copula and with $\blambda$, $\etam$ and $\etam^{\star}$ marginalized out is
\begin{align*}
    \Pr(&\bz,\bs,\bs^{\star},\blambda_{0},\alpha,\phi,\bmu,\bsigma^{2},\bb,\btheta\mid\bx;\bc) \\
    =& \int_{\Lambda} \int_{H} \int_{H^{\star}} \Pr(\bz,\bs,\bs^{\star},\blambda,\blambda_{0},\alpha,\etam,\phi,\etam^{\star},\bmu,\bsigma^{2},\bb,\btheta\mid\bx;\bc) d\blambda d\etam d\etam^{\star} \\
    \propto& \Pr(\bx\mid\bz,\bmu,\bsigma^{2}) \Pr(\blambda_{0}) \Pr(\alpha) \Pr(\phi) \Pr(\bmu) \Pr(\bsigma^{2}) \\
    &\times \int_{\Lambda} \Pr(\bz\mid\bs,\bs^{\star},\blambda) \Pr(\blambda\mid\blambda_{0},\alpha) d\blambda \\
    &\times \int_{H} \Pr(\bs\mid\etam) \Pr(\etam\mid\phi) d\etam \\
    &\times \int_{H^{\star}} \Pr(\bs^{\star}\mid\bs,\etam^{\star}) \Pr(\etam^{\star}\mid\bs) d\etam^{\star},
\end{align*}
where $\blambda\in\Lambda$, $\etam\in H$, $\etam^{\star}\in H^{\star}$, and
\begin{align*}
    \Pr(&\bx\mid\bz,\bmu,\bsigma^{2})\Pr(\blambda_{0})\Pr(\alpha)\Pr(\phi)\Pr(\bmu)\Pr(\bsigma^{2}) \\
    =& \prod_{\ell=1}^{d}\prod_{i=1}^{n} \TN(x_{\ell,i};\mu_{z_{\ell,i}},\sigma_{z_{\ell,i}}^{2}, [A,B]) \times \prod_{\ell=1}^{d} \Dir_{K}(\blambda_{\ell,0};\alpha_{0}/K) \\
    &\times \Ga(\alpha;a_{\alpha},b_{\alpha}) \times \Ga(\phi;a_{\phi},b_{\phi}) \times \prod_{k=1}^{K} \TN(\mu_{k}; m_{0},s^{2}_{0},[A,B]) \IG(\sigma_{k}^{2};a_{\sigma},b_{\sigma}).
\end{align*}
Focusing on the integrals, we have
\footnotesize
\begin{align*}
    \int_{\Lambda} \Pr(\bz\mid\bs,\bs^{\star},\blambda)\Pr(\blambda\mid\blambda_{0},\alpha) d\blambda &=
    \int_{\Lambda} \left\{ \prod_{\ell=1}^{d}\prod_{i=1}^n \Cat\left(z_{\ell,i};\blambda_{\ell,\wt{s}_{\ell,\bc_{i}}}\right) \times \prod_{\ell=1}^{d} \prod_{k^{\star}=1}^{K_{\ell}^{\star}} \Dir_{K}(\blambda_{\ell,k^{\star}};\alpha\blambda_{\ell,0}) \right\} d\blambda \\
    = \prod_{\ell=1}^{d} \prod_{k^{\star}=1}^{K_{\ell}^{\star}} &\left\{ \frac{\Gamma(\alpha)}{\prod_{k=1}^{K}\Gamma(\alpha\lambda_{\ell,0}(k))} \int_{\Delta^{K-1}} \prod_{k=1}^K \lambda_{\ell,k^{\star}}(k)^{\alpha\lambda_{\ell,0}(k)+\sum_{i=1}^{n}\Ind\{\wt{s}_{\ell,\bc_{i}}=k^{\star}\}\Ind\{z_{\ell,i}=k\}-1} d\blambda_{\ell,k^{\star}} \right\} \\
    &= \prod_{\ell=1}^{d} \prod_{k^{\star}=1}^{K_{\ell}^{\star}} \left\{ \frac{\Gamma(\alpha)}{\prod_{k=1}^{K}\Gamma(\alpha\lambda_{\ell,0}(k))} \cdot \frac{\prod_{k=1}^{K}\Gamma(\alpha\lambda_{\ell,0}(k)+n_{\ell,k^{\star}}(k))}{\Gamma(\alpha+n_{\ell,k^{\star}})} \right\}, \\[.5cm]
    \int_{H} \Pr(\bs\mid\etam)\Pr(\etam\mid\phi)d\etam &= \int_{H}\left\{ \prod_{\ell=1}^{d}\prod_{h=1}^p\prod_{c_{h}=1}^{d_{h}}\Cat(s_{\ell,h}^{(c_{h})};\etam_{\ell,h}) \times \prod_{\ell=1}^{d}\prod_{h=1}^p \Dir_{d_{h}}(\etam_{\ell,h};\phi/d_{h}) \right\} d\etam \\
    &= \prod_{\ell=1}^{d}\prod_{h=1}^p \left\{ \frac{\Gamma(\phi)}{\prod_{q_h=1}^{d_{h}}\Gamma(\phi/d_{h})} \int_{\Delta^{d_{h}-1}} \prod_{q_h=1}^{d_{h}} \eta_{\ell,h}(q_h)^{\phi/d_{h}+\sum_{c_{h}=1}^{d_{h}}\Ind\{s_{\ell,h}^{(c_{h})}=q_h\}-1}  d\etam_{\ell,h} \right\} \\
    &= \prod_{\ell=1}^{d}\prod_{h=1}^p \left\{ \frac{\Gamma(\phi)}{\prod_{q_h=1}^{d_{h}}\Gamma(\phi/d_{h})} \cdot \frac{\prod_{q_{h}=1}^{d_{h}}\Gamma(\phi/d_{h}+m_{\ell,h}(q_{h}))}{\Gamma(\phi+d_{h})} \right\}, \\[.5cm]
    \int_{H^{\star}} \Pr(\bs^{\star}\mid\bs,\etam^{\star})\Pr(\etam^{\star}\mid\bs)d\etam^{\star} &= 
    \int_{H^{\star}}\left\{ \prod_{\ell=1}^{d}\prod_{\bk\in\S_{\ell}}\Cat_{K_{\ell}^{\star}}(s_{\ell,\bk}^{\star};\etam^{\star}_{\ell}) \times \prod_{\ell=1}^{d} \Dir_{K_{\ell}^{\star}}(\etam^{\star}_{\ell};\phi^{\star}/K^{\star}_{\ell}) \right\} d\etam^{\star} \\
    &= \prod_{\ell=1}^{d} \left\{ \frac{\Gamma(\phi^{\star})}{\prod_{k^{\star}=1}^{K_{\ell}^{\star}}\Gamma(\phi^{\star}/K_{\ell}^{\star})} \int_{\Delta^{K_{\ell}^{\star}-1}} \prod_{k^{\star}=1}^{K^{\star}_{\ell}} \eta_{\ell}^{\star}(k^{\star})^{\phi^{\star}/K_{\ell}^{\star}+\sum_{\bk\in\S_{\ell}}\Ind\{s_{\ell,\bk}^{\star}=k^{\star}\}-1}  d\etam_{\ell}^{\star} \right\} \\
    &= \prod_{\ell=1}^{d} \left\{ \frac{\Gamma(\phi^{\star})}{\prod_{k^{\star}=1}^{K_{\ell}^{\star}}\Gamma(\phi^{\star}/K_{\ell}^{\star})} \cdot \frac{\prod_{k^{\star}=1}^{K_{\ell}^{\star}}\Gamma(\phi^{\star}/K_{\ell}^{\star}+m^{\star}_{\ell}(k^{\star}))}{\Gamma(\phi^{\star}+\prod_{h=1}^{p}K_{\ell,h})} \right\}.
\end{align*}
\normalsize
The joint distribution thus becomes
\begin{align*}
    \Pr(&\bz,\bs,\bs^{\star},\blambda_{0},\alpha,\phi,\bmu,\bsigma^{2},\bb,\btheta\mid\bx;\bc) \\
    \propto& \prod_{\ell=1}^{d}\prod_{i=1}^{n} \TN(x_{\ell,i};\mu_{z_{\ell,i}},\sigma_{z_{\ell,i}}^{2}, [A,B]) \times \prod_{\ell=1}^{d} \Dir_{K}(\blambda_{\ell,0};\alpha_{0}/K) \\
    &\times \Ga(\alpha;a_{\alpha},b_{\alpha}) \times \Ga(\phi;a_{\phi},b_{\phi}) \times \prod_{k=1}^{K} \TN(\mu_{k}; m_{0},s^{2}_{0},[A,B]) \IG(\sigma_{k}^{2};a_{\sigma},b_{\sigma}) \\
    &\times \prod_{\ell=1}^{d} \prod_{k^{\star}=1}^{K_{\ell}^{\star}} \left\{ \frac{\Gamma(\alpha)}{\prod_{k=1}^{K}\Gamma(\alpha\lambda_{\ell,0}(k))} \cdot \frac{\prod_{k=1}^{K}\Gamma(\alpha\lambda_{\ell,0}(k)+n_{\ell,k^{\star}}(k))}{\Gamma(\alpha+n_{\ell,k^{\star}})} \right\} \\
    &\times \prod_{\ell=1}^{d}\prod_{h=1}^p \left\{ \frac{\Gamma(\phi)}{\prod_{q_h=1}^{d_{h}}\Gamma(\phi/d_{h})} \cdot \frac{\prod_{q_{h}=1}^{d_{h}}\Gamma(\phi/d_{h}+m_{\ell,h}(q_{h}))}{\Gamma(\phi+d_{h})} \right\} \\
    &\times \prod_{\ell=1}^{d} \left\{ \frac{\Gamma(\phi^{\star})}{\prod_{k^{\star}=1}^{K_{\ell}^{\star}}\Gamma(\phi^{\star}/K_{\ell}^{\star})} \cdot \frac{\prod_{k^{\star}=1}^{K_{\ell}^{\star}}\Gamma(\phi^{\star}/K_{\ell}^{\star}+m^{\star}_{\ell}(k^{\star}))}{\Gamma(\phi^{\star}+\prod_{h=1}^{p}K_{\ell,h})} \right\}.
\end{align*}

\vskip 30pt
\subsection{Estimating the Main Quantities of Interest}\label{sec: sm estimating main quantities}
We introduce here the main quantities of interest and describe how their posterior point estimates are obtained from the MCMC samples. 

In particular, we focus on (i) conditional marginal and joint densities of selected response coordinates under different covariate combinations, (ii) corresponding unconditional marginal densities, (iii) point estimates of the partition structures arising from the partition-based tensor factorization, and (iv) correlation matrix.

In what follows, $\B$ denotes the set of $|\mathcal{B}|$ MCMC samples retained after burn-in and thinning, and we use the superscript $(b)$ to denote the value assumed by a latent quantity at the $b\th$ MCMC sample.

\subsubsection*{Conditional Marginal Densities}
The conditional marginal density of the $\ell\th$ response coordinate conditional on the covariate combination $\bc$ evaluated at $x\in[A,B]$ is estimated as
\bse
\widehat{f}_{x_{\ell}\mid\bc}(x;\bc) = \frac{1}{|\mathcal{B}|}\sum_{b\in\B} \left\{ \sum_{k=1}^{K} \widehat{\lambda}^{(b)}_{\ell,\wt{s}_{\ell,\bc}^{(b)}}(k) \TN\left(x; \mu^{(b)}_{k}, \sigma^{2(b)}_{k}, [A,B]\right) \right\}.
\ese 
A posterior point estimate for the conditional marginal density $\widehat{f}_{x_{\ell}\mid\bc}$ is then obtained evaluating $\widehat{f}_{x_{\ell}\mid\bc}(x;\bc)$ on an equi-spaced grid $(\widetilde{x}_{1},\ldots,\widetilde{x}_{G})$ of $G$ points over the interval $[A,B]$.

\subsubsection*{Conditional Joint Densities}
The conditional joint density of the subset $\mathcal{L}\subseteq\{1,\ldots,d\}$ of the response coordinates conditional on the covariate combination $\bc$ evaluated at $\bx\in[A,B]^{|\mathcal{L}|}$ is estimated as
\bse
\widehat{f}_{\bx_{\mathcal{L}}\mid\bc}(\bx;\bc) = \frac{1}{|\mathcal{B}|}\sum_{b\in\B} \left\{
C^{(b)}(\bx) \prod_{\ell\in\mathcal{L}} \sum_{k=1}^{K} \widehat{\lambda}^{(b)}_{\ell,\wt{s}_{\ell,\bc}^{(b)}}(k) \TN\left(x_{\ell}; \mu^{(b)}_{k}, \sigma^{2(b)}_{k}, [A,B]\right)
\right\},
\ese
where $C^{(b)}(\bx)$ is the density of a $|\mathcal{L}|$-variate Gaussian copula with correlation matrix $\bR^{(b)}_{\mathcal{L},\mathcal{L}}=(R^{(b)}_{\ell',\ell''})_{{\ell',\ell''}\in\mathcal{L}}$.
A posterior point estimate for the conditional joint density $\widehat{f}_{\bx_{\mathcal{L}}\mid\bc}$ is then obtained evaluating $\widehat{f}_{\bx_{\mathcal{L}}\mid\bc}(\bx;\bc)$ on an equi-spaced grid $(\widetilde{x}_{1},\ldots,\widetilde{x}_{G})^{|\mathcal{L}|}$ of $G^{|\mathcal{L}|}$ points over $[A,B]^{|\mathcal{L}|}$.
In this work, we are particularly interested in the bivariate case since it can be easily visualized using a contour plot.

\subsubsection*{Unconditional Marginal and Joint Densities}
Let $\widehat{w}_{\bc}=\sum_{i=1}^{n}\Ind(\bc_{i}=\bc)/n\in[0,1]$ be an empirical estimate of the proportion of units with covariate combination $\bc$, $w_{\bc}$, the unconditional marginal density of the $\ell\th$ response coordinate, $f_{x_{\ell}}=\sum_{\bc\in\mathcal{C}}w_{\bc}f_{x_{\ell}\mid\bc}$, is estimated as weighted average of the estimated conditional marginal densities $\widehat{f}_{x_{\ell}\mid\bc}$;
similarly, the unconditional joint density for the subset $\mathcal{L}$ of the response coordinates, $f_{\bx_{\mathcal{L}}}=\sum_{\bc\in\mathcal{C}}w_{\bc}f_{\bx_{\mathcal{L}}\mid\bc}$, is estimated as weighted average of the conditional joint densities $\widehat{f}_{\bx_{\mathcal{L}}\mid\bc}$,
\bse
\widehat{f}_{x_{\ell}} = \sum_{\bc\in\mathcal{C}} \widehat{w}_{\bc} \widehat{f}_{x_{\ell}\mid\bc}, \quad
\widehat{f}_{\bx_{\mathcal{L}}} = \sum_{\bc\in\mathcal{C}} \widehat{w}_{\bc} \widehat{f}_{\bx_{\mathcal{L}}\mid\bc}.
\ese

\subsubsection*{Partitions}
A posterior point estimate for the first-layer partitions can be obtained by independently estimating each partition $\bs_{\ell,h}$ using loss function-based algorithms, such as \cite{dahl_SearchAlgorithmsLoss_2022}. 
However, this approach is not suitable for the second layer, as the partition elements may vary during the MCMC procedure. 
Instead, we estimate the Maximum-A-Posteriori (MAP) partition that appears most frequently among the MCMC samples. 
For coherence, MAP estimates of both partition layers, denoted $\wh{\bs}_{\ell,h}$ and $\wh{\bs}^{\star}_{\ell}$, are obtained.

We can further define a partition over the covariate combinations and obtain its posterior point estimates using $\wh{\bs}_{\ell,h}$ and $\wh{\bs}^{\star}_{\ell}$; specifically, two covariate combinations $\bc'$ and $\bc''$ belong to the same cluster if $\wh{\wt{s}}_{\ell,\bc'} = \wh{\wt{s}}_{\ell,\bc''}$, i.e., if
$$\wh{s}^{\star}_{\ell,\wh{s}_{\ell,1}^{(c'_{1})},\ldots,\wh{s}_{\ell,p}^{(c'_{p})}} = \wh{s}^{\star}_{\ell,\wh{s}_{\ell,1}^{(c''_{1})},\ldots,\wh{s}_{\ell,p}^{(c''_{p})}}.$$

\subsubsection*{Conditional Marginal Densities Conditional on Reference Partitions}
A posterior point estimate of the $k^{\star th}=1,\ldots,K^{\star}_{\ell}$ conditional marginal density identified by a reference configuration $(\bs_{\ell}, \bs_{\ell}^{\star})$ of the partition layers is computed as the average of the $\widehat{f}_{x_{\ell}\mid\bc}$ with $\wt{s}_{\ell,\bc} = k^{\star}$, for $\ell=1,\ldots,d$. An interesting case is the MAP estimate $(\wh{\bs}_{\ell}, \wh{\bs}_{\ell}^{\star})$ as a reference configuration since it allows us to link an estimated marginal density to each group of covariate combinations identified by the model. 

\subsubsection*{Correlation Matrix}
A posterior point estimate of the correlation matrix $\bR$ is estimated as
$\wh{\bR} = \frac{1}{|\mathcal{B}|}\sum_{b\in\mathcal{B}} \bR^{(b)}$,
where $\bR^{(b)}$ is computed using $\bb^{(b)}$ and $\btheta^{(b)}$.

\section{Hyperparameter Choices}
We fit our model using the MCMC algorithm described in Section 4 of the manuscript.
We use $K=20$ mixture components and $K^{\star}_{1}=\ldots=K^{\star}_{d}=K^{\star}=20$ as the maximum number of densities for each coordinate of the response for the NHANES dietary data and Scenario B in the simulation study, while we use $K=10$ and $K^{\star}=10$ for Scenario A.
As prior parameters, we use $(a_{\alpha},b_{\alpha})=(2,0.5)$, $(a_{\phi},b_{\phi})=(2,0.5)$, $\alpha_{0}=1$, $(a_{\sigma},b_{\sigma})=(2,0.5)$, $\phi^{\star}=1$, and $(m_{0},s_{0})$ are set to the mean and standard deviation of the rescaled response. 
We further consider $\sigma^{2}_{\mu}=0.5^{2}$ and $\sigma^{2}_{\sigma}=0.5^{2}$ in the M-H steps for the atom update, and $\sigma^{2}_{\alpha}=0.5^{2}$ and $\sigma^{2}_{\phi}=0.5^{2}$ as starting values for the variances in the proposal distributions of $\alpha$ and $\phi$.

\clearpage
\section{NHANES Dietary Data: Additional Results}

\subsection{Posterior Checks}
We conduct a series of posterior checks to evaluate the goodness of fit of our proposed flower model on the NHANES dietary dataset. 
Specifically, we assess performance at both the covariate level and the copula level. 
The covariate-level checks verify whether the estimated conditional marginal densities align with the empirical distributions and match the model-identified groups, while the copula-level checks evaluate how effectively the model captures cross-response dependence through the copula and shared atoms.

\begin{figure}[!ht]
    \centering
    \begin{subfigure}[t]{0.49\textwidth}
        \centering
        \includegraphics[width=\textwidth, trim=0cm 0cm 0cm 0cm, clip=true]{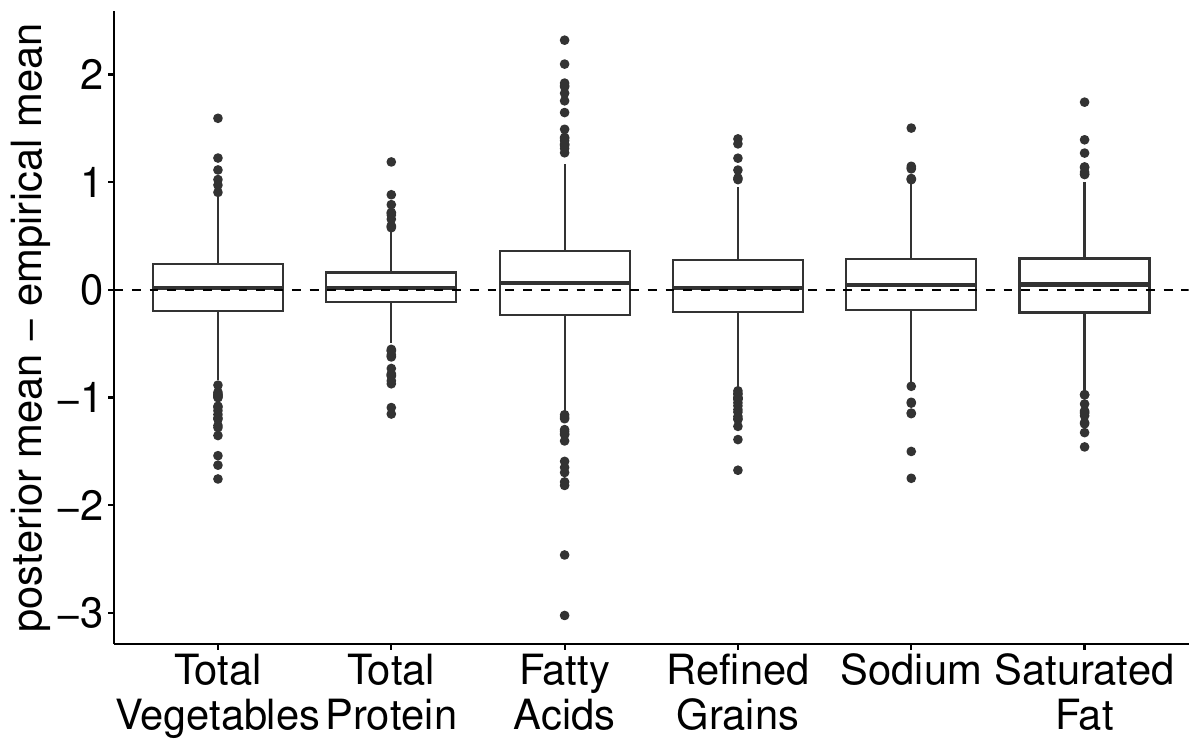}
    \end{subfigure}
    \begin{subfigure}[t]{0.49\textwidth}
        \centering
        \includegraphics[width=\textwidth, trim=0cm 0cm 0cm 0cm, clip=true]{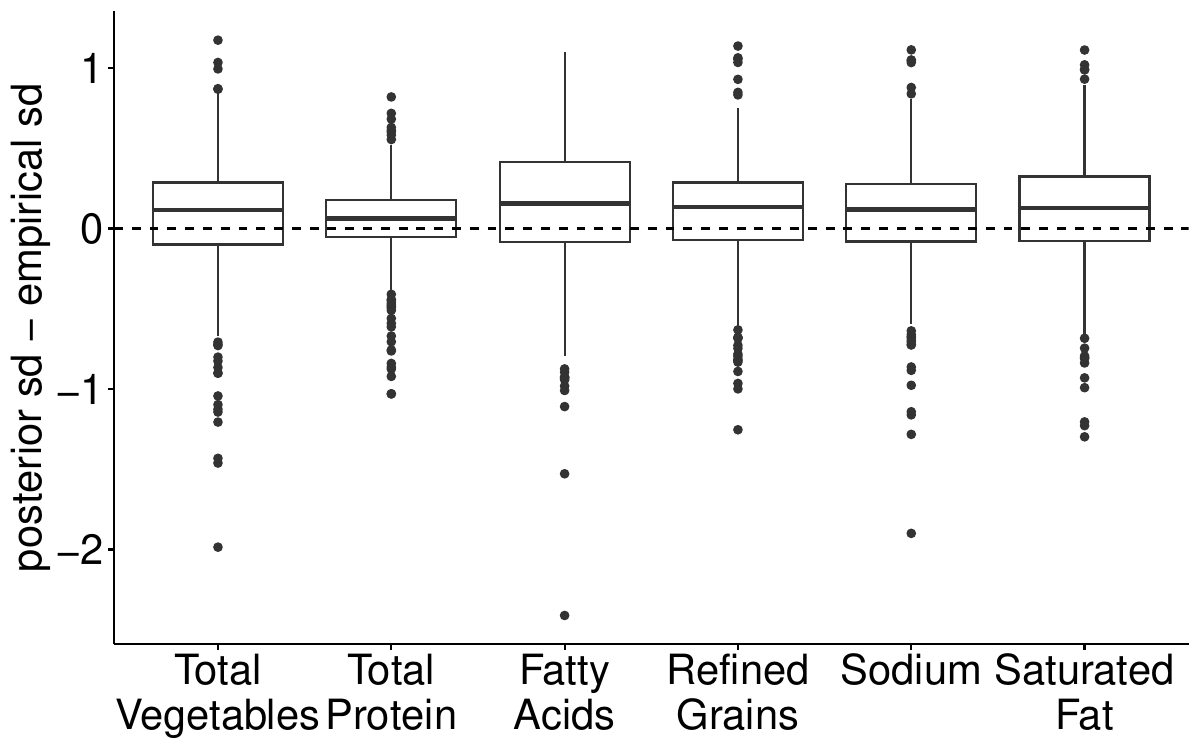}
    \end{subfigure}
    \caption{Coordinate-specific boxplots of the difference between the posterior estimate summary and the empirical summary, computed over the observed covariate combinations $\bc\in\C$: mean on the left, standard deviation on the right. 
    The dashed horizontal line marks zero difference.
    }
    \label{fig:nhanes_check_boxplot_summaries}
\end{figure}

First, we focus on the original covariate combinations $\bc\in\C$, and compare empirical summaries with their estimated counterparts for each response coordinate $\ell$.
Specifically, we compute the sample means and standard deviations as
\bse
\bar{x}_{\ell,\bc} = \frac{1}{|\{i:\bc_{i}=\bc\}|}\sum_{i:\bc_{i}=\bc} x_{\ell,i}, \qquad
s_{x,\ell,\bc} = \sqrt{ \frac{1}{|\{i:\bc_{i}=\bc\}|-1} \sum_{i:\bc_{i}=\bc} (x_{\ell,i} - \bar{x}_{\ell,\bc})^{2} },
\ese
and their estimated counterparts, specifically their posterior means, as
\bse
\frac{1}{|\mathcal{B}|}\sum_{b\in\mathcal{B}}\sum_{k=1}^{K} \hat{\lambda}^{(b)}_{\ell,\wt{s}^{(b)}_{\ell,\bc}}(k) m^{(b)}_{k}, \qquad
\frac{1}{|\mathcal{B}|}\sum_{b\in\mathcal{B}} \sqrt{\sum_{k=1}^{K} \hat{\lambda}^{(b)}_{\ell,\wt{s}^{(b)}_{\ell,\bc}}(k) \left\{ v^{(b)}_{k} + \left(m_{k}^{(b)} - \bar{m}^{(b)}\right)^{2}\right\} },
\ese
where $m_{k}^{(b)}$ and $v_{k}^{(b)}$ are the expected value and variance of the $k\th$ mixture component at the $b\th$ iteration of the MCMC, and $\bar{m}^{(b)} = \sum_{k=1}^{K}\hat{\lambda}^{(b)}_{\ell,\wt{s}^{(b)}_{\ell,\bc}}(k) m_{k}^{(b)}$; these are computed as
\bse
&m_{k}^{(b)} = \mu_{k}^{(b)} + \sigma_{k}^{(b)} \frac{\varphi\left(A_{k}^{(b)}\right) - \varphi\left(B_{k}^{(b)}\right)}{\Phi\left(B_{k}^{(b)}\right) - \Phi\left(A_{k}^{(b)}\right)}, \quad
v_{k}^{(b)} = \sigma_{k}^{2(b)} \left[ 1 + \frac{A_{k}^{(b)}\varphi\left(A_{k}^{(b)}\right) - B_{k}^{(b)}\varphi\left(B_{k}^{(b)}\right)}{\Phi\left(B_{k}^{(b)}\right) - \Phi\left(A_{k}^{(b)}\right)} - \left\{\frac{\varphi\left(A_{k}^{(b)}\right) - \varphi\left(B_{k}^{(b)}\right)}{\Phi\left(B_{k}^{(b)}\right) - \Phi\left(A_{k}^{(b)}\right)}\right\}^{2}\right],
\ese
with $A_{k}^{(b)} = (A - \mu_{k}^{(b)}) / \sigma_{k}^{(b)}$ and $B_{k}^{(b)} = (B - \mu_{k}^{(b)}) / \sigma_{k}^{(b)}$, where $\varphi(\cdot)$ denotes the probability density function of a standard normal distribution.
The differences between these quantities are summarized as coordinate-specific boxplots in \autoref{fig:nhanes_check_boxplot_summaries}. 
Overall, the results demonstrate good model agreement: estimated and empirical mean differences are centered near zero for three of the six response coordinates, with the remaining three showing slight overestimation. 
By contrast, the estimated standard deviations tend to be mildly overestimated across coordinates, consistent with model-based clustering, which incorporates extra variance to aggregate subgroups with differing underlying means and spreads.

\begin{figure}[!ht]
    \begin{center}
    \includegraphics[width=\textwidth, trim=0cm 0cm 0cm 0cm, clip=true]{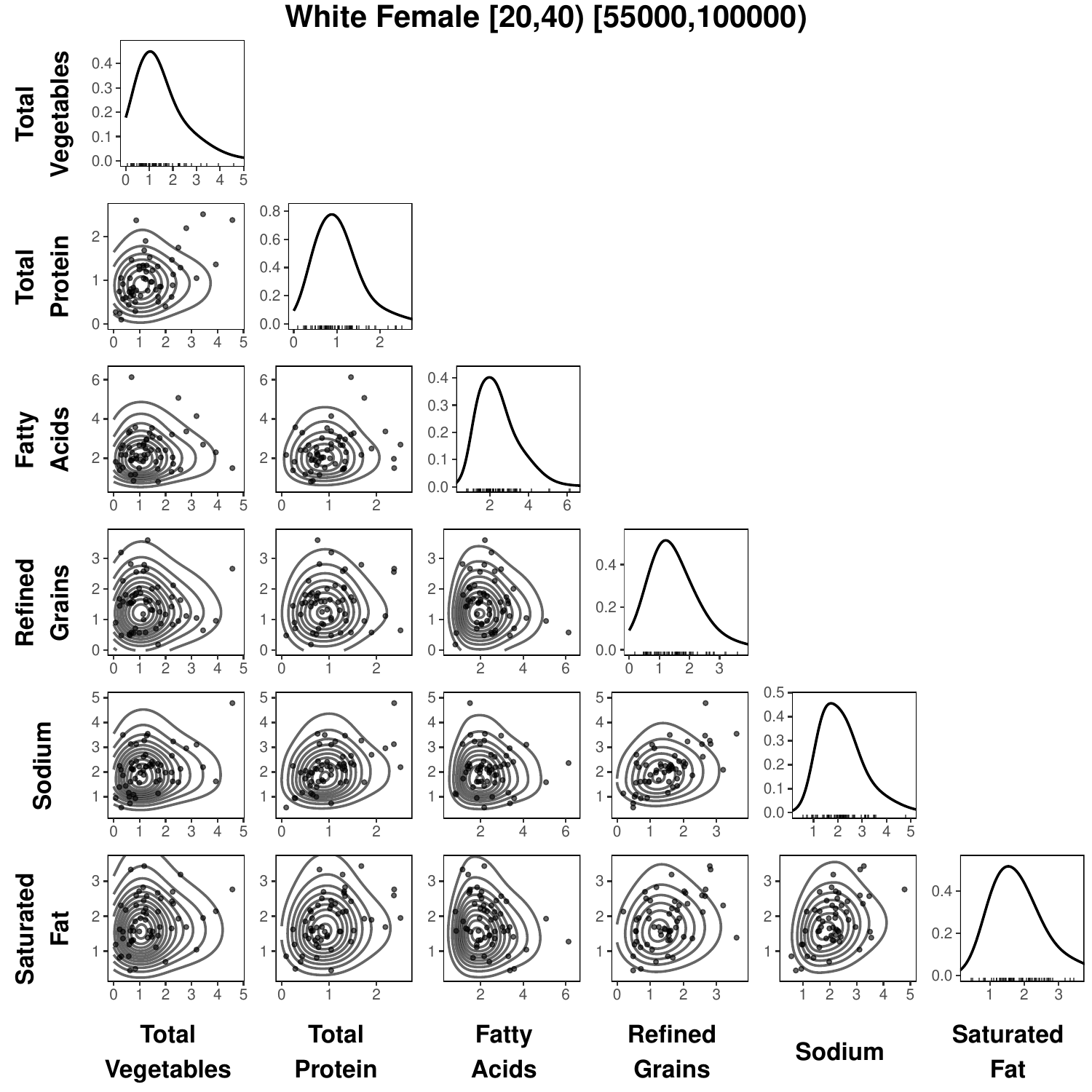}
    \end{center}
    \caption{Estimated conditional densities for White females aged $[20,40)$ earning between $\$55{,}000$ and $\$100{,}000$. The diagonal shows the $d=6$ estimated conditional marginal densities, with the corresponding observations as a rug on the horizontal axis; the lower triangle shows the $d(d-1)/2=15$ estimated conditional bivariate densities as contour plots, with the corresponding observations superimposed.}
    \label{fig:nhanes_check_cc}
\end{figure}

Focusing on a representative covariate combination, we evaluate goodness of fit at both the marginal and pairwise joint levels by comparing the estimated densities directly against the empirical data. 
For example, \autoref{fig:nhanes_check_cc} displays the $d=6$ estimated conditional marginal densities along the main diagonal and the $d(d-1)/2=15$ conditional bivariate densities off-diagonal for White females aged $[20,40)$ earning between $\$55{,}000$ and $\$100{,}000$. 
Aside from a few extreme outliers, the observed data points concentrate well within the high-density regions of the model estimates. 
Repeating this graphical inspection across major covariate profiles provides strong qualitative validation of overall model fit.

\begin{figure}[!ht]
    \centering
    \includegraphics[width=\textwidth, trim=0cm 0cm 0cm 0cm, clip=true]{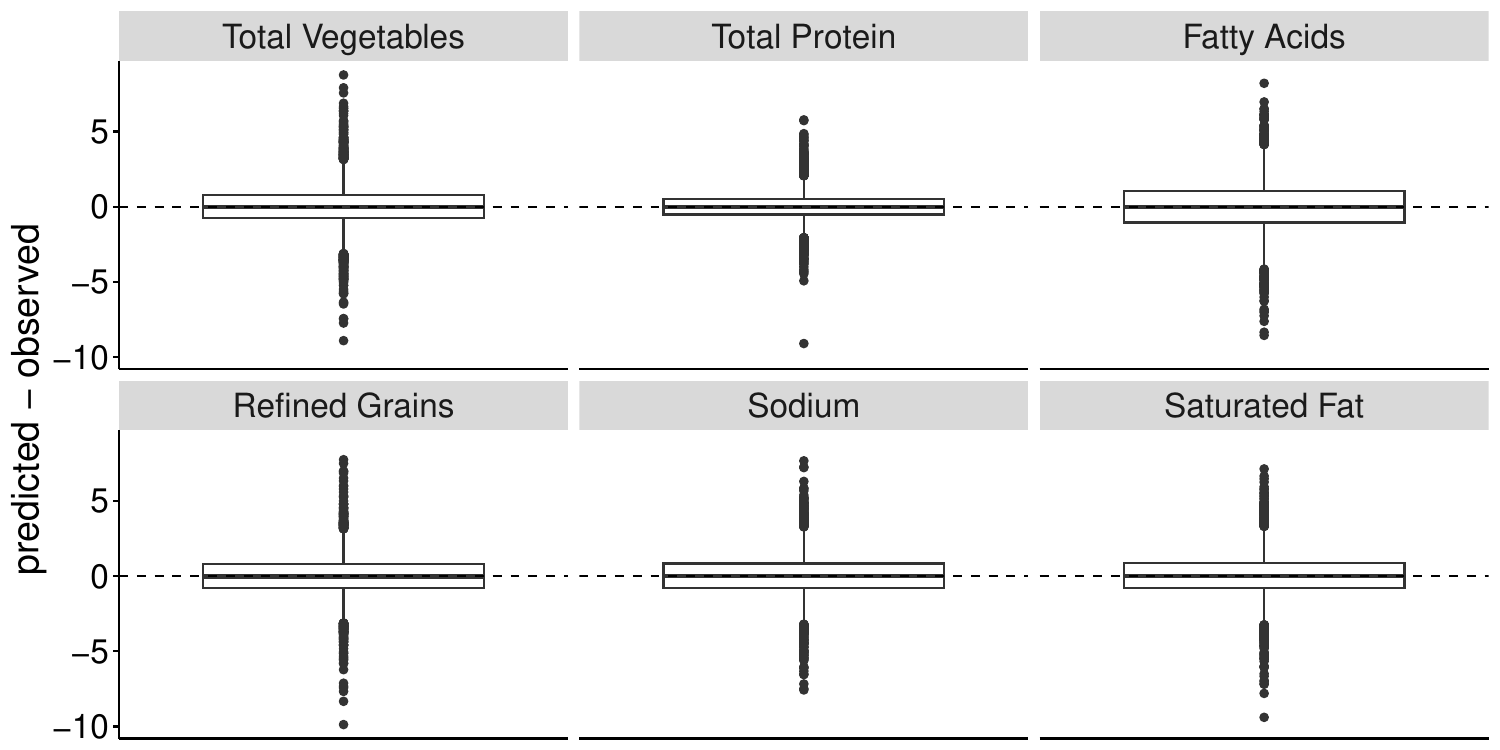}
    \caption{Coordinate-specific boxplots of the unit-level differences $\wt{x}_{\ell,i}-x_{\ell,i}$ between a dataset replicated from the posterior predictive distribution on the observed covariate design and the observed NHANES data. 
    The dashed horizontal line marks zero difference.
    }
    \label{fig:nhanes_check_boxplot_predictive}
\end{figure}

Moving our attention to the whole joint distribution $f_{\bx\mid\bc}(\cdot\,;\bc)$, we evaluate the posterior predictive distribution by generating a new dataset $\wt{\bx}_{1}, \ldots, \wt{\bx}_{n}$ with covariate vectors matching those observed, enabling pairwise comparisons between observed and simulated units. 
Sampling from the posterior predictive distribution requires two inputs: the posterior point estimates of the conditional marginal densities $\wh{f}_{x_{\ell}\mid\bc}(x;\bc)$ for each response coordinate $\ell=1,\ldots,d$ and covariate combination $\bc\in\C$, and the posterior point estimate of the correlation matrix $\wh{\bR}$. 
Both are estimated as described in Section \ref{sec: sm estimating main quantities}.
For each unit $i=1,\ldots,n$, we generate $\wt{\bx}_{i}=(\wt{x}_{1,i},\ldots,\wt{x}_{d,i})\trans$ using a standard approach to obtain a realization of a Gaussian-copula-distributed random variable:
\begin{enumerate}
    \item sample $\bx^{\Delta}_{i}\sim\MVN_{d}(\bzero_{d},\wh{\bR})$,
    \item set $\wt{x}_{\ell,i}=\wh{F}_{x_{\ell}\mid\bc}^{-1}(\Phi(x^{\Delta}_{\ell,i}); \bc_{i})$,
\end{enumerate}
where $\wh{F}_{x_{\ell}\mid\bc}(\cdot\,; \bc_{i})$ is the cumulative distribution function corresponding to $\wh{f}_{x_{\ell}\mid\bc}(\cdot\,;\bc)$, obtained by numerical integration over the grid on which the densities are evaluated.
The differences $\wt{x}_{\ell,i}-x_{\ell,i}$ are shown as coordinate-specific boxplots in \autoref{fig:nhanes_check_boxplot_predictive}.
Results indicate a good fit: the differences for all coordinates are centered and symmetric around zero, indicating no systematic over- or under-prediction of individual dietary intakes.

\subsection{Posterior Distribution of the Number of Clusters}
\begin{figure}[!ht]
    \centering
    \begin{subfigure}[t]{0.98\textwidth}
        \centering
        \includegraphics[width=\textwidth, trim=0cm 0cm 0cm 0cm, clip=true]{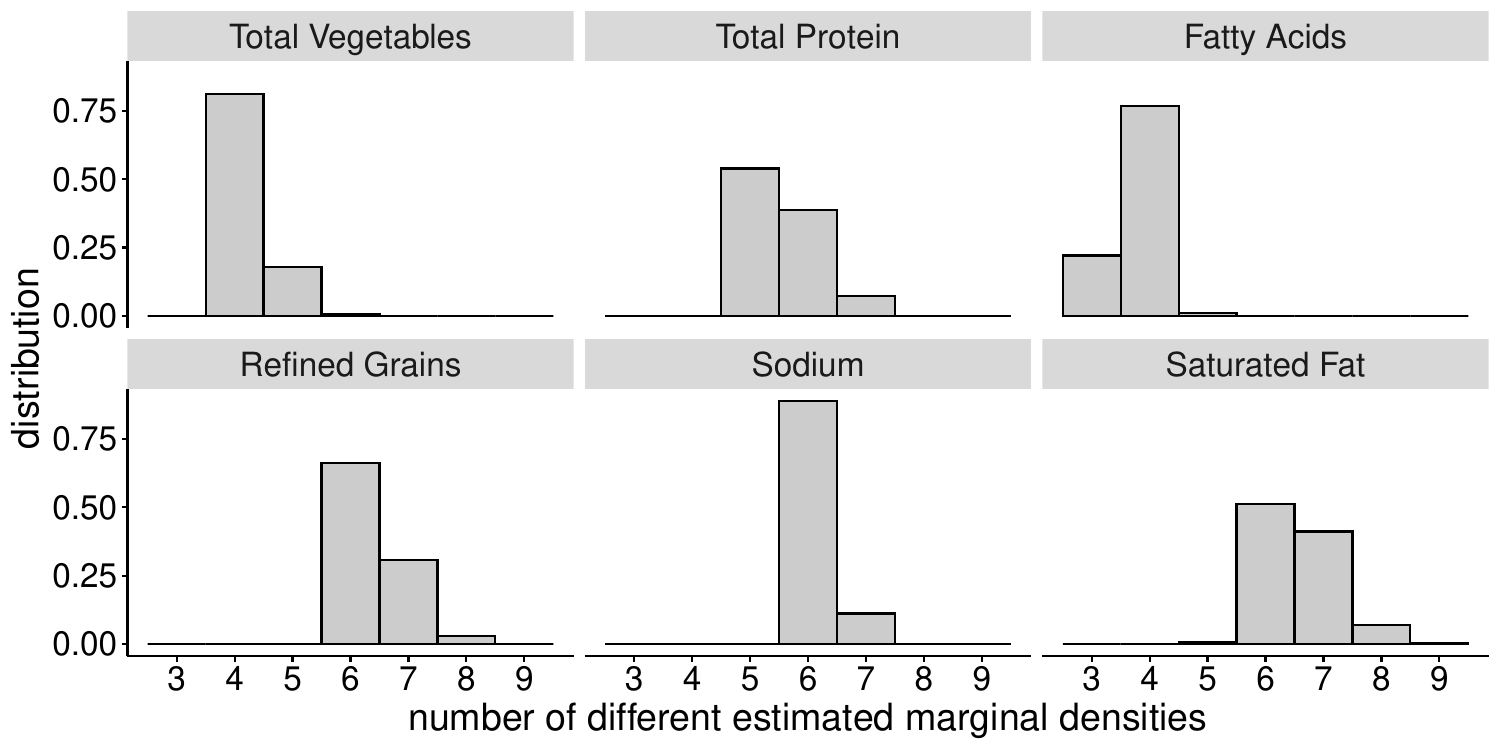}
        \caption{Posterior distribution of the number of different conditional marginal densities $f_{x_{\ell}\mid\bc}(\cdot\,;\bc)$, which corresponds to the number of clusters in each second-layer partition $\bs_{\ell}^{\star}$.}
        \label{fig:nhanes_ndens}
    \end{subfigure}
    \begin{subfigure}[t]{0.98\textwidth}
        \centering
        \includegraphics[width=\textwidth, trim=0cm 0cm 0cm 0cm, clip=true]{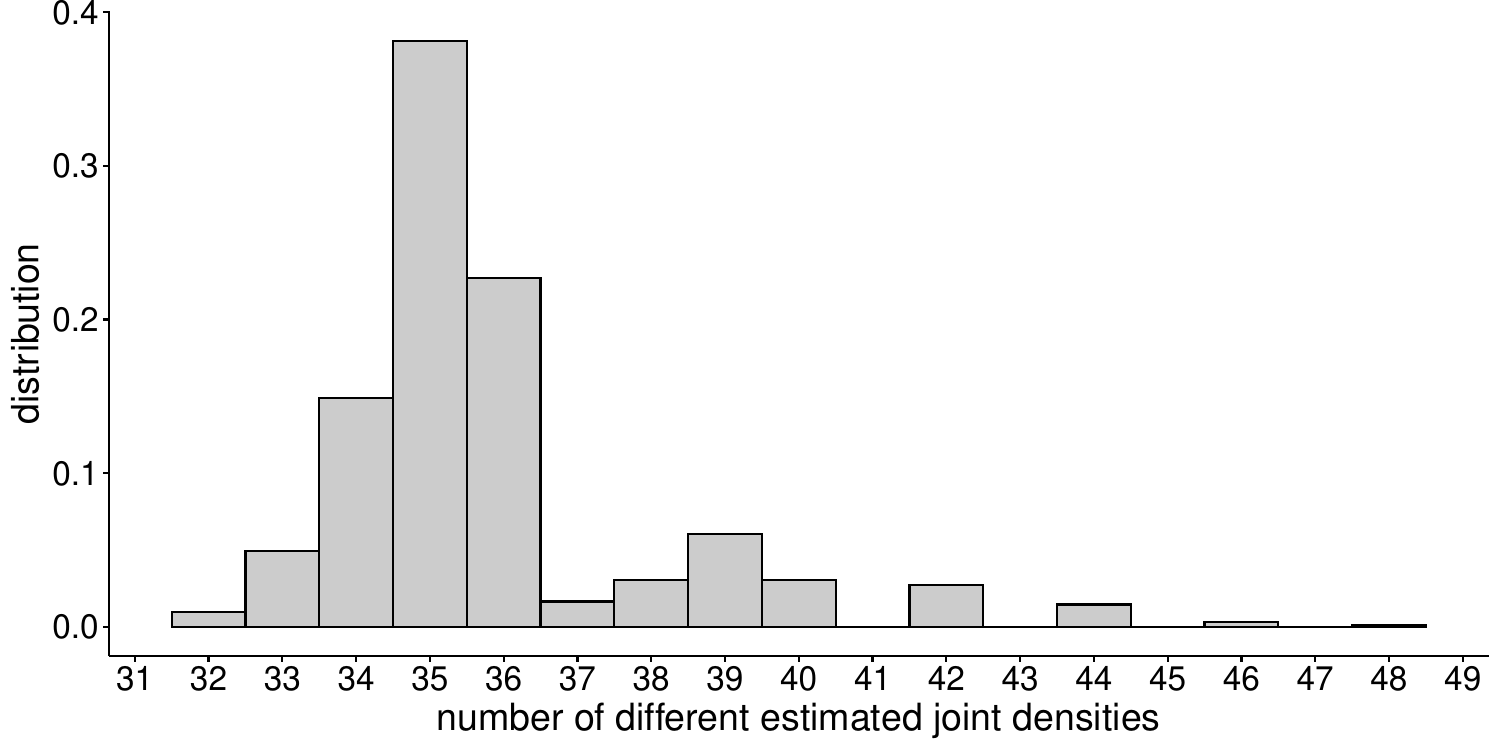}
        \caption{Posterior distribution of the number of different conditional joint densities $f_{\bx\mid\bc}(\cdot\,;\bc)$.}
        \label{fig:nhanes_ndens_joint}
    \end{subfigure}
    \caption{Posterior distribution of the number of different conditional marginal densities (top) and different conditional joint distributions (bottom).}
\end{figure}

First, consider the posterior distribution of the number of distinct conditional marginal densities $f_{x_{\ell}\mid\bc}(\cdot\,;\bc)$, which are shown in Figure 5 of the main paper. 
For each response coordinate $\ell$, the number of distinct $f_{x_{\ell}\mid\bc}(\cdot\,;\bc)$'s equals the size of the second-layer partition $\bs^{\star}_{\ell}$ induced over the joint covariate space $\mathcal{C}$. 
\autoref{fig:nhanes_ndens} shows these distributions; all concentrate between 3 and 9, far below the 504 distinct covariate level combinations.

Next, we consider the posterior distribution of the number of distinct conditional joint densities $f_{\bx\mid\bc}(\cdot\,;\bc)$, which equals the number of clusters over the joint covariate space $\mathcal{C}$ formed by the Cartesian product of the $d$ coordinate-specific partitions of $\mathcal{C}$ induced by our model. 
Specifically, two covariate profiles share the same joint density if and only if they belong to the same second-layer cluster across all $d$ coordinates. 
\autoref{fig:nhanes_ndens_joint} presents this distribution; it ranges between 32 and 48, concentrating mainly between 33 and 36. 

\clearpage
\section{Simulation Study - Additional Details}
We describe here in greater detail the simulations we performed to assess the performance of our proposed flower model and compare it with other alternative approaches.

To generate a dataset, we have to generate a $p \times n$ covariate matrix $\bc$, and a $d \times n$ response matrix $\bx$, where $n$ is the number of units, $d$ is the dimensionality of the multivariate response, and $p$ is the number of categorical covariates. 
Therefore, each column of these matrices refers to a different statistical unit.

We considered two simulation scenarios:
\vspace*{-10pt}
\begin{enumerate}[label=\Alph*.,itemsep=-5pt] 
\item a simpler setup featuring a 3-dimensional response and 5 covariates, where the conditional marginal densities display distinct shapes with good separation,  
\item 
a more complex setup designed to emulate the key characteristics of the NHANES dietary dataset. 
\end{enumerate}
\vspace*{-10pt}
Covariate values were sampled randomly in Scenario A, whereas they were fixed to match the observed NHANES covariate profiles in Scenario B.
The multivariate response is assumed to be distributed as the model introduced in (4), (11), and (7) without the common atoms, leading to $f_{\bx\mid\bc}(\bx_{i};\bc_{i}) = C(\bx_{i})\prod_{\ell=1}^{d}f_{x_{\ell}\mid\bc}(x_{\ell,i};\bc_{i})$ with 
\bse
& C(\bx_{i}) = |\bR|^{-\frac{1}{2}} \exp\left\{-\frac{1}{2}\by_{i}\trans(\bR^{-1}- \bI_{d})\by_{i}\right\}, \\ 
& f_{x_{\ell}\mid\bc}(x_{\ell,i};\bc_{i}) = \sum_{k=1}^{K} \lambda_{\ell,\wt{s}_{\ell,\bc_{i}}}(k) \TN\left(x_{\ell,i}; \mu_{\ell,k}, \sigma^{2}_{\ell,k}, [A,B]\right).
\ese
In Scenario A, we sampled the core vector elements $\blambda_{\ell,k^{\star}}$ from a symmetric Dirichlet distribution, and fixed the correlation matrix $\bR$, the atoms $(\bmu,\bsigma^2)$ and the partitions $(\bs,\bs^{\star})$; in Scenario B, we use the posterior point estimates of $\bR$ and $f_{x_{\ell}\mid\bc}(\cdot\,;\bc)$.
In both scenarios, we generated $\bx_{i}$ using the standard approach to obtain a realization of a Gaussian-copula-distributed random variable:
\begin{enumerate}
    \item sample $\bx^{\Delta}_{i}\sim\MVN_{d}(\bzero_{d},\bR)$,
    \item set $x_{\ell,i}=F_{x_{\ell}\mid\bc}^{-1}(\Phi(x^{\Delta}_{\ell,i}); \bc_{i})$,
\end{enumerate}
where $F_{x_{\ell}\mid\bc}(\cdot\,; \bc_{i})$ is the cumulative distribution function corresponding to $f_{x_{\ell}\mid\bc}(\cdot\,; \bc_{i})$. 

\subsection{Scenario A}
We simulated datasets composed of $n\in\{1000,2000,3000\}$ 3-dimensional responses in the presence of $p$ covariates. To generate the covariate matrix $\bc$, we assumed that the observed levels for any covariate $h$ and unit $i$ are randomly selected, i.e., $\Pr(c_{h,i}=c_{h})= 1/d_{h}$ for all $(h,i)$. 
To generate the response matrix $\bx$, we used the proposed model, i.e., responses are assumed to be distributed as in (4), (11), and (7) without the common atoms
\bse
f_{\bx\mid\bc}(\bx_{i};\bc_{i}) = |\bR|^{-\frac{1}{2}} \exp\left\{-\frac{1}{2}\by_{i}\trans(\bR^{-1}- \bI_{d})\by_{i}\right\} \prod_{\ell=1}^{d} \sum_{k=1}^{K} \lambda_{\ell,\wt{s}_{\ell,\bc_{i}}}(k) \TN\left(x_{\ell,i}; \mu_{\ell,k}, \sigma^{2}_{\ell,k}, [A,B]\right).
\ese
The core vector elements are sampled from a symmetric Dirichlet distribution, $\blambda_{\ell,k^{\star}}\sim\Dir_{K}(2/K)$ for all $(\ell,k^{\star})$, while the remaining quantities are fixed as $[A, B]=[0,10]$,
\bse
\hskip -0.5cm ~\bR = \left(\begin{array}{ccc}
    1 & 0.7 & 0.7^2 \\
      & 1 & 0.7 \\
      &  & 1 
\end{array} \right),
\quad
\bmu =  \left(\begin{array}{c}
    \bmu_{1}\trans \\
    \bmu_{2}\trans \\
    \bmu_{3}\trans
\end{array} \right) 
= 
\left(\begin{array}{cccc}
    1 & 2 & 3 & 5 \\
    1 & 2 & 4 & 5 \\
    2 & 3 & 4 & 5
\end{array} \right),
\ese
and $\sigma_{\ell,k}=0.75$ for all $(\ell,k)$.
To evaluate model performance across varying covariate dimensions, we consider three sub-cases:
\vspace*{-10pt}
\begin{enumerate}[label=\alph*.,itemsep=-5pt] 
\item 
\textbf{Reference:} $p=5$ covariates with 6, 2, 4, 5, and 3 levels, respectively. 
\item 
\textbf{Increased levels:} $p=5$ covariates matching the reference sub-case, but with one additional level added to each covariate.
\item 
\textbf{Increased dimension:} $p=6$ covariates, incorporating an additional 4-level covariate alongside those of the reference sub-case.
\end{enumerate}
\vspace*{-10pt}
We further consider two sub-settings for each sub-case, for a total of 6 possible configurations for the covariates and their partitions under Scenario A, where we change the number of clusters in the partitions and, as a result, the number of true densities.

Within the same sub-setting, the three different sub-cases are designed to have different true partitions $\bs$ in the first layer but share the same partition $\bs^{\star}$ in the second layer, as $\bs$ induces the same coordinate-specific partitions over the joint covariate space in all sub-cases.
We report $\bs^{\star}$ here first, deferring $\bs$ to later in the section. 
For the two sub-cases with $p=5$, in the first sub-setting $\bs^{\star}$ is set to
\begin{center}
    \begin{tabular}{lll}
       $s^{\star}_{1,(1,1,1,1,1)}=1$, \quad &  $s^{\star}_{2,(1,1,1,1,1)}=1$, \quad & $s^{\star}_{3,(1,1,1,1,1)}=1$, \\
       $s^{\star}_{1,(2,1,1,1,1)}=2$, \quad &  $s^{\star}_{2,(1,2,1,1,1)}=2$, \quad & $s^{\star}_{3,(1,1,2,1,1)}=2$, \\
       $s^{\star}_{1,(2,2,1,1,1)}=2$, \quad &  $s^{\star}_{2,(1,1,1,2,1)}=2$, \\
       $s^{\star}_{1,(1,2,1,1,1)}=3$, \quad &  $s^{\star}_{2,(1,2,1,2,1)}=3$, \\
                                      &  $s^{\star}_{2,(1,1,1,3,1)}=3$, \\
                                      &  $s^{\star}_{2,(1,2,1,3,1)}=4$, \\
    \end{tabular}
\end{center}
while in the second sub-setting, it is set to
\begin{center}
    \begin{tabular}{lll}
       $s^{\star}_{1,(1,1,1,1,1)}=1$, \quad & $s^{\star}_{2,(1,1,1,1,1)}=1$, \quad & $s^{\star}_{3,(1,1,1,1,1)}=1$, \\
       $s^{\star}_{1,(2,1,1,1,1)}=2$, \quad & $s^{\star}_{2,(1,2,1,1,1)}=2$, \quad & $s^{\star}_{3,(2,1,1,1,1)}=2$, \\
       $s^{\star}_{1,(1,2,1,1,1)}=3$, \quad & $s^{\star}_{2,(1,1,2,1,1)}=3$, \quad & $s^{\star}_{3,(3,1,1,1,1)}=1$, \\
       $s^{\star}_{1,(2,2,1,1,1)}=4$, \quad & $s^{\star}_{2,(1,2,2,1,1)}=3$, \quad & $s^{\star}_{3,(1,1,1,2,1)}=3$, \\
       $s^{\star}_{1,(1,1,2,1,1)}=5$, \quad & $s^{\star}_{2,(1,1,3,1,1)}=4$, \quad & $s^{\star}_{3,(2,1,1,2,1)}=1$, \\
       $s^{\star}_{1,(2,1,2,1,1)}=6$, \quad & $s^{\star}_{2,(1,2,3,1,1)}=4$, \quad & $s^{\star}_{3,(3,1,1,2,1)}=2$, \\
       $s^{\star}_{1,(1,2,2,1,1)}=7$, \quad & $s^{\star}_{2,(1,1,1,2,1)}=5$, \quad & $s^{\star}_{3,(1,1,1,3,1)}=3$, \\
       $s^{\star}_{1,(2,2,2,1,1)}=8$, \quad & $s^{\star}_{2,(1,2,1,2,1)}=5$, \quad & $s^{\star}_{3,(2,1,1,3,1)}=1$, \\
                                     & $s^{\star}_{2,(1,1,2,2,1)}=6$, \quad & $s^{\star}_{3,(3,1,1,3,1)}=2$, \\
                                     & $s^{\star}_{2,(1,2,2,2,1)}=6$, \\
                                     & $s^{\star}_{2,(1,1,3,2,1)}=7$, \\
                                     & $s^{\star}_{2,(1,2,3,2,1)}=7$, \\
                                     & $s^{\star}_{2,(1,1,1,3,1)}=8$, \\
                                     & $s^{\star}_{2,(1,2,1,3,1)}=8$, \\
                                     & $s^{\star}_{2,(1,1,2,3,1)}=9$, \\
                                     & $s^{\star}_{2,(1,2,2,3,1)}=9$, \\
                                     & $s^{\star}_{2,(1,1,3,3,1)}=10$, \\
                                     & $s^{\star}_{2,(1,2,3,3,1)}=10$, \\
    \end{tabular}
\end{center}
For $p=6$, we designed $\bs^{\star}$ such that its number of elements did not change; however, the tensor has an additional dimension of one.
For instance, $s^{\star}_{2,(1,2,3,2,1)}=7$ becomes $s^{\star}_{2,(1,2,3,2,1,1)}=7$.

Between the $p=5$ and $p=6$ settings, the latter is more complex due to both the larger number of true outcome densities and the higher-dimensional second-layer partition allocation tensor $\bs^{\star}$ required to induce the intricate joint covariate space partition.

\subsubsection*{Sub-Case (a): Reference}
For the first sub-case with $p=5$ covariates, we simulated 6, 2, 4, 5, and 3 levels, respectively, for a total of $\prod_{h=1}^{p}d_{h}=720$ covariate level combinations.
For each covariate $h$, the coordinate-specific partitions of the covariate levels $\C_{h}$ across the $d=3$ outcomes are collected into a $d \times d_{h}$ matrix, $\bs_{(h)}$. 
The two columns of $\bs_{(h)}$ matrices below display these partitions across the two simulation sub-settings under the reference $p=5$ case.
\bse
&
\bs_{(1)} =
\left(\begin{array}{cccccc}
    1 & 1 & 1 & 2 & 2 & 2 \\
    1 & 1 & 1 & 1 & 1 & 1 \\
    1 & 1 & 1 & 1 & 1 & 1 \\
\end{array} \right),
\qquad
\bs_{(1)} =
\left(\begin{array}{cccccc}
    1 & 1 & 1 & 2 & 2 & 2 \\
    1 & 1 & 1 & 1 & 1 & 1 \\
    1 & 1 & 2 & 2 & 3 & 3 \\
\end{array} \right),
\\
&
\bs_{(2)} =
\left(\begin{array}{cc}
    1 & 2 \\
    1 & 2 \\
    1 & 1 \\
\end{array} \right),
\qquad
\bs_{(2)} =
\left(\begin{array}{cc}
    1 & 2 \\
    1 & 2 \\
    1 & 1 \\
\end{array} \right),
\\
&
\bs_{(3)} =
\left(\begin{array}{cccc}
    1 & 1 & 1 & 1 \\
    1 & 1 & 1 & 1 \\
    1 & 1 & 2 & 2 \\
\end{array} \right),
\qquad
\bs_{(3)} =
\left(\begin{array}{cccc}
    1 & 1 & 2 & 2 \\
    1 & 1 & 2 & 3 \\
    1 & 1 & 1 & 1 \\
\end{array} \right), \\
\\
&
\bs_{(4)} =
\left(\begin{array}{ccccc}
    1 & 1 & 1 & 1 & 1 \\
    1 & 2 & 2 & 2 & 3 \\
    1 & 1 & 1 & 1 & 1 \\
\end{array} \right),
\qquad
\bs_{(4)} =
\left(\begin{array}{ccccc}
    1 & 1 & 1 & 1 & 1 \\
    1 & 1 & 1 & 2 & 3 \\
    1 & 2 & 2 & 3 & 3 \\
\end{array} \right),
\\
&
\bs_{(5)} =
\left(\begin{array}{ccc}
    1 & 1 & 1 \\
    1 & 1 & 1 \\
    1 & 1 & 1 \\
\end{array} \right),
\qquad
\bs_{(5)} =
\left(\begin{array}{ccc}
    1 & 1 & 1 \\
    1 & 1 & 1 \\
    1 & 1 & 1 \\
\end{array} \right).
\ese
For instance, in both settings, the second covariate belongs to the set of the most influential covariates for the first two coordinates of $\bx$ but not the last one; on the other hand, the fifth covariate is never influential for any coordinate of $\bx$.

\subsubsection*{Sub-Case (b): Higher Number of Levels}
For the second sub-case with $p=5$, we consider the same covariates as the reference sub-case but with one additional level each, resulting in $p=5$ covariates with 7, 3, 5, 6, and 4 levels, respectively, for a total of $\prod_{h=1}^{p}d_{h}=2520$ covariate level combinations.
In both sub-settings, the additional level is assigned to the same true cluster as the last level in the reference covariate.
For instance, $\bs_{1,1}=(1,1,1,2,2,2)$ becomes $\bs_{1,1}=(1,1,1,2,2,2,2)$. 

\subsubsection*{Sub-Case (c): Higher Number of Covariates}
In the third and final sub-case, we consider an additional covariate with 4 levels, resulting in $p=6$ covariates with 6, 2, 4, 5, 3, and 4 levels, respectively, for a total of $\prod_{h=1}^{p}d_{h}=2880$ covariate level combinations.
In both sub-settings, the sixth newly included covariate is never influential.

\subsubsection*{Results for Scenario A}
We compare our flower model with two univariate and one multivariate density regression methods in terms of Integrated Squared Error (ISE) between the posterior estimates of the conditional marginal densities and their true counterparts, and Adjusted Rand Index (ARI) between the posterior point estimates of the partitions over the covariate level combinations and the corresponding truths.
Results are reported in \autoref{tab:ISE_scenario123} and \autoref{tab:ARI_scenario123}. 
In terms of ISE, the proposed flower model and BoxCoxTree consistently outperform all competing approaches, with the flower model achieving the best performance in all configurations except one. 
In terms of ARI, our flower model achieves the top performance across all sub-cases under the first sub-setting, and under the first sub-case within the second sub-setting.

\begin{sidewaystable}
    \footnotesize
    \centering
    \begin{subtable}{\linewidth}
        \centering
        \caption{Results for the first sub-case (reference).}
        \begin{tabular}{c|ccc|ccc}
            & \multicolumn{3}{c|}{sub-setting 1} & \multicolumn{3}{c}{sub-setting 2} \\
            model & $n=1000$ & $n=2000$ & $n=3000$ & $n=1000$ & $n=2000$ & $n=3000$ \\
            \hline
            PYMM & 0.0348 (0.0044) & 0.0367 (0.0157) & 0.0365 (0.0135) & 0.0633 (0.0076) & 0.0627 (0.0102) & 0.0612 (0.0038) \\ 
            BoxCoxTree & 0.0066 (0.0010) & 0.0072 (0.0013) & 0.0071 (0.0015) & 0.0198 (0.0034) & 0.0134 (0.0095) & 0.0090 (0.0114) \\ 
            MCTM-Lm & 0.0375 (0.0089) & 0.0363 (0.0104) & 0.0357 (0.0107) & 0.0585 (0.0093) & 0.0629 (0.0098) & 0.0627 (0.0073) \\ 
            MCTM-BoxCox & 0.0237 (0.0039) & 0.0227 (0.0035) & 0.0217 (0.0033) & 0.0477 (0.0100) & 0.0500 (0.0101) & 0.0498 (0.0086) \\ 
            MCTM-Colr & 0.0216 (0.0042) & 0.0202 (0.0033) & 0.0192 (0.0033) & 0.0457 (0.0092) & 0.0474 (0.0093) & 0.0478 (0.0083) \\ 
            FlowerModel & \textbf{0.0010 (0.0007)} & \textbf{0.0008 (0.0002)} & \textbf{0.0004 (0.0002)} & \textbf{0.0096 (0.0024)} & \textbf{0.0040 (0.0015)} & \textbf{0.0013 (0.0030)} \\ 
        \end{tabular}
        \label{tab:ISE_scenario1}
    \end{subtable}
    
    \vspace{0.5cm}
    
    \begin{subtable}{\linewidth}
        \centering
        \caption{Results for the second sub-case (higher number of levels).}
        \begin{tabular}{c|ccc|ccc}
            & \multicolumn{3}{c|}{sub-setting 1} & \multicolumn{3}{c}{sub-setting 2} \\
            model & $n=1000$ & $n=2000$ & $n=3000$ & $n=1000$ & $n=2000$ & $n=3000$ \\
            \hline
            PYMM & 0.0362 (0.0121) & 0.0352 (0.0130) & 0.0354 (0.0133) & 0.0540 (0.0124) & 0.0536 (0.0110) & 0.0535 (0.0136) \\ 
            BoxCoxTree & 0.0085 (0.0008) & 0.0094 (0.0024) & 0.0088 (0.0022) & 0.0113 (0.0031) & 0.0095 (0.0008) & 0.0086 (0.0017) \\ 
            MCTM-Lm & 0.0303 (0.0046) & 0.0295 (0.0044) & 0.0282 (0.0052) & 0.0462 (0.0147) & 0.0438 (0.0151) & 0.0434 (0.0167) \\ 
            MCTM-BoxCox & 0.0189 (0.0055) & 0.0182 (0.0036) & 0.0172 (0.0058) & 0.0360 (0.0162) & 0.0357 (0.0170) & 0.0347 (0.0178) \\ 
            MCTM-Colr & 0.0177 (0.0056) & 0.0163 (0.0048) & 0.0139 (0.0054) & 0.0349 (0.0163) & 0.0345 (0.0166) & 0.0336 (0.0175) \\ 
            FlowerModel & \textbf{0.0014 (0.0006)} & \textbf{0.0008 (0.0002)} & \textbf{0.0006 (0.0002)} & \textbf{0.0055 (0.0045)} & \textbf{0.0052 (0.0020)} & \textbf{0.0056 (0.0027)} \\ 
        \end{tabular}
        \label{tab:ISE_scenario2}
    \end{subtable}
    
    \vspace{0.5cm}
    
    \begin{subtable}{\linewidth}
        \centering
        \caption{Results for the third sub-case (higher number of covariates).}
        \begin{tabular}{c|ccc|ccc}
            & \multicolumn{3}{c|}{sub-setting 1} & \multicolumn{3}{c}{sub-setting 2} \\
            model & $n=1000$ & $n=2000$ & $n=3000$ & $n=1000$ & $n=2000$ & $n=3000$ \\
            \hline
            PYMM & 0.0354 (0.0071) & 0.0397 (0.0048) & 0.0401 (0.0050) & 0.0596 (0.0210) & 0.0603 (0.0112) & 0.0602 (0.0089) \\ 
            BoxCoxTree & 0.0066 (0.0034) & 0.0068 (0.0027) & 0.0070 (0.0025) & 0.0099 (0.0008) & 0.0089 (0.0006) & 0.0081 (0.0005) \\ 
            MCTM-Lm & 0.0345 (0.0091) & 0.0334 (0.0058) & 0.0340 (0.0049) & 0.0542 (0.0127) & 0.0527 (0.0128) & 0.0527 (0.0107) \\ 
            MCTM-BoxCox & 0.0173 (0.0089) & 0.0148 (0.0066) & 0.0145 (0.0054) & 0.0446 (0.0165) & 0.0432 (0.0148) & 0.0421 (0.0129) \\ 
            MCTM-Colr & 0.0168 (0.0082) & 0.0136 (0.0062) & 0.0131 (0.0049) & 0.0438 (0.0167) & 0.0415 (0.0144) & 0.0406 (0.0125) \\ 
            FlowerModel & \textbf{0.0013 (0.0002)} & \textbf{0.0007 (0.0002)} & \textbf{0.0004 (0.0001)} & \textbf{0.0049 (0.0036)} & \textbf{0.0034 (0.0008)} & \textbf{0.0036 (0.0071)} \\
        \end{tabular}
        \label{tab:ISE_scenario3}
    \end{subtable}
    \caption{Results for the three sub-cases of Scenario A. Median ISEs (interquartile ranges in parentheses) between the posterior estimates of the conditional marginal densities and the corresponding truths, computed on 5 replications.}
    \label{tab:ISE_scenario123}
\end{sidewaystable}

\begin{sidewaystable}
    \footnotesize
    \centering
    \begin{subtable}{\linewidth}
        \centering
        \caption{Results for the first sub-case (reference).}
        \begin{tabular}{c|ccc|ccc}
            & \multicolumn{3}{c|}{sub-setting 1} & \multicolumn{3}{c}{sub-setting 2} \\
            model & $n=1000$ & $n=2000$ & $n=3000$ & $n=1000$ & $n=2000$ & $n=3000$ \\
            \hline
            BoxCoxTree & 0.8142 (0.0354) & 0.8877 (0.0741) & 0.8877 (0.0278) & 0.5907 (0.1061) & 0.6105 (0.1488) & 0.7408 (0.1696) \\ 
            FlowerModel & \textbf{0.9437 (0.0271)} & \textbf{0.9165 (0.0464)} & \textbf{0.9437 (0.0563)} & \textbf{0.7620 (0.0106)} & \textbf{0.7624 (0.1179)} & \textbf{0.8193 (0.1067)} \\ 
        \end{tabular}
        \label{tab:ARI_scenario1}
    \end{subtable}
    
    \vspace{0.5cm}
    
    \begin{subtable}{\linewidth}
        \centering
        \caption{Results for the second sub-case (higher number of levels).}
        \begin{tabular}{c|ccc|ccc}
            & \multicolumn{3}{c|}{sub-setting 1} & \multicolumn{3}{c}{sub-setting 2} \\
            model & $n=1000$ & $n=2000$ & $n=3000$ & $n=1000$ & $n=2000$ & $n=3000$ \\
            \hline
            BoxCoxTree & 0.8443 (0.0279) & 0.8468 (0.1055) & 0.9077 (0.0000) & \textbf{0.6730 (0.1933)} & \textbf{0.7461 (0.0699)} & \textbf{0.8208 (0.0544)} \\ 
            FlowerModel & \textbf{0.9274 (0.0771)} & \textbf{1.0000 (0.0000)} & \textbf{1.0000 (0.0000)} & 0.6573 (0.0330) & 0.6721 (0.0764) & 0.7129 (0.1245) \\ 
        \end{tabular}
        \label{tab:ARI_scenario2}
    \end{subtable}
    
    \vspace{0.5cm}
    
    \begin{subtable}{\linewidth}
        \centering
        \caption{Results for the third sub-case (higher number of covariates).}
        \begin{tabular}{c|ccc|ccc}
            & \multicolumn{3}{c|}{sub-setting 1} & \multicolumn{3}{c}{sub-setting 2} \\
            model & $n=1000$ & $n=2000$ & $n=3000$ & $n=1000$ & $n=2000$ & $n=3000$ \\
            \hline
            BoxCoxTree & 0.8470 (0.2145) & 0.8602 (0.0422) & 0.8602 (0.0307) & \textbf{0.6998 (0.0956)} & 0.7847 (0.0702) & \textbf{0.7915 (0.0925)} \\ 
            FlowerModel & \textbf{0.9438 (0.1569)} & \textbf{0.9438 (0.0562)} & \textbf{1.0000 (0.0000)} & 0.6700 (0.1318) & \textbf{0.8052 (0.1267)} & 0.7611 (0.0865) \\ 
        \end{tabular}
        \label{tab:ARI_scenario3}
    \end{subtable}
    \caption{Results for the three sub-cases of Scenario A. Median ARIs (interquartile ranges in parentheses) between the posterior estimates of the partitions of covariate level combinations and the corresponding truths, computed on 5 replications.}
    \label{tab:ARI_scenario123}
\end{sidewaystable}

\newpage
\FloatBarrier
\subsection{Scenario B}
We simulated a dataset with $n=6087$ units, 
$d=6$ dimensional responses, and $p=4$ covariates with 2, 7, 6, and 6 levels, respectively. 
We set the covariate matrix $\bc$ to that observed in the NHANES dietary data, and generated the multivariate responses from 
\bse
f_{\bx\mid\bc}(\bx_{i};\bc_{i}) = |\bR|^{-\frac{1}{2}} \exp\left\{-\frac{1}{2}\by_{i}\trans(\bR^{-1}- \bI_{d})\by_{i}\right\} \prod_{\ell=1}^{d} f_{x_{\ell}\mid\bc}(x_{\ell,i};\bc_{i}),
\ese
where $f_{x_{\ell}\mid\bc}(x_{\ell,i};\bc_{i})$ are set to their posterior point estimates conditional on the MAP estimates of the partition layers, and $\bR$ is set to its posterior point estimate rounded to two decimal places, which is
\bse
\hskip -0.5cm ~\bR = \left(\begin{array}{cccccc}
    1.00 &  0.25 &  0.07 &  0.11 &  0.36 &  0.26 \\ 
         &  1.00 &  0.15 &  0.18 &  0.57 &  0.41 \\ 
         &       &  1.00 & -0.12 & -0.06 & -0.43 \\ 
         &       &       &  1.00 &  0.60 & 0.51 \\ 
         &       &       &       &  1.00 & 0.74 \\ 
         &       &       &       &       & 1.00 \\ 
\end{array} \right).
\ese

\subsubsection*{Results for Scenario B}
\begin{figure}[!ht]
    \begin{center}
    \includegraphics[width=\textwidth, trim=0cm 0cm 0cm 0cm, clip=true]{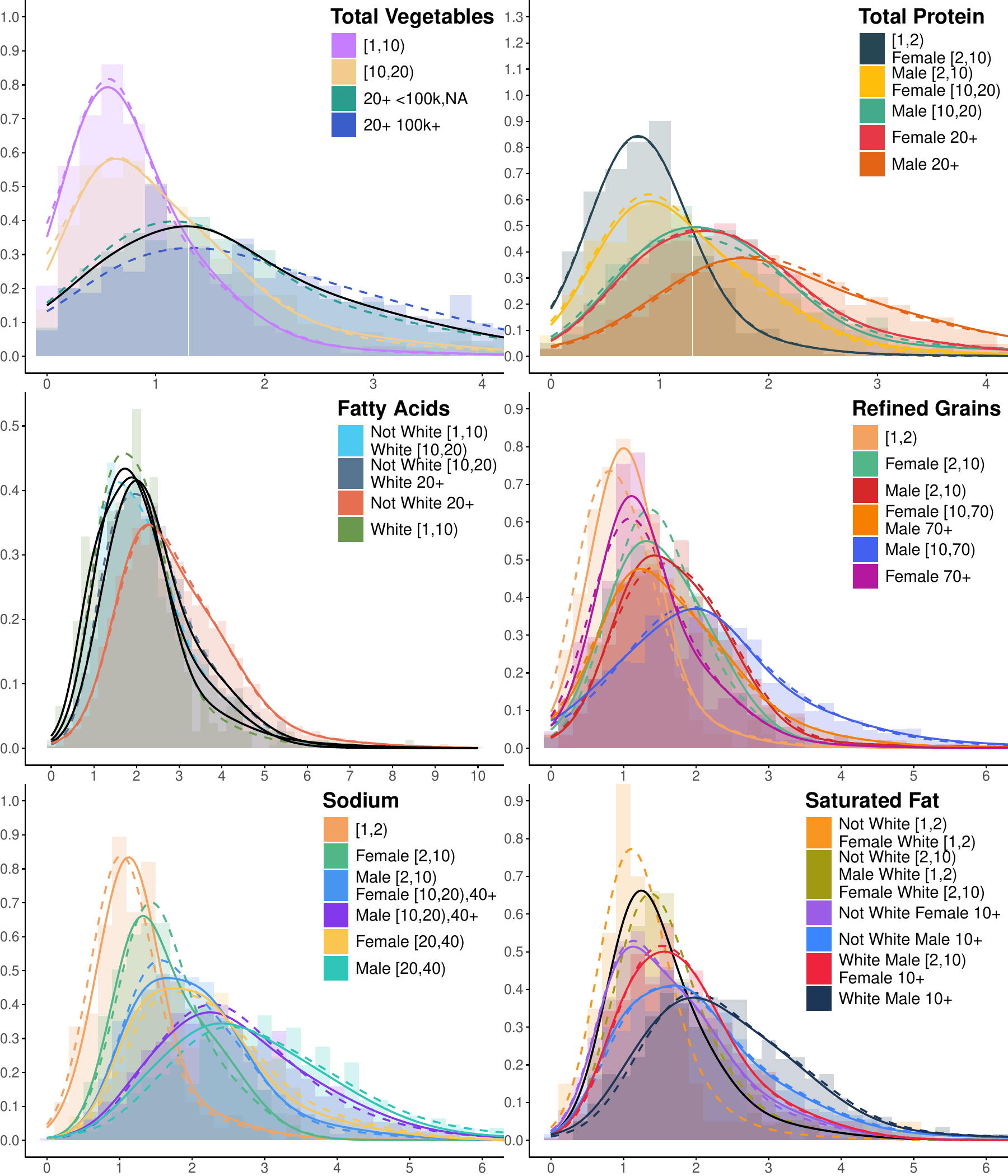}
    \end{center}
    \caption{Results for simulated data in Scenario B. The panels, one for each coordinate of the response, compare the estimated conditional densities for the estimated partitions of the covariate level combinations (solid lines) and the true conditional densities for the true partitions of the covariate level combinations (dashed lines), overlaid on histograms representing the corresponding empirical distributions. 
    Each color refers to a different true underlying distribution.
    Black is used to denote estimated densities of incorrectly identified groups, which typically happens when the model collapses highly similar true distributions or misassigns lower-prevalence groups in the estimated partitions.}
    \label{fig:sim_marginals_est_nhanes}
\end{figure}

\begin{figure}[!ht]
    \begin{center}
    \includegraphics[width=\textwidth, trim=0cm 0cm 0cm 0cm, clip=true]{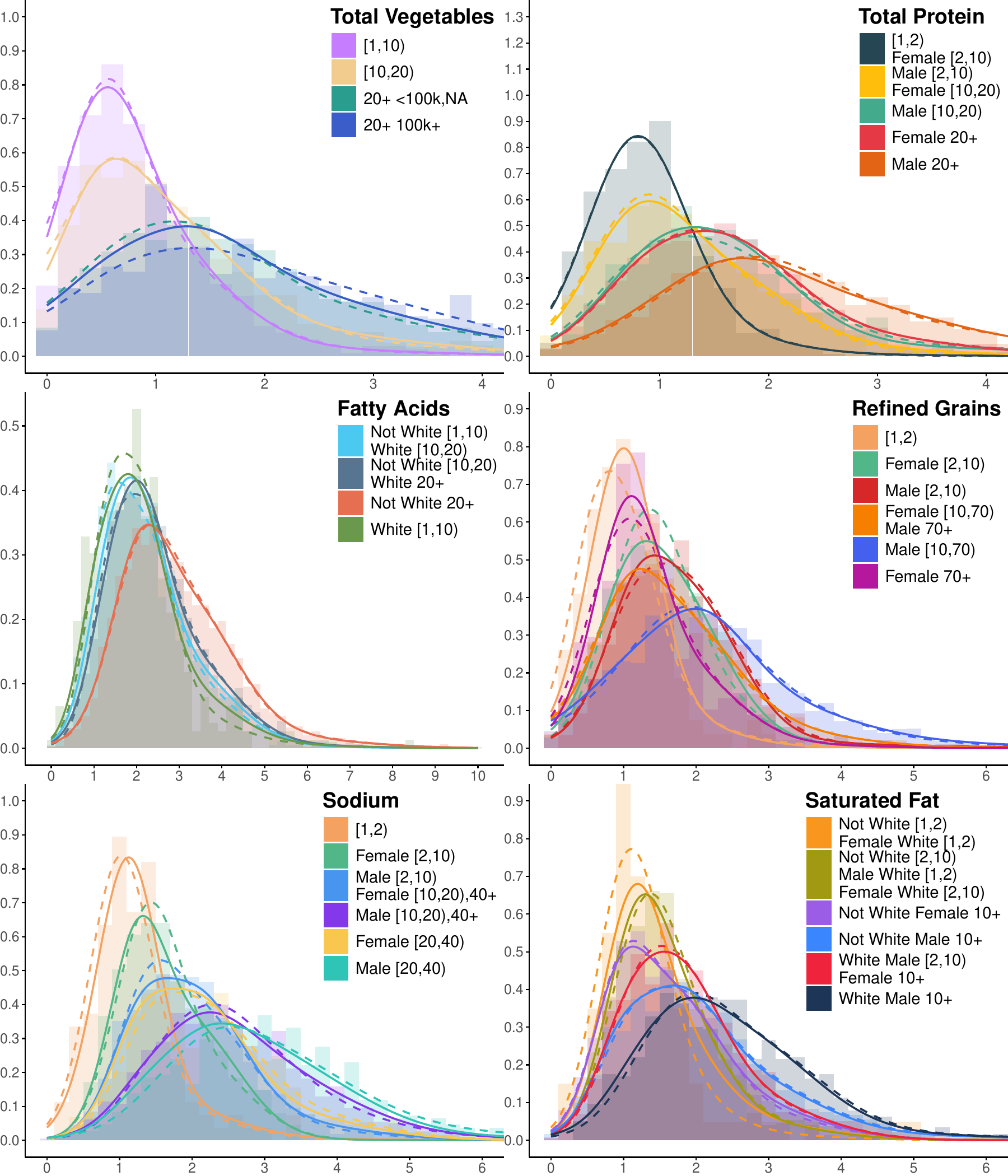}
    \end{center}
    \caption{Results for simulated data in Scenario B. The panels, one for each response coordinate, compare the estimated conditional densities (solid lines) and the corresponding truths (dashed lines) for different true partitions of the covariate-level combinations, overlaid on histograms of the corresponding empirical distributions. 
    Each color refers to a different true underlying distribution.
    }
    \label{fig:sim_marginals_true_nhanes}
\end{figure}

\autoref{fig:sim_marginals_est_nhanes} and \autoref{fig:sim_marginals_true_nhanes} show a comparison between the estimated conditional marginal distributions and the true conditional marginals under Scenario B.
While \autoref{fig:sim_marginals_est_nhanes} shows the estimated conditional marginal densities for each group identified by the model, \autoref{fig:sim_marginals_true_nhanes} shows the estimated conditional marginals for the true groups.
Posterior estimates of $f_{x_{\ell}\mid\bc}(\cdot\,;\bc)$ given $(\wh{\bs}_{\ell}, \wh{\bs}_{\ell}^{\star})$, or $(\bs_{\ell}, \bs_{\ell}^{\star})$, for any coordinate $\ell$ are shown as solid lines, while the true distributions used to simulate the responses are shown as dashed lines. 
Each color identifies a different true marginal distribution, and the same color is assigned to all estimated densities referring to the same group of covariate combinations.
While the responses are now synthetic, the covariates coincide with those observed in the NHANES dietary data, so we maintain labels consistent with the empirical data to provide a familiar context for describing and interpreting the model outputs. 
In both figures, the legends report the true groups used to generate the data, not those identified by the model.

Overall, we observe a good recovery of the underlying true densities. 
As expected in such a high-dimensional setting, some discrepancies arise from the complexity of reconstructing a large number of disparate densities. 
This usually happens when the model collapses true distributions with very similar shapes or misallocates lower-prevalence groups in the estimated partitions. 
For instance, in \autoref{fig:sim_marginals_est_nhanes}, certain true groups appear merged into single ``black'' estimated densities, 
whereas in \autoref{fig:sim_marginals_true_nhanes}, which displays estimates conditioned on the true partitions, these merged components (e.g., Saturated Fat) resolve into distinct colored curves, suggesting that while our model identified separate groups during MCMC sampling, their close empirical similarities made consistent separation challenging. 
Conversely, for nearly indistinguishable groups like those for Total Vegetables and Fatty Acids, the model appropriately assigns a single density to both. 
The high ARI between the true and the estimated partitions indicates that these differences are limited.

\clearpage
\section{Speed and Scalability}
We focus here on the time and memory complexity of our implementation.

\subsubsection*{Time Complexity}
\autoref{tab:time_complexity_noncoord} and \autoref{tab:time_complexity_coord} show the time complexity of each update within an MCMC iteration, expressed in terms of $O(\cdot)$.
\autoref{tab:time_complexity_noncoord} collects the quantities that are shared across coordinates, while \autoref{tab:time_complexity_coord} collects the coordinate-specific ones, reporting the cost for both the single coordinate $\ell$ and the total cost for all $d$ coordinates.
In what follows, we denote the number of elements in the $\ell\th$ second-layer partition $\bs^{\star}_{\ell}$ by $K_{\ell} = \prod_{h=1}^{p}K_{\ell,h}$.

\begin{table}[ht!]
    \centering
    \begin{tabular}{cc!{\hspace{1em}\vrule\hspace{1em}}cc}
         Quantity & $O(\cdot)$ & Quantity & $O(\cdot)$ \\
         \hline
         $\phi$ &  $d \cdot \sum_{h=1}^{p}d_{h}$ & $\by$ & $d\cdot n\cdot(p+K)$ \\
         $\alpha$ & $K \sum_{\ell=1}^{d} K^{\star}_{\ell}$ & $\by\by\trans$ & $d^{2}\cdot n$ \\
         $\bmu$ & $d\cdot n + K$ & $\mathbf{b}$ & $d^{4}$ \\
         $\bsigma^{2}$ & $d\cdot n + K$ & $\boldsymbol{\theta}$ & $d^{5}$ \\
    \end{tabular}
    \caption{Time complexity of the updates of quantities shared across coordinates.}
    \label{tab:time_complexity_noncoord}
\end{table}
\begin{table}[ht!]
    \footnotesize
    \centering
    \begin{tabular}{c|c|c}
        Quantity & \multicolumn{2}{c}{$O(\cdot)$} \\\cline{2-3}
        & coordinate & total \\
        \hline
        $\bz_{\ell}$ & $n (p+K)$ & $d \cdot n \cdot (p+K)$ \\
        $(\bs_{\ell,h},\bs^{\star}_{\ell})$ &
            $p\sum_{h=1}^{p}d_{h} + p\,K_{\ell}^{2} + p K^{\star}_{\ell} K + \sum_{h=1}^{p}d_{h}^{2} + n p^{2}$ &
            $\displaystyle\sum_{\ell=1}^{d} \left( p\sum_{h=1}^{p}d_{h} + p\,K_{\ell}^{2} + p K^{\star}_{\ell} K + \sum_{h=1}^{p}d_{h}^{2} + n p^{2} \right)$ \\
        $\bs^{\star}_{\ell}$ &
            $n p + K_{\ell}\,K^{\star}_{\ell} K$ &
            $\displaystyle d \cdot n p + K\sum_{\ell=1}^{d}K_{\ell}\,K^{\star}_{\ell}$ \\
        $\blambda_{\ell,0}$ &
            $n + K^{\star}_{\ell} K$ &
            $\displaystyle d\cdot n + K\sum_{\ell=1}^{d}K^{\star}_{\ell}$ \\
        $\blambda_{\ell}$ &
            $K^{\star}_{\ell} K$ &
            $\displaystyle K\sum_{\ell=1}^{d}K^{\star}_{\ell}$ \\
    \end{tabular}
    \caption{Time complexity of the updates of coordinate-specific quantities, for a single coordinate $\ell$ and for all $d$ coordinates. The joint update of $(\bs_{\ell,h},\bs^{\star}_{\ell})$ is repeated for $h=1,\ldots,p$ and the cost accounts for all $p$ repetitions.}
    \label{tab:time_complexity_coord}
\end{table}
Since all updates are executed sequentially, the total per-iteration cost is simply given by the sum of the individual costs.
After simplifying, the total cost for updating the quantities related to the $\ell\th$ coordinate is
\bse
O\!\left( n\,(p^{2}+K)
+ p\sum_{h=1}^{p} d_{h}
+ \sum_{h=1}^{p} d_{h}^{2}
+ p\,K_{\ell}^{2}
+ p\,K\,K^{\star}_{\ell}
+ K\,K_{\ell}\,K^{\star}_{\ell} \right),
\ese
while the total cost for an MCMC iteration is
\bse
O\Big( d\,n\,(p^{2}+K) + d^{2} n + d\,p \sum_{h=1}^{p}d_{h}
+ d \sum_{h=1}^{p} d_{h}^{2}
+ p\sum_{\ell=1}^{d}K_{\ell}^{2}
+ p\,K\sum_{\ell=1}^{d}K^{\star}_{\ell}
+ K\sum_{\ell=1}^{d}K_{\ell}\,K^{\star}_{\ell}
+ d^{5} \Big).
\ese
Notably, the covariates enter only through $p$, $\sum_{h=1}^{p}d_{h}$ and $\sum_{h=1}^{p}d^{2}_{h}$, and never through the number of covariate combinations $\prod_{h=1}^{p}d_{h}$, implying that the computational cost of the proposed implementation depends on the number of covariate levels rather than their product.
Additionally, the sample size $n$ enters only linearly and is not influenced by the size $K_{\ell}$ of the second-layer partitions.

\subsubsection*{Memory Complexity}
\autoref{tab:memory_state} reports the memory requirements of the quantities sampled by the MCMC algorithm, i.e., the size of each state of the Markov chain, while \autoref{tab:memory_working} reports the main auxiliary objects that are required to run the sampler.
\begin{table}[ht!]
    \centering
    \begin{tabular}{cc!{\hspace{1em}\vrule\hspace{1em}}cc}
         Quantity & Size & Quantity & Size \\
         \hline
         $\phi$ & $1$ & $\bz$ & $d\cdot n$ \\
         $\alpha$ & $1$ & $\blambda_{0}$ & $d\cdot K$ \\
         $\bmu$ & $K$ & $\bs$ & $\displaystyle d\cdot \sum_{h=1}^{p}d_{h}$ \\
         $\bsigma^{2}$ & $K$ & $\bs^{\star}$ & $\displaystyle\sum_{\ell=1}^{d} K_{\ell}$ \\
         $\bb$ & $d - 1$ & $\wh{\blambda}$ & $\displaystyle K\sum_{\ell=1}^{d} K^{\star}_{\ell}$ \\
         $\btheta$ & $(d^{2}-3d+2)/2$ & & \\
    \end{tabular}
    \caption{Memory requirements of the quantities being sampled (one instance).}
    \label{tab:memory_state}
\end{table}
\begin{table}[ht!]
    \footnotesize
    \centering
    \begin{tabular}{c|c|c}
        Quantity & Role & Size \\
        \hline
        $K_{\ell,h}, K^{\mathrm{(new)}}_{\ell,h}$ & number of clusters in current and proposed first-layer partitions & $d\cdot p$ \\
        $m_{\ell,h}(q_{h})$, $m^{\mathrm{(new)}}_{\ell,h}(q_{h})$ & list of $p$ count matrices of dimension $d\times d_{h}$ & $\displaystyle d\cdot\sum_{h=1}^{p}d_{h}$ \\[-0.3cm]
        & based on current and proposed quantities & \\
        $m^{\star}_{\ell}(k^{\star})$, $m^{\star\mathrm{(new)}}_{\ell}(k^{\star})$ & list of $d$ count vectors of length $K_{\ell}^{\star}$ & $\displaystyle\sum_{\ell=1}^{d}K^{\star}_{\ell}$ \\[-0.3cm]
        & based on current and proposed quantities & \\
        $n_{\ell,k^{\star}}(k)$, $n^{\mathrm{(new)}}_{\ell,k^{\star}}(k)$ & list of $d$ count matrices of dimension $K_{\ell}^{\star}\times K$ & $\displaystyle K\sum_{\ell=1}^{d}K^{\star}_{\ell}$ \\[-0.3cm]
        & based on current and proposed quantities & \\
        \hdashline\noalign{\vskip 1pt}
        $\bs^{\mathrm{(new)}}_{\ell}$ & proposed first-layer partitions (shared across coordinates) & $\sum_{h=1}^{p}d_{h}$ \\
        $\bs^{\star\mathrm{(new)}}_{\ell}$ & proposed second-layer partition (shared across coordinates) & $\displaystyle O\left(\prod_{h=1}^{p}K_{\ell,h}\right)$ \\
        $\bmu^{\mathrm{(new)}}$, $\bsigma^{2\mathrm{(new)}}$ & proposed common atoms & $K$ \\
         \hdashline\noalign{\vskip 1pt}
        $\by$ & copula latent variables & $d\cdot n$ \\
        $\by\by\trans$ & cross-product of copula latent variables & $d^{2}$ \\
        $\bR^{-1}$, $\{\bR^{\mathrm{(new)}}\}^{-1}$ & current and proposed copula correlation matrices & $d^{2}$ \\
        $\bb^{\mathrm{(new)}}$ & proposed copula parameter vector & $d-1$ \\
        $\btheta^{\mathrm{(new)}}$ & proposed copula parameter vector & $(d^{2}-3d+2)/2$ \\
    \end{tabular}
    \caption{Memory requirements of the main auxiliary objects required by the MCMC algorithm.}
    \label{tab:memory_working}
\end{table}

The component allocation variables $\bz$ and the copula latent variables $\by$, both of size $d\cdot n$, are the only quantities that scale with the sample size $n$; all remaining quantities scale with the number of covariate levels $\sum_{h=1}^{p}d_{h}$, the number of mixture components $K$, the maximum number of conditional marginal densities $K_{\ell}^{\star}$, or with the dimension $K_{\ell}$ of the second-layer partitions.
Importantly, no object scales with the number of covariate combinations $\prod_{h=1}^{p}d_{h}$, and the size of the second-layer partitions $\bs_{\ell}^{\star}$ varies as the number of clusters in the first-layer partitions changes, as described in Figure 4 of the manuscript.
Overall, the memory required to run the sampler is
\bse
O\!\left( dn
+ d^{2}
+ d\sum_{h=1}^{p}d_{h}
+ K\sum_{\ell=1}^{d}K^{\star}_{\ell}
+ \sum_{\ell=1}^{d}K_{\ell} \right).
\ese

Finally, we note that the quantities $(\bs,\bs^{\star},\bmu,\bsigma^{2},\bb,\btheta,\wh{\blambda})$, which are sufficient to compute estimates of the conditional marginal densities, do not include $\bz$ and $\by$.
Therefore, the memory required to store their posterior samples is independent of the sample size $n$.

\subsubsection*{Results for Simulations}
\autoref{tab:TIME_sim} reports the elapsed time to run the MCMC and estimate the conditional marginal densities for all sub-cases and sub-settings, measured in seconds.
The elapsed time grows approximately linearly with the sample size $n$, which is the dominant factor, and it increases with the characteristics of the covariates; on the other hand, it is essentially unaffected by the complexity of the latent partition structure.

\autoref{tab:MEMORY_sim} reports the memory required to store posterior samples from $(\bs,\bs^{\star},\bmu,\bsigma^{2},\bb,\btheta,\wh{\blambda})$, which are necessary to estimate posterior point estimates of the conditional densities for all sub-cases and sub-settings, measured in Megabytes (MB).
The memory is independent of the sample size and is governed mainly by the characteristics of the covariates through the dimension of the first-layer partitions. In a minor way, a more complex latent structure further increases the needed memory through the second-layer partitions.

\begin{table}[ht]
    \centering
    \begin{subtable}{\linewidth}
        \centering
        \caption{Elapsed time (in seconds) to run the MCMC and estimate the conditional marginal densities.}
        \begin{tabular}{r|ccc|ccc}\hline
            & \multicolumn{3}{c|}{sub-setting 1} & \multicolumn{3}{c}{sub-setting 2} \\ \cline{2-7}
            & $n=1000$ & $n=2000$ & $n=3000$ & $n=1000$ & $n=2000$ & $n=3000$ \\ 
            \hline
            Sub-case a & 407 (10) & 761 (28) & 1104 (23) & 415 (20) & 775 (11) & 1125 (40) \\ 
            Sub-case b & 513 (1) & 895 (22) & 1254 (83) & 521 (14) & 877 (12) & 1227 (42) \\ 
            Sub-case c & 551 (6) & 897 (21) & 1251 (3) & 543 (1) & 906 (19) & 1317 (60) \\ 
            \hline
        \end{tabular}
        \label{tab:TIME_sim}
    \end{subtable}
    
    \vspace{0.3cm}
    
    \begin{subtable}{\linewidth}
        \centering
        \caption{Memory (in MB) required to store posterior samples from $(\bs,\bs^{\star},\bmu,\bsigma^{2},\bb,\btheta,\wh{\blambda})$.}
        \begin{tabular}{r|ccc|ccc}\hline
            & \multicolumn{3}{c|}{sub-setting 1} & \multicolumn{3}{c}{sub-setting 2} \\ \cline{2-7}
            & $n=1000$ & $n=2000$ & $n=3000$ & $n=1000$ & $n=2000$ & $n=3000$ \\ 
            \hline
            Sub-case a & 11.0 (0.0) & 10.9 (0.1) & 11.0 (0.2) & 11.2 (0.1) & 11.3 (0.1) & 11.4 (0.1) \\ 
            Sub-case b & 11.2 (0.0) & 11.2 (0.0) & 11.2 (0.1) & 11.5 (0.0) & 11.7 (0.2) & 11.6 (0.0) \\ 
            Sub-case c & 11.6 (0.0) & 11.6 (0.0) & 11.6 (0.2) & 12.0 (0.1) & 11.9 (0.0) & 11.9 (0.2) \\ 
        \end{tabular}
        \label{tab:MEMORY_sim}
    \end{subtable}
    \caption{Results for simulated data. Top: Median elapsed time (interquartile ranges in parentheses) to run the MCMC and estimate the conditional marginal densities, computed on 5 replications. Bottom: Median memory (interquartile ranges in parentheses) required to store posterior samples from $(\bs,\bs^{\star},\bmu,\bsigma^{2},\bb,\btheta,\wh{\blambda})$, computed on 5 replications.}
    \label{tab:TIME_MEMORY_sim}
\end{table}

\clearpage
\section{Other Approaches}
Table \ref{tab:models_mix} provides a list of relevant univariate and multivariate density regression frameworks.
Within each block of the table, these are reported in order of appearance in the manuscript; the horizontal dashed line in the last block separates methods mentioned in the manuscript from those not mentioned.
For each method, we detail the types of covariates supported (`covariates' column), whether variable selection is possible (`vs'), and the availability of an implementation (`code').
For multivariate models, we further specify whether each method explicitly employs a copula (`\checkmark') or can be seen as such (`(\checkmark)'), and whether the marginal distributions are suitable for density estimation (`density').
The nature of covariates is described as C=continuous, D=discrete, and d=discrete but included as dummy variables.
For univariate mixture models, mostly employing Gaussian kernels, we additionally specify how covariates enter the model: via the mixture weights (`weight' sub-column) and the kernel mean (`mean') and variance (`variance').
Regarding the implementation, `SM' denotes code provided within the supplementary materials, while `Author' indicates that code may be accessible upon request, as specified in the corresponding manuscript. For publicly hosted repositories, the direct website URL is provided. 

\begin{table}
    \scriptsize
    \centering
    \begin{tabular}{lcccccc}
        \textbf{Univariate mixture models} & name & \multicolumn{3}{c}{covariates} & vs & code \\
         &  & mean & variance & weights &  &  \\
        \hline
        \cite{wood_BayesianMixtureSplines_2002} &  & Cd &  & Cd &  &  \\
        \cite{geweke_SmoothlyMixingRegressions_2007} & SMR & Cd &  & Cd &  &  \\
        \cite{villani_RegressionDensityEstimation_2009} & SAGM & Cd & Cd & Cd & \checkmark &  \\
        \cite{nott_RegressionDensityEstimation_2012} &  & Cd & Cd & Cd & \checkmark & SM \\
        \cite{tran_SimultaneousVariableSelection_2012} &  & Cd & Cd & Cd & \checkmark & Author  \\
        \cite{gelfand_BayesianNonparametricSpatial_2005} &  & Cd &  &  &  &  \\
        \cite{deiorio_ANOVAModelDependent_2004} & ANOVA DDP & D &  &  &  & \texttt{\href{https://web.ma.utexas.edu/users/pmueller/prog.html\#ddpanova}{ddpanova}} \\
        \cite{deiorio_BayesianNonparametricNonproportional_2009} & LINEAR DDP & Cd &  &  &  & \texttt{\href{https://www.jstatsoft.org/article/view/v040i05}{DPpackage}} \\
        \cite{griffin_OrderBasedDependentDirichlet_2006} & order-based DDP & C &  & C &  &  \\
        \cite{dunson_BayesianDensityRegression_2007} & WMDP & C &  & C &  & \\
        \cite{dunson_KernelStickBreakingProcesses_2008} & KSBP & C &  & C &  &  \\
        \cite{dunson_NonparametricBayesianModels_2011} & dependent PSBP &  &  & C &  &  \\
        \cite{chung_NonparametricBayesConditional_2009} & PSBP & C &  & C & \checkmark & \href{https://github.com/david-dunson/probit-stick-breaking}{GitHub} \\
        \cite{park_BayesianGeneralizedProduct_2010} & GPPM & Cd &  & CD &  &  \\
        \cite{corradin_BNPmixPackageBayesian_2021} & PYMM & Cd &  &  &  & \texttt{\href{https://www.jstatsoft.org/article/view/v100i15}{BNPmix}} \\
        \hline
        &  &  &  &  &  &  \\
        \textbf{Univariate non-mixture models} & name & \multicolumn{3}{c}{covariates} & vs & code \\
        \hline
        \cite{muller_BayesianCurveFitting_1996} & WDDP & \multicolumn{3}{c}{C}  &  & \texttt{\href{https://www.jstatsoft.org/article/view/v040i05}{DPpackage}} \\
        \cite{taddy_BayesianNonparametricApproach_2010} &  & \multicolumn{3}{c}{CD} &  &  \\
        \cite{norets_BayesianModelingJoint_2012} &  & \multicolumn{3}{c}{CD} &  &  \\
        \cite{tokdar_BayesianDensityRegression_2010} & spLGP & \multicolumn{3}{c}{C} &  & \href{https://www2.stat.duke.edu/~st118/Software/}{Website} \\
        \cite{payne_ConditionalDensityEstimation_2020} &  & \multicolumn{3}{c}{C} &  & \href{https://github.com/richardbayes/bayes-cde}{GitHub} \\
        \cite{jara_ClassMixturesDependent_2011} & DTFP & \multicolumn{3}{c}{C} &  & \texttt{\href{https://www.jstatsoft.org/article/view/v040i05}{DPpackage}} \\
        \cite{trippa_MultivariateBetaProcess_2011} & MMPT & \multicolumn{3}{c}{CD} &  &  \\
        \cite{kundu_LatentFactorModels_2014} &  & \multicolumn{3}{c}{CD} &  &  \\
        \cite{hothorn_ConditionalTransformationModels_2014} & CTM & \multicolumn{3}{c}{CD} &  & \texttt{\href{https://cran.r-project.org/web/packages/mlt/index.html}{mlt}} \\
        \cite{hothorn_PredictiveDistributionModeling_2021} &  & \multicolumn{3}{c}{CD} & \checkmark & \texttt{\href{https://cran.r-project.org/web/packages/trtf/index.html}{trtf}} \\
        \cite{shen_AdaptiveBayesianDensity_2016} &  & \multicolumn{3}{c}{C} & \checkmark &  \\
        \cite{barrientos_FullyNonparametricRegression_2017} & LDBPP & \multicolumn{3}{c}{Cd}  &  & \texttt{\href{https://www.jstatsoft.org/article/view/v040i05}{DPpackage}} \\
        \cite{ma_RecursivePartitioningMultiscale_2017} & cond-OPT & \multicolumn{3}{c}{CD} & \checkmark  & \texttt{\href{https://github.com/MaStatLab/PTT/tree/master}{PTT}} \\
        \cite{horiguchi_TreePerspectiveStickBreaking_2025} &  & \multicolumn{3}{c}{C} &  &  \\
        \cite{orlandi_DensityRegressionBayesian_2021} & DR-BART & \multicolumn{3}{c}{Cd} & \checkmark & \texttt{\href{https://github.com/vittorioorlandi/drbart}{drbart}} \\
        \cite{li_AdaptiveConditionalDistribution_2023} & SBART-DS & \multicolumn{3}{c}{Cd} & \checkmark & SM \\
        \cite{lohse_ContinuousOutcomeLogistic_2017} & Colr & \multicolumn{3}{c}{CD} &  & \texttt{\href{https://cran.r-project.org/web/packages/mlt/index.html}{mlt}} \\
        \hline
        &  &  &  &  &  &  \\
        \textbf{Multivariate models} & name & covariates & copula & density & vs & code \\
        \hline
        \cite{dao_FlexibleMultivariateRegression_2021} & MRDE-MN & Cd &  & \checkmark & \checkmark & Author \\
        \cite{pitt_EfficientBayesianInference_2006} &  & Cd & \checkmark &  &  &  \\
        \cite{song_JointRegressionAnalysis_2009} & VGLM/GCR & Cd & \checkmark &  & \checkmark &  \\
        \cite{yee_VectorGeneralizedLinear_2015} & VGAM & Cd & (\checkmark) &  & \checkmark & \texttt{\href{https://cran.r-project.org/web/packages/VGAM/index.html}{VGAM}} \\
        \cite{kock_TrulyMultivariateStructured_2024} & TM-SADR & Cd & \checkmark &  &  & SM \\
        \cite{klein_MultivariateConditionalTransformation_2022} & MCTM & Cd & (\checkmark) & \checkmark &  &    \texttt{\href{https://cran.r-project.org/web/packages/tram/index.html}{tram}} \\
        \cite{sarkar_BayesianSemiparametricCovariate_2022} &  & D & \checkmark & \checkmark & \checkmark & SM \\
        \hdashline\noalign{\vskip 1pt}
        \cite{zellner_EfficientMethodEstimating_1962} & SUR & Cd & (\checkmark) &  &  & \texttt{\href{https://cran.r-project.org/web/packages/systemfit/index.html}{systemfit}} \\
        \cite{wang_SparseSeeminglyUnrelated_2010} & Sparse SUR & Cd & (\checkmark) &  & \checkmark &  \\
        \cite{klein_BayesianStructuredAdditive_2015} & GAMLSS & Cd &  &  & \checkmark & \texttt{\href{https://cran.r-project.org/web/packages/BayesX/index.html}{BayesX}} \\
        \cite{alexopoulos_BayesianVariableSelection_2021} & BVSGCR & Cd & \checkmark &  & \checkmark & \texttt{\href{https://github.com/lb664/BVS4GCR}{BVS4GCR}} \\
        \cite{muschinski_CholeskybasedMultivariateGaussian_2024} &  & Cd &  &  &  & \texttt{\href{https://github.com/meteosimon/mvnchol}{mvnchol}} \\     
        \cite{gioia_AdditiveCovarianceMatrix_2024} &  & Cd &  &  & \checkmark & \href{https://zenodo.org/records/13330081}{Zenodo} \\
        \hline
        Proposed approach & Flower Model & D & \checkmark & \checkmark & \checkmark & SM \\
        \hline
    \end{tabular}
    \caption{
    Relevant approaches for density regression, highlighting their supported covariate types (`covariates'), variable selection capability (`vs'), copula structure (`copula'), suitability for marginal density estimation (`density'), and code availability (`code').
    Nature of covariates is described as C=continuous, D=discrete, d=discrete but included as dummy variables; `(\checkmark)' denotes models that are not copula-based but can be seen as such.
    Approaches with `SM' have code provided as part of the supplementary materials; `Author' indicates code may be accessible upon request; web links point to publicly hosted code repositories.
    Within each block, methods are reported in order of appearance in the manuscript.}
    \label{tab:models_mix}
\end{table}

\clearpage
\putbib[references,references_implementation,references_nutrition]
\end{bibunit}

\end{document}